# Олександр ВАШКІВ

# ОСНОВНІ ВИРОБНИЧІ ФОНДИ ПІДПРИЄМСТВ ВАНТАЖНОГО АВТОТРАНСПОРТУ: ПРОБЛЕМИ ЕФЕКТИВНОГО ВИКОРИСТАННЯ

**Вашків О. Основні виробничі фонди підприємств вантажного автотранспорту: проблеми ефективного використання.** – Тернопіль: Економічна думка, 1999. – 172 с.

У мронографії аналізуються величина і структура основних виробничих фондів вантажних автотранспортних підприємств та особливості їх відтворення в умовах трансформації суспільно-економічних відносин. На базі статистичного матеріалу вантажних АТП Івано-Франківської, Львівської та Тернопільської областей з'ясовується вплив виробничих факторів на рівень фондовіддачі основних фондів, виявляються резерви її підвищення. Вмотивовано необхідність адаптивних якісних змін в управлінні і реалізації виробничого потенціалу та активізації інновацій у сфері вантажних автоперевезень, дано науково обгрунтовані рекомендації щодо підвищення ефективності використання основних виробничих фондів вантажних автогосподарств в сучасних економічних умовах.

This work investigates the structure of basic industrial funds of cargo motor transport enterprises and peculiarities of the processes of their reproduction in the conditions of social and economic relations transformation. On the basis of statistic data of cargo-motor transport enterprises of Ivano-Frankivsk, Lviv and Ternopil regions the author investigates the effect of production factors on the level of capital productivity of the basic funds, he determines reserves of its increase. The author motivates the necessity of adaptive qualitative changes in the management and realization of industrial potential and innovations activization in the sphere of cargo motor transportations, scientiffically grounded recommendations for efficiency increase of the usage of basic industrial funds of cargo motor transportation enterprises in modern economic conditions are provided in this work.

# ВСТУП

Одним із важливих напрямів підвищення ефективності суспільного виробництва в умовах переходу України до ринкових відносин є здійснення збалансованої економічної політики. Екномічні перетворення в державі спрямовані на розв'язання конкретних завдань стабілізації та відновлення економічного зростання, пошук шляхів інтенсифікації виробництва, більш раціональне використання виробничого потенціалу.

Матеріальною передумовою зростання продуктивності суспільної праці, збільшення обсягів виробництва сукупного суспільного продукту і національного доходу України є підвищення ефективності використання основних виробничих фондів.

Специфіка і труднощі розвитку українського суспільства в посткомуністичну епоху вимагають постійного і наполегливого опрацювання концептуальних засад та практичних кроків у реформуванні економіки, впровадженні нетрадиційних форм і методів оновлення та розвитку виробничої бази підприємств різних форм власності, модифікації їх основних виробничих фондів. В сучасних умовах особливо зростає, набуваючи специфічних рис, діяльність вантажного автотранспорту. Об'єктивно постає проблема ефективного використання його основних виробничих фондів, які, власне, і досліджуються у даній роботі на основі статистичних матеріалів про результати діяльності 97 підприємств вантажного автотранспорту Івано-Франківської, Львівської та Тернопільської областей.

Проблема підвищення ефективності використання основних виробничих фондів вантажного автомобільного транспорту через особливості виробничого процесу цієї галузі та її роль у розвитку народного господарства України набуває сьогодні особливої актуальності. Автомобільний транспорт України, зв'язуючи всі галузі народного господарства в єдиний комплекс, перетворився нині у найбільш масовий вид транспорту, який за обсягом перевезень вантажів і пасажирів не має собі рівних. Від його діяльності в значній мірі залежить ефективність роботи всіх інших галузей народного господарства.

Питання ефективного використання основних виробничих фондів є актуальним, а стосовно підприємств вантажного автотранспорту недостатньо дослідженим. Опубліковані праці, переважна більшість з яких є принагідними, не дають підстав робити висновки про вичерпаність задекларованої проблеми. Вчені-економісти (Настенко А., Васіна Т., Бородюк В., Орлов П., Остапенко В., Рудченко О., Омельянчик Н., Тютюн М., Супруненко В., Фукс А. та ін.) єдині в розумінні необхідності її розв'язання, проте усі спроби, здійснені на сьогоднішній день, не можна вважати достатніми на рівні теорії, а тим більш практики. До того ж майже всі праці (Благоразумової Н., Бистрицької А., Кривошапова І., Каплана Т., Голованенка С., Губіна В., Голобородкіна Б., Попченка Я., Луцкера Г. та інших.) в галузі вантажного автотранспорту, що донедавна вважались фундаментальними, на сьогоднішній день певною мірою втратили свою актуальність.

Важливість і актуальність проблеми мобілізації всієї сукупності технічних, організаційних та економічних заходів щодо підвищення ефективності використання основних виробничих фондів значно зростає в сучасних умовах. Перед практикою постають нові завдання, що вимагають розробки та впровадження комплексу заходів, спрямованих на підвищення фондовіддачі. Нерозробленість питань щодо ефективності використання основних виробничих фондів вантажних АТП в умовах перехідного періоду, необхідності впровадження інновацій, диверсифікації методів та засобів управління транспортним процесом є очевидною. Тому не зважаючи на всі об'єктивні і суб'єктивні чинники підвищення ефективності використання основних виробничих фондів, з точки зору теорії і практики поставлена проблема є актуальною і потребує свого розв'язання.

Розроблені у монографії теоретичні питання, що стосуються вдосконалення класифікації основних фондів, методології побудови головного оціночного показника роботи вантажного автотранспорту, коефіцієнтів оновлення та вибуття основних фондів, способу нарахування амортизації рухомого складу, систематизації принципів удосконалення транспортного обслуговування мають певну наукову і практичну цінність. Їх впровадження та використання збагатить наукову методологію вивчення ефективності використання основних виробничих фондів, сприятиме прискоренню інноваційних процесів, отриманню більш достовірних даних про діяльність автомобільного транспорту.

Виявлені у результаті дослідження резерви можуть бути покладені в основу екстраполяції і прогнозування зростання фондовіддачі основних виробничих фондів автомобільного транспорту на перспективу. Коефіцієнти множинної регресії дають можливість об'єктивно оцінити виробничу діяльність окремих АТП, співставити показники фондовіддачі між окремими автогосподарствами, в цілому по областях чи регіонах. Разом з тим вони дозволяють виявити ступінь коливності рівнів фондовіддачі, зумовлених неоднаковою ефективністю використання наявних виробничих чинників окремими автотранспортними підприємствами.

# РОЗДІЛ 1
# ОСНОВНІ ВИРОБНИЧІ ФОНДИ АВТОПІДПРИЄМСТВ ВАНТАЖНОГО ТРАНСПОРТУ І ОСОБЛИВОСТІ ЇХ ВІДТВОРЕННЯ В УМОВАХ РЕФОРМУВАННЯ ЕКОНОМІКИ УКРАЇНИ

## 1.1. Реструктуризація виробничого потенціалу автотранспортної системи України в умовах становлення ринкових відносин

Особливістю нинішнього етапу економічного розвитку України є її перехідний стан від командно-адміністративної економіки до економіки ринкового типу. Одним із важливих напрямів підвищення ефективності суспільного виробництва в цих умовах є здійснення збалансованої економічної політики, спрямованої на розв'язання конкретних завдань стабілізації та відновлення економічного зростання, пошуки шляхів інтенсифікації виробництва, більш раціонального використання виробничого потенціалу, повсюдної економії всіх видів ресурсів, підвищення якості роботи.

Матеріальною передумовою зростання продуктивності суспільної праці, збільшення обсягів виробництва сукупного суспільного продукту і національного доходу України є підвищення ефективності використання основних виробничих фондів.

Проблема підвищення ефективності використання основних виробничих фондів вантажного автомобільного транспорту через особливості виробничого процесу цієї галузі та її роль у розвитку народного господарства України набуває сьогодні особливої актуальності. Автомобільний транспорт України перетворився нині у найбільш масовий вид транспорту, який за обсягом перевезень вантажів і пасажирів не має собі рівних. Тому цілком зрозумілим і виправданим є постійний інтерес представників економічної науки до проблеми ефективного використання основних виробничих фондів загалом і у сфері вантажного автотранспорту зокрема. Ця проблема досліджується на рівні ґрунтовних монографічних праць (Ребров Ю.В. «Проблемы использования основных производственных фондов», Лемпа С.В., Полторигін В.К. «Повышение эффективности использования основных производственных фондов», Волков Ф.М., Кашкін І.Т. «Повышенние эффективности основных производственных фондов в процессе интенсификации»), так і спеціальних журнальних статей Настенка А., Васіної Т., Бородюка В., Орлова П., Остапенка В., Рудченка О., Омельянчика Н., Тютюна М., Супруненка В., Фукса А. та ін.

Усвідомлення необхідності і пошук шляхів ефективного використання основних виробничих фондів у галузі вантажного автотранспорту у різний час перебували в полі зору таких вчених, як Благоразумова Н.І., Бистрицька А.К., Кривошапов І.Г. («Эффективность использования производственных фондов автомобильного транспорта»), Каплан Т.Л. («Пути повышения эффективности использования основных фондов на автомобильном транспорте»), Голованенко С.Л., Губін В.А., Голобородкін Б.М. («Эффективность использования материально-технической базы автотранспорта»), Попченко Я.А., Луцкер Г.Д. («Пути повышения эффективности грузовых автомобильных перевозок»). Ці вчені досліджували ефективність використання основних виробничих фондів в умовах планової економіки.

З повагою ставлячись до висновків, зроблених авторами згаданих праць, мусимо, проте, враховувати і фактор часу, і, що значно важливіше, зміну суспільно-економічних відносин, яка змінила й умови функціонування вантажного автотранспорту в Україні. Причому цей процес далекий від стабілізації ще й сьогодні. Саме тому об'єктивно виникає проблема нового підходу у використанні основних виробничих фондів. Опираючись на досягнення представників як вітчизняної, так і зарубіжної наукової думки у цій сфері, мусимо зважати на означені чинники. Проблеми ефективного використання основних фондів на сьогодні гостро стоять у всіх сферах економіки України.

Суттєвий спад обсягів промислового виробництва в Україні логічно призвів до неминучого скорочення обсягів вантажних перевезень. Аналіз сучасного парку вантажних автомобілів України [див., напр., 81; 161; 201; 203], що склався внаслідок технічної політики колишнього державтопрому, засвідчує серед його основних недоліків низькі технічні характеристики (надійність, паливна економічність, екологічність) та неоптимальність структури за вантажопідійма́льністю та типами рухомого складу.

Автомобілі вантажопідіймальністю 2,0 – 5,0 т складають сьогодні більшу частину автопарку, тоді як оптимальна їх кількість не повинна перевищувати 20 %. Не вистачає автомобілів вантажопідіймальністю до 2,0 т: їх сьогодні лише 18 % при потребі – 40 %. У розвинених країнах автомобілі цього класу складають основну частину рухомого складу. У Франції, наприклад, їх 88 % від загальної кількості парку, в Японії – 82 %, у США – 85 %, Великобританії – 75%, Італії – 68 %, Німеччині – 60 % [див.: 201, с. 2].

Невідповідність складу вантажного автотранспорту сучасній структурі попиту на автоперевезення виявляється і за типом кузова. Бортові автомобілі в Україні складають 39,2 % при потребі – 25 %, самоскидні – 35 %, а потрібно – 30 %, фургони – 14,3 %, що на 12 % менше від потреби, рефрижератори біля 1 % при нинішній потребі – 20 %.

Повільно здійснюється дизелізація вантажного автомобільного парку. Так, для автомобілів вантажопідіймальністю понад 2,0 т таких лише 22 %, тоді як у Німеччині і Франції вантажівки такого класу майже всі обладнані дизельними двигунами.

Враховуючи те, що транспорт в Україні споживає 30 % усього рідкого палива, то переведення автомобілів на дизельні двигуни дозволить знизити споживання нафти у 4 – 7 разів [див.: там само, с. 2 – 3].

Необхідність реформуваня виробничої бази та удосконалення методів організації транспортного обслуговування зумовлена й інтеграційними процесами, які відбуваються у сфері транспорту та геополітичним і соціально-економічним положенням України. Активізація взаємовигідного співробітництва з іншими країнами є для України важливим питанням, оскільки вона знаходиться на перетині головних транспортних напрямків Євроазійського континенту, а сформовані широтні і меридіальні транспортні шляхи мають тисячолітню історію. Однак, забезпеченість України автомобільними шляхами ще низька. «Щоб досягти забезпеченості автошляхами рівня європейських країн, необхідно побудувати близько 200 тис. км шляхів, у т. ч. 60 % у сільській місцевості» [161, с. 31]. Частково цьому сприятиме реалізація прийнятої 12.03.1997 р. Кабінетом

Міністрів України Програми створення і функціонування національної мережі міжнародних транспортних коридорів [див.: 188, ст. 588; 70, с.33]. Користь від будівництва нових автомагістралей для України більш, ніж очевидна. Це – до 1 мільйона нових робочих місць; у сфері автоперевезень підвищиться швидкість руху транзитних вантажівок, яка сьогодні не перевищує 16 км/год., тоді як у Європі – 72 км/год. З огляду на те, що Україна за дослідженням англійського інституту РЕНДЕЛ, проведеним у 1995 р., має найвищий серед держав Європи коефіцієнт транзитності території, який дорівнює 3,11 (у Польщі він складає 2,72), реалізація цієї програми дозволить Україні додатково отримати 6 – 7 млрд. американських доларів прибутку від транзиту [див.: 11, с. 4]. «Транспортна система України, – як слушно зауважив М. Габрель, – має розвиватись, трансформуючись із регіональної (якою вона була у складі народно-господарського комплексу колишнього СРСР) в організаційно, фінансово і юридично самостійну національну систему, що тісно інтегрується в загальноєвропейські транспортні комунікації» [45, с. 120]. Виходячи з цих реалій, поліпшення транспортних зв'язків України з іншими державами, підвищення рівня конкурентоспроможності вітчизняних автопідприємств на міжнародному ринку транспортних послуг і подолання залежності України від іноземних перевізників є наріжним каменем сучасної політики держави у сфері вантажних автоперевезень. Одним з перших кроків у цьому напрямку було затвердження постановою Кабінету Міністрів України від 11.07.1994 р. № 475 «Положення про національного автомобільного перевізника». Серед його основних вимог, дотримання яких є необхідним для здобуття автотранспортним підприємством статусу національного перевізника, є вимога щодо підвищення якісних характеристик автомобілів. Відповідно до ст. 2 цього положення не менше 50 % рухомого складу автогосподарств повинні становити автомобілі, що відповідають міжнародним технічним вимогам [див.: 170]. На забезпечення належного рівня виконання транспортної роботи вітчизняними автопідприємствами спрямовані заходи Уряду України, що знайшли своє відображення в прийнятті ряду законодавчих та нормативних актів, серед яких вдосконалення системи ліцензування внутрішніх і міжнародних автоперевезень вантажів [див.: 184], створення безпечних умов для перевезень автомобільним транспортом [див.: 172; 182] та інші [див.: 87; 103].

На сьогодні в Україні за даними Асоціації міжнародних автоперевізників кількість офіційно зареєстрованих перевізників перевищує 790 (у 1991 р. їх було понад 40). Найбільшу частку учасників (87,1 %) складають невеликі автопідприємства, парк рухомого складу яких нараховує від 1 до 20 вантажівок [див.: 58, с. 79]. «В основному це приватні компанії, що працюють за довгостроковими угодами з постійними замовниками, причому немалу частку серед них займають фірми – представництва крупних міжнародних корпорацій, дочірні підприємства іноземних компаній або українські підприємства зі 100 %-ним іноземним капіталом» [50, с. 72]. Однак серед них слабо представлені крупні підприємства, автопарк яких складає понад 50 вантажівок (2,7 %) [див.: 58, с. 79]. В основному це колишні державні підприємства, що на сьогодні акціоновані, і які успадкували не тільки застарілу матеріальну базу, але й негативний авторитет «Совтрансавто». Господарська діяльність останнього «не була не тільки досконалою, але й достатньо ефективною, причому як у структурному, так і в організаційно-управлінському відношеннях» [211, с. 80]. Як наслідок, частка вантажів, перевезених у міжнародному сполученні досліджуваними автогосподарствами Івано-Франківської, Львівської та Тернопільської областей, за підсумками 1996 р. заледве перевищує 1 % від загального обсягу перевезень.

Поряд з тим існуюча щільність зовнішньоекономічних зв'язків далеко не завжди є свідченням їх доцільності і раціональності. Так, наприклад, з позицій сьогодення простежується неефективність підходів у колишньому СРСР щодо місця транспорту в економічній системі і принципів формування транспортних зв'язків. І це не тільки прорахунки, які є практично неминучими за умови централізації планування. Основна їх причина крилась у відмежованості і негнучкості правової бази перевезень, з допомогою якої суворо контролювались організаційно-технологічні зв'язки транспортного комплексу. «Міжнародні автомобільні перевезення в СРСР монопольно протягом останніх двадцяти п'яти років здійснювались через систему «Совтрансавто» Міністерства автомобільного транспорту РРФСР. Підприємства інших республік лише залучались до перевезень» [99, с. 4].

Структурні зрушення, які сьогодні спостерігаються в транспортній системі України, значною мірою зумовлені структурною перебудовою автомобільного транспорту, де найбільш активно відбувався процес приватизації.

Зараз в системі вантажних автоперевезень країни поряд з державними підприємствами з'явилися автогосподарства, власниками яких є акціонери, окремі громадяни, орендарі. В основному «перевагу надано акціонерним товариствам відкритого чи закритого типу» [2, с. 20]. В Івано-Франківській, Львівській і Тернопільській областях питома вага вантажних автомобілів, що знаходяться у колективній власності, зросла протягом 1993 – 1997 рр. з 36,5 до 65,9 %; у державній – скоротилася з 56,8 до 20,2 % (див. рис. 1.1.1).

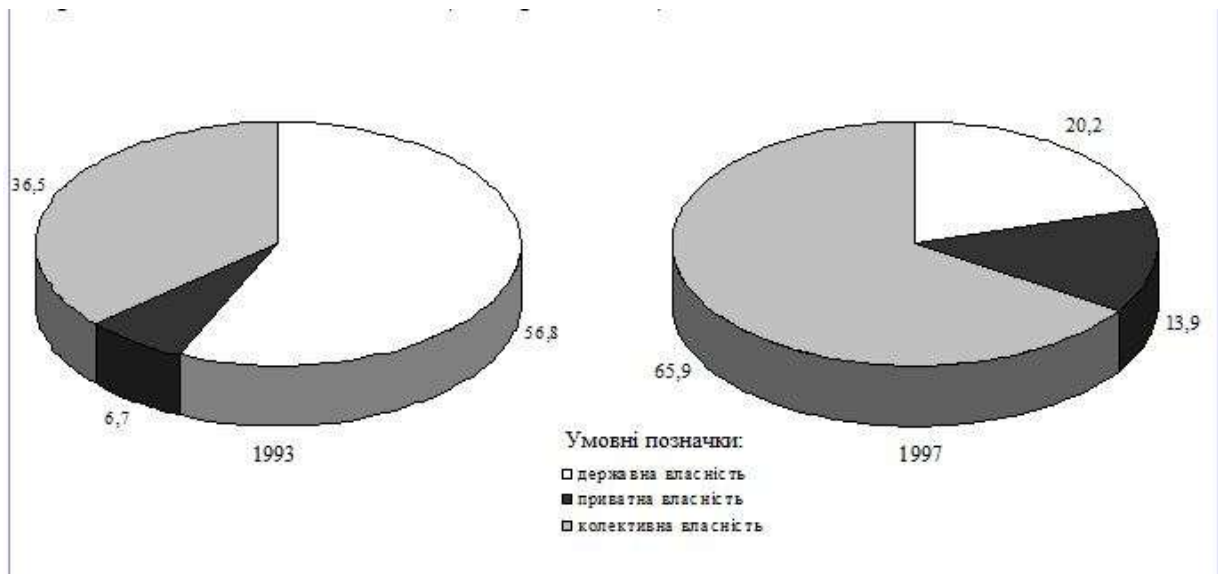

Рис. 1.1.1. Структурна характеристика парку вантажних автомобілів за формами власності в Івано-Франківській, Львівській та Тернопільській областях, %

Зокрема, лише в Тернопільській області кількість вантажівок, що перебувають у приватній власності, протягом цього ж періоду зросла з 1257 до 3423 одиниць. Всі ці процеси «значно розширили конкурентне середовище на ринку транспортних послуг, хоча проте і не покращили їх якості» [69, с. 3]. Державна програма приватизації автомобільного транспорту, яка за задумом повинна була кардинально оздоровити галузь, поки що з цим завданням не справилась.

Приватизовані та державні автотранспортні підприємства сьогодні працюють малоефективно. Автогосподарства зіткнулися з проблемами адаптації до ринкових відносин. Для автопідприємств характерною стала наявність вільних, незавантажених замовниками потужностей. Разом з тим сама по собі зміна форми власності чи то загалом в економіці України, чи у сфері вантажних автомобільних перевезень не забезпечує підвищення ефективності виробництва, оскільки здійснювана «приватизація не вирішує головного завдання, яке бачиться у поєднанні праці і власності» [76, с. 43] і сам процес приватизації «мало пов'язаний з якісною зміною управління об'єктами, що приватизуються» [52, с. 60]. В результаті «колективні інтереси в цих умовах втратили свої пріоритети і перестали бути стимулом для поліпшення виробничо-господарської діяльності» [23, с. 55]. У більшості випадків приватизовані АТП не змінюють кардинальним чином своєї економічної поведінки. Це стосується тарифної, маркетингової, виробничої політики. Не ставлять собі за мету максимізацію прибутку і його реінвестування. Тобто, можна стверджувати, що в силу об'єктивних і суб'єктивних причин на сьогодні «мікроекономіка лишається переважно соціалістичною – з браком властивих для ринкових (відповідно до змін) систем реакцій пристосування відносних цін, співвідношень попиту і пропозиції тощо» [210, с. 65]. Економічні результати роздержавлення і приватизації в автотранспортній галузі мало чим відрізняються від загальних результатів в Україні. І хоча за попередніми підсумками 1997 р. «вперше вдалося припинити спад обсягів вантажних перевезень і навіть одержати деякий приріст, близько 1 %», що відбулося «в першу чергу, завдяки розширенню ринку автомобільних послуг» [173, с. 6]. Та це ще не є свідченням докорінних якісних змін в управлінні та реалізації виробничого потенціалу вантажних автотранспортних підприємств України.

Підвищення ефективності транспортного процесу можливе за рахунок заміни застарілого рухомого складу і обладнання на нове, використання більш прогресивних технологій і форм обслуговування, поглиблення маркетингових досліджень, вдосконалення управління. Усі ці процеси є тривалими в часі і потребують значних додаткових інвестицій, а, отже, їх вирішення вимагає більш глибокого усвідомлення та врахування тих чинників, що формують результативність виробничо-господарської діяльності.

У розрізі цього кола питань одним з чинників, а одночасно й матеріальною основою у забезпеченні ефективності використання потужностей вантажних АТП є основні виробничі фонди. Найбільш повне їх використання, і в першу чергу – рухомого складу, устаткування майстерень та зон технічного обслуговування, а також виробничих площ сприяє поліпшенню кількісних і якісних характеристик вантажних перевезень без додаткових капітальних вкладень. Підвищення рівня використання основних виробничих фондів позначається позитивно в усіх сферах діяльності автогосподарств, охоплюючи найрізноманітніші її аспекти: від соціально-психологічних до виробничо-господарських.

При вивченні питань підвищення ефективності використання основних виробничих фондів одним з найбільш важливих завдань є аналіз величини їх вартості. Звертаючись до нього, слід враховувати вплив макроекономічного середовища на формування величини загальної вартості основних фондів. У сучасних умовах безпосереднє здійснення такого аналізу утруднюється існуючою в Україні невідповідністю між реальною і фактичною вартістю основних засобів. Це, в свою чергу, стало однією з головних причин порушення інвестиційних процесів та відтворення основних фондів.

Так, на кінець 1990 р. загальна вартість основних виробничих фондів досліджуваних автотранспортних підприємств Тернопільської області (в перерахунку на долари США) становила за офіційним курсом Національного банку України (1 амер. дол. = 0,64 руб.) 63,53 млн. амер. дол., а за ринковим і обмінним інтуристівським курсом (1 амер. дол. = 4 руб.) – 10,16 млн. амер. дол., що, на думку українського економіста В. Бородюка, і відповідало приблизно їх реальній вартості [див.: 26, с. 8 – 9].

Проте на кінець 1992 р. загальна вартість основних виробничих фондів з урахуванням їх переоцінки становила за офіційним курсом Національного банку України (1 амер. дол. = 637 укр. крб.) уже 0,95 млн. амер. дол., а за неофіційним ринковим (1 амер. дол. = 750 укр. крб.) – 0,807 млн. амер. дол., тобто при розрахунках за офіційним курсом вона зменшилася порівняно з 1990 р. майже у 66,9 раза, а за ринковим курсом – у 12,6 раза, що не може не викликати сумніву в її об'єктивності.

На кінець 1993 р. загальна вартість основних виробничих фондів вказаних вантажних автотранспортних підприємств з урахуванням переоцінки станом на 1 серпня цього ж року за офіційним курсом Національного банку України (1 амер. дол. = 12610 укр. крб.) становила 2,76 млн. амер. дол., а за ринковим (1 амер. дол. = 32000 укр. крб.) – 1,1 млн. амер. дол. Отже, за офіційним курсом вона збільшилася порівняно з 1992 р. у 2,9 раза; за неофіційним – у 1,36 раза. Однак, зіставляючи з даними 1990 р., загальна вартість основних виробничих фондів АТП Тернопільської області зменшилася відповідно у 23 та у 9,2 раза за офіційним і ринковим курсами долара США.

Відтак, найменше значення величини загальної вартості основних виробничих фондів (у доларах США) склалося на початок 1995 р. (див. рис. 1.1.2), оскільки від серпня 1993 р. до лютого 1995 р. індексація їх балансової вартості не здійснювалася.

«Заниження реальної вартості основних фондів і несистематичне проведення їх індексації в умовах інфляції призвели до істотного погіршення їх відтворення у зв'язку з недорахуванням значних обсягів амортизації та невиконанням нею своїх функцій» [26, с. 9]. Детальніше це питання розглядається у третьому параграфі цього розділу.

В умовах загального спаду виробництва різко скорочується потреба в перевезеннях вантажів, зростає незатребуваність рухомого складу автогосподарств. Утримувати надлишкові виробничі потужності в умовах складного фінансового становища стає недоцільно. За підсумками 1996 р. парк вантажівок автотранспортних підприємств України скоротився на чверть [див.: 217, с. 8]. Скорочення реального валового внутрішнього продукту на кінець 1997 р. сягнуло 31,1 % від рівня 1990 р. і зумовлене цим зменшення обсягів перевезень до 27,6 % від рівня того ж року (див. табл. 1.1.1 і рис.1.1.3) стало однією з головних причин, в результаті якої загальна вартість основних виробничих фондів в порівняльних цінах протягом вказаного періоду знизилася на 35 – 40 %.

Величина вартості основних виробничих фондів автотранспортних підпри-

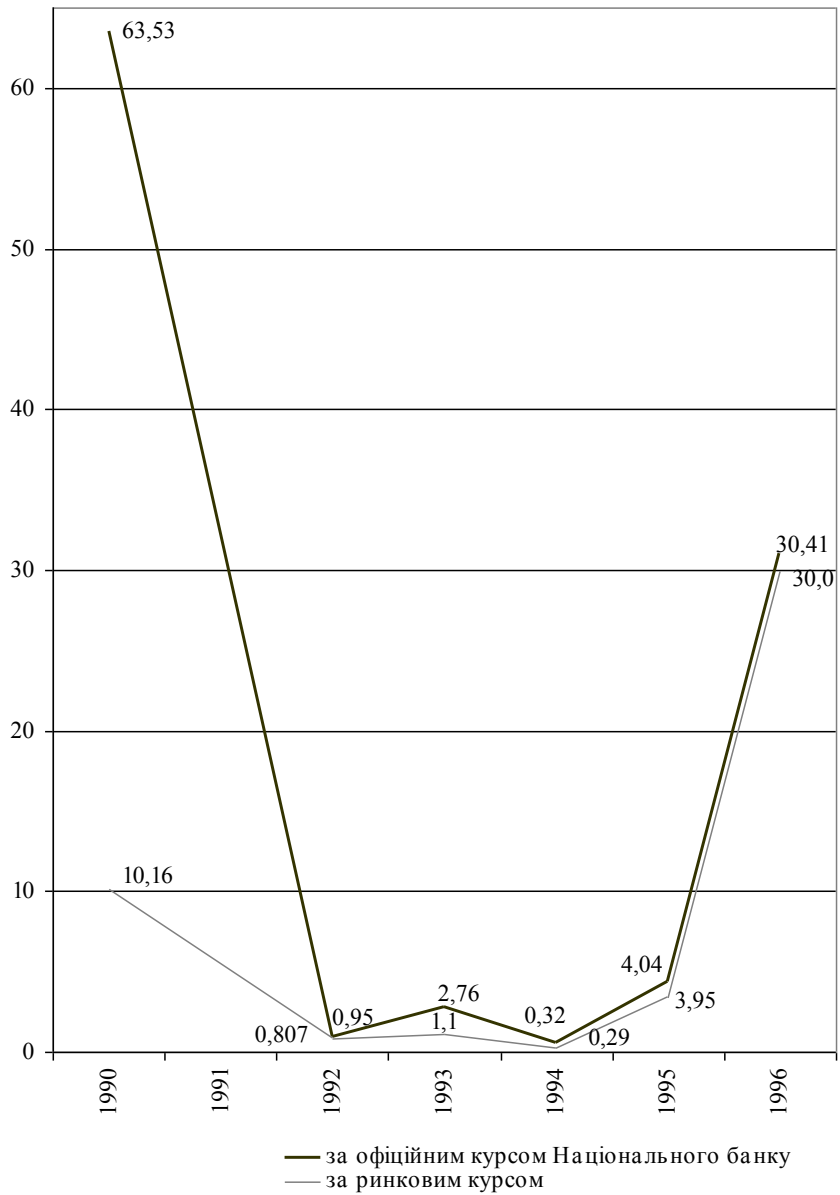

Рис. 1.1.2. Коливання величини вартості основних виробничих фондів вантажних автотранспортних підприємств Тернопільської області у перерахунку на долари США за офіційним та ринковим курсами ємств формується під впливом різних чинників. До них можна віднести: кількість і вартість одиниць рухомого складу, виробничого і допоміжного устаткування, будівель, споруд, наявність виробничих потужностей автотранспортних підприємств для організації технічного обслуговування, ремонту і зберігання автомобільного парку тощо.

**Таблиця 1.1.1**

Динаміка реального валового внутрішнього продукту, обсягів перевезень вантажів та вантажообігу, % (1990 = 100 %)

| Показник | Роки | | | | | | | | | | | |
|---|---|---|---|---|---|---|---|---|---|---|---|---|
| | 1960 | 1970 | 1980 | 1985 | 1990 | 1991 | 1992 | 1993 | 1994 | 1995 | 1996 | 1997 |
| Темпи приросту реального ВВП* | ... | ... | ... | ... | 100,0 | 90,5 | 82,1 | 60,1 | 36,0 | 33,5 | 31,4 | 31,3 |
| Темпи приросту перевезень вантажів в Україні | 34,3 | 62,5 | 89,7 | 96,5 | 100,0 | 98,1 | 76,0 | 57,4 | 38,2 | 32,8 | 25,6 | 27,6 |
| в т. ч. по областях: Івано-Франківська | 34,9 | 69,2 | 111,6 | 119,5 | 100,0 | 94,1 | 81,3 | 68,2 | 47,4 | 43,2 | 29,5 | 29,0 |
| Львівська | 25,3 | 60,1 | 88,6 | 98,9 | 100,0 | 90,5 | 73,0 | 59,7 | 40,6 | 27,3 | 18,8 | 19,1 |
| Тернопільська | 28,2 | 58,3 | 87,8 | 101,2 | 100,0 | 89,4 | 74,7 | 49,1 | 37,4 | 35,4 | 26,0 | 26,1 |
| Темпи приросту вантажообігу | 22,2 | 52,2 | 88,0 | 89,8 | 100,0 | 98,5 | 81,1 | 71,8 | 44,3 | 38,9 | 27,9 | 27,7 |

\* До 1990 р. валовий внутрішній продукт не розраховувався.

Транспортні засоби є основним знаряддям праці в процесі отримання транспортної продукції. Незначна питома вага всіх інших основних фондів пояснюється частково характером виробництва транспортної продукції, яка не потребує ніяких інших засобів виробництва, крім рухомого складу.

Вартість виробничих основних фондів (без вартості транспортних засобів) в розрахунку на один середньосписковий автомобіль автотранспортного підприємства є показником, за допомогою якого характеризують їх технічну оснащеність. Середній рівень цього показника, що розрахований на базі різних автотранспортних підприємств, дає можливість зіставляти їх оснащеність.

Дані про вартість основних виробничих фондів (без вартості транспортних засобів) в розрахунку на один середньосписковий автомобіль вантажних автотранспортних підприємств Івано-Франківської, Львівської і Тернопільської областей, наведені в табл. 1.1.2, показують, що найбільш високий рівень технічної оснащеності спостерігається в автотранспортних підприємствах Тернопільської області, де у 1996 р. він складав в середньому 16,893 тис. грн., і який в 1,87 та

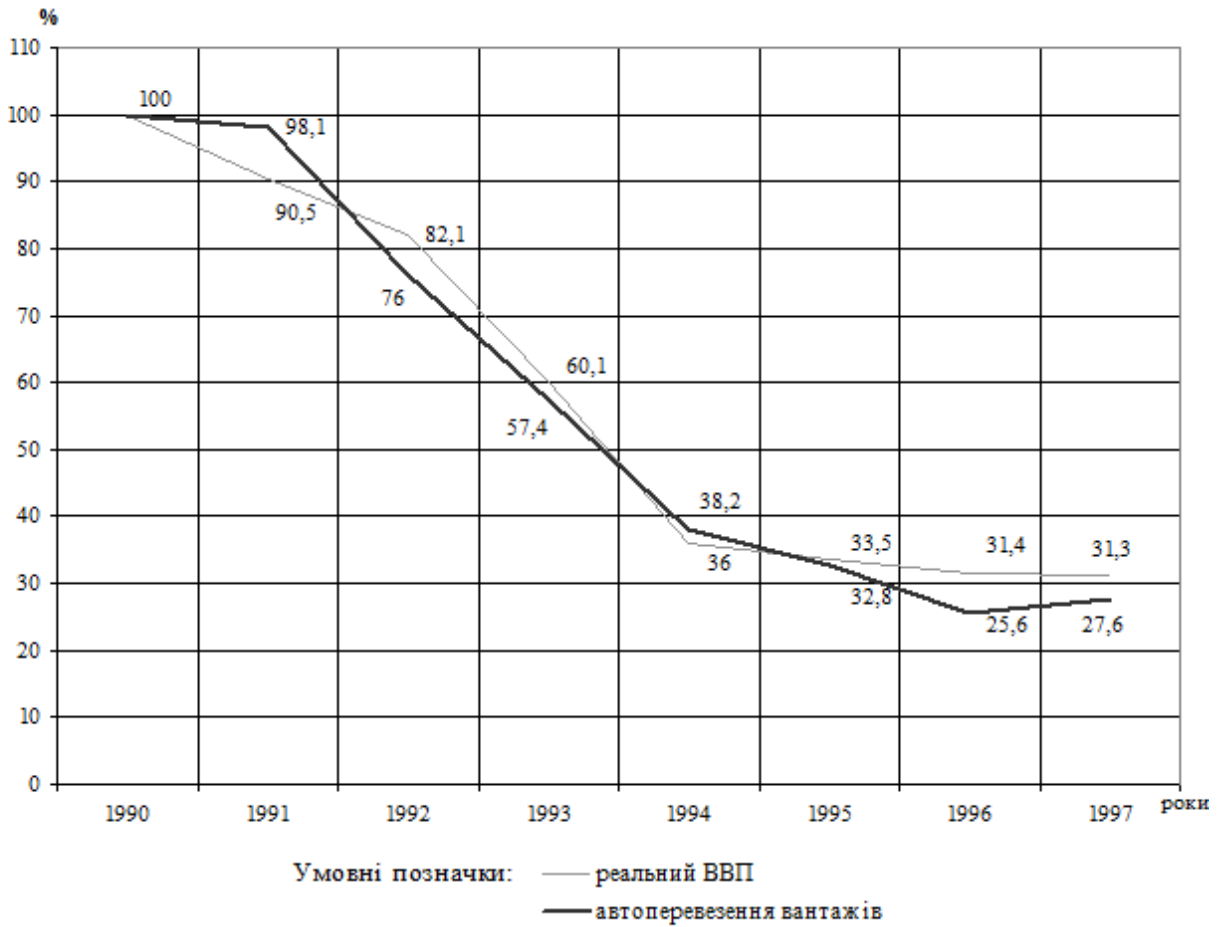

Рис. 1.1.3. Динаміка реального валового внутрішнього продукту та обсягів автоперевезень вантажів

**Таблиця 1.1.2**

Оснащеність вантажних автотранспортних підприємств
Івано-Франківської, Львівської та Тернопільської областей у 1996 р.,
тис. грн. в розрахунку на один вантажний автомобіль

| Область | В середньому по областях | Групи автогосподарств залежно від величини вартості основних фондів, тис. грн. | | | |
|---|---|---|---|---|---|
| | | до 1000 | 1001-4000 | 4001-7000 | 7001 і більше |
| Івано-Франківська | 9,006 | 3,205 | 14,844 | 10,571 | 17,290 |
| Львівська | 10,834 | 4,849 | 11,428 | 14,985 | 12,450 |
| Тернопільська | 16,839 | 6,170 | 22,517 | 47,007 | 10,158 |
| По регіону | 11,827 | 4,787 | 17,514 | 17,011 | 12,456 |

у 1,55 раза вищий від аналогічних в Івано-Франківській і Львівській областях. Це свідчить про високу концентрацію основних виробничих фондів та значні потенційні можливості автогосподарств області.

Звернемо увагу на те, що збільшення вартості основних виробничих фондів (без вартості транспортних засобів) на один середньосписковий автомобіль в автотранспортних підприємствах свідчить про ріст вартості пасивної частини основних фондів, а отже, підвищення технічної оснащеності автотранспортних підприємств, внаслідок чого постійно збільшуються виробничі потужності по ремонту і технічному обслуговуванню автомобілів, поліпшуються умови праці робітників, підвищується рівень механізації виробничих процесів.

Необхідно зауважити, що показник вартості основних фондів на один автомобіль залежить від ряду чинників, основними з яких є:

1. Середня кількість автомобілів, причепів і напівпричепів автотранспортних підприємств.

2. Середня вантажопідіймальність автомобіля.

3. Середньодобовий пробіг одного автомобіля.

4. Коефіцієнт випуску автомобілів на лінію і тривалість перебування їх на лінії.

5. Ступінь зношеності рухомого складу.

6. Спосіб організації технічного обслуговування і ремонту рухомого складу.

7. Відносна кількість місць стоянок автомобілів у закритих приміщеннях.

8. Організація і спосіб виконання будівельно-монтажних робіт.

9. Величина цін на будівельні матеріали, виробниче і силове устаткування.

На сьогодні немає достатньо обґрунтованої методики визначення нормативу вартості основних фондів на один автомобіль, внаслідок чого його величина береться з проектів автотранспортних підприємств, що недостатньою мірою відображає конкретні умови його роботи.

Природньо, що різний рівень оснащеності АТП забезпечує різний рівень ефективності використання основних фондів.

Аналізуючи динаміку розвитку основних виробничих фондів вантажних автотранспортних підприємств Івано-Франківської, Львівської та Тернопільської областей, неважко помітити, що зміни величини основних фондів різних автопідприємств, а тим більше різних галузей проходять непропорційно до змін кількості автомобілів у них. Це відбувалося через ряд причин, однією з яких є ступінь зношеності рухомого складу.

Зниження рівня оновлення рухомого складу вимагає від автотранспортних підприємств експлуатувати автомобілі понад амортизаційний строк їх служби, в той же час підвищення рівня зношеності автомобілів призводить до збільшення обсягу робіт, пов'язаних із їх технічним оглядом і ремонтом. Це, в свою чергу, вимагає розширення виробничих потужностей майстерень, профілакторіїв і утеплених стоянок, що приводить до зростання вартості основних виробничих фондів.

Питання впливу зношення і відтворення основних фондів на динаміку обсягу і структури основних виробничих фондів розглядаються у наступних параграфах.

## 1.2. Особливості структури основних виробничих фондів вантажних автотранспортних підприємств і її формування в умовах перехідної економіки

Показники динаміки основних виробничих фондів автотранспортних підприємств дають характеристику їхніх кількісних змін. Щодо якісних змін, які відбулись в основних фондах за певний період, то їх можна простежити за допомогою вивчення структури. Характеризуючи кількісно матеріально-речові елементи основних фондів і співвідношення між ними, ми розкриваємо структуру, аналіз якої дозволяє виявляти закономірності кількісних змін цих елементів і основних фондів в цілому.

Вивчення структури основних виробничих фондів автотранспортних підприємств проводиться на основі прийнятої офіційною статистикою класифікації, яка передбачає залежно від їх функціональної ролі в процесі виробництва поділ на такі групи:

I. Будинки; II. Споруди; III. Передавальні пристрої; IV. Машини і устаткування: силові машини і устаткування, вантажно-розвантажувальні

механізми, вимірювальні та регулювальні прилади, пристрої та лабораторне устаткування, обчислювальна техніка, засоби зв'язку і інші машини та устаткування; V. Транспортні засоби; VI. Інструменти; VII. Виробничий інвентар та приладдя; VIII. Господарський інвентар; IX. Інші основні фонди.

Наведену класифікацію можна використовувати при розв'язанні різних загальних завдань економічного дослідження.

Однак, вона не дозволяє детально вивчити основні фонди з точки зору їхньої ролі у виробничому процесі, створює певні труднощі в техніко-економічних розрахунках при вивченні показників технічної оснащеності підприємств і ефективності використання основних фондів. Її позиції не чітко визначають технологічну структуру основних фондів у вигляді співвідношення активної і пасивної частин, не забезпечують можливості вивчати галузеве походження і функціональний склад основних виробничих фондів. Перешкодою для розв'язання цих завдань є наявність у діючій класифікації укрупнених груп основних фондів. Так, у групу «Споруди», наприклад, крім об'єктів, що забезпечують умови виробництва (сховища, огорожі), включають і такі, які виконують активні функції основних фондів (функції знарядь праці), наприклад, гідроспоруди, газові і нафтові свердловини та інші.

Всі об'єкти, що входять у дану групу, є, як правило, результатом будівельно-монтажних робіт, виготовлені переважно з продукції чорної металургії та промисловості будівельних матеріалів. В цьому розумінні вони однорідні. Щодо їх призначення і речово-натурального складу, то вони надзвичайно різноманітні. До цієї групи включають і гірничопрохідні споруди (шахти, кар'єри, свердловини, тощо), і гідроспоруди (греблі, канали, дамби), і транспортні споруди (дороги залізничні і автомобільні, естакади тощо), і споруди, що виконують технічні функції, необхідні для виробничого процесу (сушильні печі, пропарювальні камери), що, по суті, як і силові та робочі машини є активною частиною основних фондів. В цю ж групу включають і об'єкти, що забезпечують умови виробництва (сховища, огорожі).

Наявність в групах діючої класифікації різних елементів основних фондів не дозволяють вивчати співвідношення їх активної і пасивної частин, технологічної структури, галузевого походження та прогресивних тенденцій.

На нашу думку, класифікація основних фондів повинна здійснюватися за схемою, яка пропонується нижче.

Класифікація основних виробничих фондів:

I. Будівлі:

    1) цехові;

    2) адміністративні;

    3) складські.

II. Споруди:

    1) гірничопрохідні (свердловини, кар'єри);

    2) гідроспоруди (греблі, канали, дамби);

    3) транспортні (автомобільні дороги, залізничні колії, мости).

    4) споруди, що виконують функції, необхідні для виробничого процесу (димові труби, сховища).

III. Передавальне устаткування:

    1) для передачі електричної енергії (повітряні і кабельні електромережі);

    2) для передачі теплової енергії (тепломережі, паропроводи);

    3) для передачі механічної енергії (трансмісії, транспортери і ін.).

IV. Силові машини і устаткування:

    1) енерґоґенеруюче устаткування, яке перетворює потенційну енерґію природних ресурсів в необхідний вид енергії (ґенератори, турбіни, двигуни внутрішнього згоряння);

    2) енерґоперетворювальне устаткування – пристрої для зміни параметрів енергії одного і того ж виду (трансформатори і ін.);

    3) приймачі енергії (електродвигуни, електропечі).

V. Робочі машини і устаткування:

    1) для механічної обробки (металорізальні верстати, преси);

    2) для термічної обробки (домни, мартени);

    3) для електрофізичної обробки (електрозварювальні апарати);

    4) для електрохімічної обробки (електролітичні ванни тощо);

    5) для хімічної обробки (реактори, автоклави і ін.).

VI. Контрольно-вимірювальні, випробувальні і управлінські машини та прилади:

    1) контрольно-вимірювальні і випробувальні прилади та пристрої;

    2) облікові і управлінські машини.

VII. Транспортні засоби:

    1) автомобілі, автобуси;

    2) причепи (різні);

    3) локомотиви, паровози тощо;

    4) вагони, платформи і ін.;

    5) трубопроводи (для транспортування різних рідких і газоподібних речовин – газу, води, нафти, цементу тощо).

VIII. Інструменти (відбійні молотки, вібратори, електродрилі тощо).

IX. Виробничий інвентар і прилади (верстаки, універсальні контейнери).

X. Господарський інвентар (меблі, друкарські машинки тощо).

XI. Інші основні фонди (бібліотеки і ін.).

Побудована таким чином класифікація основних виробничих фондів дозволяє розмежувати будівлі цехові від будівель адміністративних та складських; споруди, що пов'язані з розробкою мінеральних і використанням водних ресурсів, транспортні споруди і споруди, що виконують технічні функції; розподілити передавальні пристрої на окремі підгрупи залежно від виду транспортованої енергії і галузевого походження засобів; виділити в групі робочих машин і устаткування підгрупи щодо методів впливу на предмет праці і галузевого походження. Кількість в ній позицій і принципи систематизації груп відповідають завданням сьогодення, полегшують вивчення структури і ефективності використання основних виробничих фондів автотранспортних підприємств.

Вивчення структури основних виробничих фондів, виявлення тенденцій її зміни найбільш повно розкривається при розгляді і аналізі відповідних процесів та їх показників у динаміці. Співвідношення між окремими елементами структури основних виробничих фондів, що формується на певний момент часу, залежить від ряду чинників. Серед них слід відзначити такі групи чинників:

1) чинники, що діють на рівні окремого підприємства (оновлення, вибуття основних фондів, технічна оснащеність, форми організації транспортно-виробничого процесу і ін.);

2) чинники внутрігалузевого характеру (регіональні особливості економіки того чи іншого географічного району, сезонність автоперевезень);

3) макроекономічні чинники (політика держави в галузі інновацій, інвестицій, зовнішньоекономічної діяльності, кредитної та податкової системи, вимоги міжнародних стандартів щодо екологічності автомобілів тощо).

Інтенсивність впливу кожного чинника підсилюється або ж послаблюється дією інших, характером їх взаємодії чи суперечностей.

Характеристика структури основних виробничих фондів досліджуваних вантажних автотранспортних підприємств Івано-Франківської, Львівської і Тернопільської областей наведена в табл. 1.2.1.

Як видно з табл. 1.2.1, найбільшу питому вагу в структурі основних виробничих фондів автотранспортних підприємств займають транспортні засоби. Така специфіка структури зумовлена тим, що корисний результат діяльності вантажного автотранспорту формується внаслідок виконання транспортної роботи під час безпосереднього перевезення вантажів. В ролі основних засобів праці виступають не машини і устаткування, як у інших галузях матеріального виробництва, а власне транспортні засоби. У 1996 р. їх частка в структурі основних виробничих фондів досліджуваних вантажних автопідприємств коливалась відповідно від 53,5 % та 54,0 % у Тернопільській і Львівській областях до 66,9 % в Івано-Франківській (табл. 1.2.2). В цілому питома вага транспортних засобів на досліджуваних АТП регіону становить 56,3 % і є дещо нижчою від аналогічного усередненого показника вантажного автотранспорту України.

Зниження питомої ваги транспортних засобів, яке спостерігалося протягом 1992 – 1996 рр., стало загальною закономірністю зміни структури їх основних фондів. Таку ситуацію зумовлюють певні причини. Починаючи з 1990 р.,

**Таблиця 1.2.1**

Порівняльна характеристика структури основних виробничих фондів вантажних АТП Івано-Франківської, Львівської, Тернопільської областей та вантажного автомобільного транспорту України за 1992 – 1996 рр., %.

| Назви елементів основних фондів | 1992 Регіон | 1992 Україна | 1993 Регіон | 1993 Україна | 1994 Регіон | 1994 Україна | 1995 Регіон | 1995 Україна | 1996 Регіон | 1996 Україна | 1996 р. в % до 1992 р. Регіон | 1996 р. в % до 1992 р. Україна |
|---|---|---|---|---|---|---|---|---|---|---|---|---|
| Основні виробничі фонди, всього: в т. ч.: | 100,0 | 100,0 | 100,0 | 100,0 | 100,0 | 100,0 | 100,0 | 100,0 | 100,0 | 100,0 | X | X |
| будівлі | 25,8 | 24,9 | 26,3 | 25,6 | 26,4 | 25,2 | 26,3 | 25,0 | 26,4 | 25,3 | 102,3 | 101,6 |
| споруди | 7,0 | 6,4 | 7,9 | 6,6 | 7,5 | 6,0 | 7,6 | 5,7 | 7,6 | 5,8 | 108,6 | 90,0 |
| передавальні пристрої | 1,4 | 1,7 | 1,3 | 1,8 | 1,2 | 1,8 | 1,4 | 1,8 | 1,3 | 1,7 | 92,9 | 100,0 |
| машини і устаткування | 6,6 | 6,9 | 6,8 | 7,0 | 6,6 | 7,1 | 6,7 | 7,0 | 6,7 | 7,0 | 101,5 | 101,4 |
| транспортні засоби | 57,3 | 57,9 | 56,0 | 56,6 | 56,6 | 58,1 | 56,4 | 58,7 | 56,3 | 58,3 | 98,3 | 100,7 |
| реманент виробничий і господарський інвентар | 1,6 | 1,5 | 1,5 | 1,6 | 1,4 | 1,0 | 1,3 | 1,0 | 1,4 | 1,1 | 87,5 | 73,3 |
| інші основні фонди | 0,3 | 0,7 | 0,2 | 0,8 | 0,3 | 0,8 | 0,3 | 0,8 | 0,3 | 0,8 | 100,0 | 114,3 |

відбувається постійне скорочення парку рухомого складу за рахунок його старіння (середній вік вантажного автомобіля на початок 1996 р. досяг 8,2 року) і вибуття, а також, частково, у зв'язку з реалізацією вантажних автомобілів стороннім організаціям та фізичним особам. Так, протягом 1992 – 1996 рр. парк рухомого складу досліджуваних вантажних автопідприємств Тернопільської області скоротився на 37 %.

Дані екстраполяції показують, що зниження питомої ваги транспортних засобів спостерігатиметься і в найближчі роки. Це пов'язано «майже з повним припиненням оновлення рухомого складу» [71, с. 1]. Вкрай незадовільний фінансовий стан вантажних автопідприємств України, збитковість автоперевезень, значна дебіторська заборгованість, заборгованість перед бюджетом та по виплаті заробітної плати, унеможливлюють придбання надто дорогих транспортних засобів.

**Таблиця 1.2.2**

Натурально-речова структура основних виробничих фондів вантажних автотранспортних підприємств Івано-Франківської, Львівської та Тернопільської областей станом на 1 січня 1997 року, %

| Назва елементів основних фондів | Івано-Франківська область | Львівська область | Тернопільська область |
|---|---|---|---|
| Основні виробничі фонди, всього: | 100,0 | 100,0 | 100,0 |
| в т. ч.: | | | |
| Будівлі | 21,1 | 26,6 | 28,8 |
| Споруди | 3,3 | 9,1 | 7,3 |
| Передавальні пристрої | 0,6 | 2,3 | 0,2 |
| Машини і обладнання, всього: | 7,7 | 5,7 | 8,0 |
| з них: | | | |
| силові машини і обладнання | 3,3 | 1,5 | 1,8 |
| робочі машини і обладнання | 3,6 | 2,7 | 4,5 |
| вантажно-розвантажувальні механізми | 0,1 | 0,8 | 0,4 |
| вимірювальні та регулювальні прилади, пристрої і лабораторне обладнання | 0,1 | 0,1 | 0,5 |
| засоби зв'язку | – | 0,1 | 0,1 |
| обчислювальна техніка | 0,6 | 0,5 | 0,7 |
| Транспортні засоби | 66,9 | 54,0 | 53,5 |
| Ремонт виробничий і господарський інвентар | 0,3 | 2,0 | 1,9 |
| Інші види основних фондів | 0,1 | 0,3 | 0,3 |

Так, за підсумками 1996 р. згідно даних Державного департаменту автомобільного транспорту України збитки від автоперевезень склали 150 млн. грн. Зокрема, у вантажних автопідприємствах Івано-Франківської області – 1,22 млн. грн., Львівської області – 4,47 млн. грн., Тернопільської області – 1,14 млн. грн. [197, с. 1]. З року в рік зростає дебіторська заборгованість, питома вага якої в структурі обігових коштів сягнула з 10 % у 1990 р. до 44 % у 1996 р., а частка грошових засобів на розрахунковому рахунку за той самий період знизилась з 31% до 2% [див.: 228 ]. «Постійне збільшення дефіциту обігових коштів в поєднанні з неплатоспроможністю споживачів послуг автотранспорту призводить до зростання заборгованості автопідприємств бюджету, Пенсійному фонду, а також виплат заробітної плати» [195, с. 5].

Відсутність внутрішніх джерел інвестування вантажного автомобільного транспорту України доповнюється згортанням ще й зовнішніх надходжень капітальних вкладень. У 1996 р. капіталовкладення в автомобільний транспорт України склали 71 млн. грн. проти 665 млн. грн. в порівняльних цінах у 1990 р. [див.: 228]. У цій ситуації розширення вантажної автостанції в с. Біла Тернопільської області є радше винятком, ніж закономірним явищем.

Припинення довго- та середньострокового банківського кредитування в Україні позбавляє автопідприємства можливості придбання капітальних активів. Що стосується короткострокових кредитів комерційних банків, наданих підприємствам транспорту України, то їх питома вага на кінець 1996 р. заледве сягнула 1,4 % від загальної суми кредитів у всіх галузях економіки [там само]. Частка кредитів, наданих власне вантажному автотранспорту, в 7 разів нижча від означеної величини, а тому відчутного впливу на поліпшення фінансового стану автопідприємств не мала.

Продовжуючи аналіз статистичних даних табл. 1.2.1, слід зазначити, що значне місце в структурі основних фондів автотранспортних підприємств відводиться будівлям і спорудам. За період з 1992 по 1996 рр. питома вага цих елементів основних фондів на відміну від транспортних засобів зросла у автотранспортних підприємствах Івано-Франківської, Львівської та Тернопільської областей майже на 4 %. Характерною особливістю цих змін є значно вищі темпи зростання вартості споруд порівняно з темпами зростання вартості будівель і становлять відповідно 8,6 % та 2,3 %.

Частка вартості будівель у структурі основних виробничих фондів вантажних автопідприємств коливається від 31,1 % в Івано-Франківській області до 28,8% в Тернопільській; споруд – від 3,3 % до 9,1 % відповідно в Івано-Франківській та Львівській областях (див. табл. 1.2.2).

Значення кількісного і якісного удосконалення структури основних фондів є беззаперечним. Воно дозволяє створити ту матеріально-технічну базу автотранспортних підприємств, яка найбільш повно і ефективно забезпечить потреби експлуатації рухомого складу. Однак на сьогодні ці питання в практичній роботі ще далеко не вирішуються. Згідно з Типовим проектом вантажних автотранспортних підприємств з чисельністю автомобілів 125 одиниць співвідношення між питомою вагою рухомого складу і іншими основними фондами повинне складати 41 % проти 59 %. Фактично воно сьогодні складає 57,2 проти 42,8 %.

Структура основних виробничих фондів автотранспортних підприємств формується під впливом конкретних умов, в яких здійснюється транспортний процес: форм організації транспортно-виробничого процесу; технічної оснащеності підприємств; типу, структури і спеціалізації рухомого складу. Так, на початок 1997 р. в автотранспортних підприємствах, середня вантажопідіймальність автомобілів яких 4 – 5 тонн, питома вага транспортних засобів у них складає 44,3%, в автотранспортних підприємствах, автопарк яких складається з автомобілів великої вантажопідіймальності (8 і більше тонн), питома вага транспортних засобів складає відповідно 57,3 % до загальної вартості основних виробничих фондів.

Наведені в табл. 1.2.3 дані підтверджують вищезгадану закономірність: чим вище значення показника середньої вантажопідіймальності рухомого складу автопідприємств, тим вагомішу частку в структурі основних виробничих фондів займають транспортні засоби.

**Таблиця 1.2.3**

Залежність структури основних виробничих фондів від величини показника середньої вантажопідіймальності вантажних АТП Івано-Франківської, Львівської та Тернопільської областей станом на 1 січня 1997 року, %

| Назва елементів основних фондів | Середня вантажопідіймальність рухомого складу, тонн | | | |
|---|---|---|---|---|
| | 2,0 – 3,9 | 4,0 – 5,9 | 6,0 – 7,9 | 8,0 і більше |
| Основні фонди, всього: | 100,0 | 100,0 | 100,0 | 100,0 |
| в т. ч.: | | | | |
| будівлі | 26,3 | 30,9 | 18,6 | 20,1 |
| споруди | 5,3 | 5,6 | 14,0 | 11,9 |
| передавальні пристрої | 0,2 | 1,2 | 3,0 | 1,9 |
| машини і обладнання | 5,1 | 8,4 | 7,6 | 3,5 |
| транспортні засоби | 61,7 | 52,9 | 56,6 | 57,3 |
| ремонт виробничий і господарський інвентар та інші види основних фондів | 1,4 | 1,0 | 0,2 | 5,3 |

Виняток із цієї закономірності становлять дані першої групи з показником середньої вантажопідіймальності від 2,0 до 3,9 тонн. У цій групі частка транспортних засобів в структурі основних виробничих фондів найвища – 61,7 %. Однак існує об'єктивне пояснення цього факту. Означену групу на 91,8 % формують вантажні АТП Укоопспілки, в яких парк автомобілів вантажопідіймальністю до 3,9 тонн складає 86,8 %. Враховуючи те, що вартість основних виробничих фондів автопідприємств Укоопспілки становить лише 17,3 % від загальної вартості основних виробничих фондів усіх досліджуваних автогосподарств регіону, ця розбіжність зникає при більш укрупненому групуванні за величиною показника середньої вантажопідіймальності (див. діаграми на рис. 1.2.1).

Залежність структури основних виробничих фондів від вантажопідіймальності рухомого складу автопідприємств знаходить своє відображення в аналогічних тенденціях, що спостерігаються у таких класифікаційних групах, як споруди та передавальні пристрої. Їх частка в структурі основних виробничих фондів тим більша, чим вище значення середньої вантажопідіймальності АТП. І навпаки, питома вага будівель, машин і обладнання знижується у відповідь на зростання показника середньої вантажопідіймальності рухомого складу.

Охарактеризована структура основних виробничих фондів відображає ті тенденції, які відбувались до початку 90-х років. Зростання питомої ваги великотоннажних автомобілів у структурі рухомого складу автопідприємств вимагало і відповідної інфраструктури, яка забезпечувала б належні умови зберігання, технічного обслуговування і ремонту автомобілів. Ось чому в структурі основних виробничих фондів таких автотранспортних підприємств значно більшу частку займають обладнані відкриті стоянки, паливозаправні колонки, естакади, сховища і ін.

Значний вплив на формування структури основних виробничих фондів мають розміри основних виробничих фондів автотранспортних підприємств (табл. 1.2.4).

Дані таблиці засвідчують те, що у найбільш дрібних автотранспортних підприємствах питома вага транспортних засобів у загальній вартості основних фондів найвища – 66 %, в крупних – 51,6 %. Отже, збільшення розміру автотранспортних підприємств з порівняно невеликим розміром основних фондів

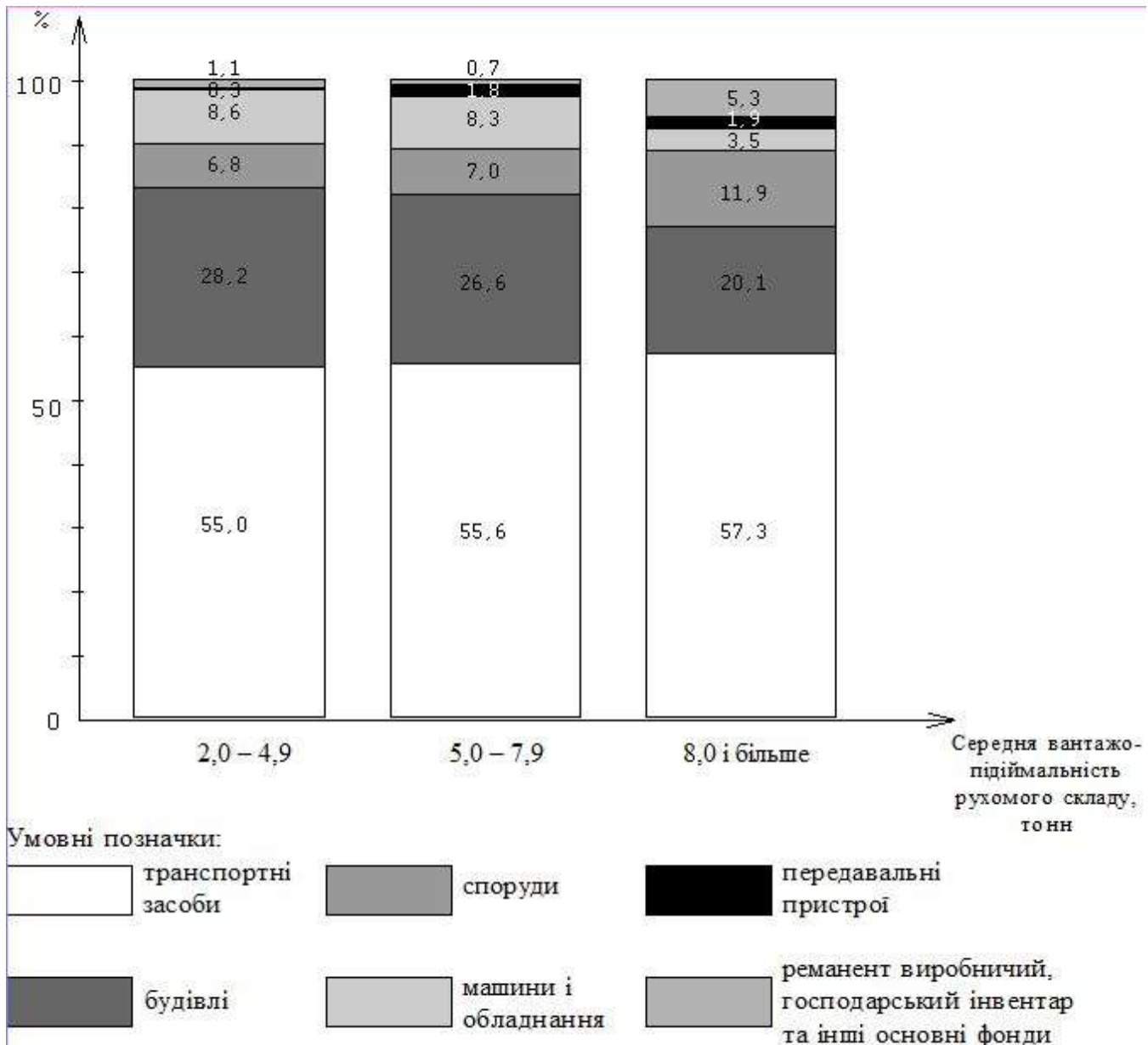

Рис. 1.2.1. Структура основних виробничих фондів вантажних АТП Івано-Франківської, Львівської і Тернопільської областей за середньою вантажопідіймальністю рухомого складу здійснювалось на початку 90-х років в першу чергу за рахунок транспортних засобів, а не за рахунок таких видів основних фондів, як будівлі і споруди. На нашу думку, це обумовлювалось об'єктивними причинами, які випливали з тогочасного рівня економічного розвитку автомобільного транспорту. Дрібні автотранспортні підприємства, які, як правило, забезпечені основними фондами, орієнтувались на придбання в першу чергу найбільш активних елементів основних

**Таблиця 1.2.4**

Залежність структури основних виробничих фондів автотранспортних підприємств від величини їх вартості станом на 1 січня 1997 року, %

| Назва елементів основних фондів | Групи підприємств за розміром основних фондів, тис. грн. | | | |
|---|---|---|---|---|
| | до 1000 | 1001 – 4000 | 4001 – 7000 | 7001 і більше |
| Основні фонди, всього:<br>в т. ч.: | 100,0 | 100,0 | 100,0 | 100,0 |
| будівлі | 23,8 | 24,3 | 27,7 | 27,8 |
| споруди | 3,3 | 5,4 | 8,1 | 10,3 |
| передавальні пристрої | 0,2 | 0,5 | 1,5 | 2,2 |
| машини і обладнання | 4,7 | 7,3 | 7,9 | 6,0 |
| транспортні засоби | 66,0 | 60,5 | 54,2 | 51,6 |
| реманент виробничий і господарський інвентар та інші основні фонди | 2,0 | 2,0 | 0,6 | 2,1 |

фондів – транспортних засобів. І лише після досягнення достатнього забезпечення транспортними засобами переходили до нарощування основних фондів за рахунок тих елементів, що забезпечують ефективну експлуатацію транспортних засобів, їх ремонт і збереження.

Таким чином, вивчення структури основних виробничих фондів автотранспортних підприємств показало, що:

1. Результатом діяльності транспорту є переміщення продукції, виробленої іншими галузями матеріального виробництва. Ця незвичайність транспортного процесу визначає особливості формування структури основних виробничих фондів автомобільного транспорту, до яких слід віднести відносно високу питому вагу активної частини основних фондів – транспортних засобів; високий ступінь залежності структури основних фондів від розміру автотранспортних підприємств, середньої вантажопідіймальності їх рухомого складу.

2. Структура основних виробничих фондів автотранспортних підприємств Івано-Франківської, Львівської, Тернопільської областей за 1992 – 1996 рр. змінилась в бік значного зниження питомої ваги активної їх частини – транспортних засобів. В основному це зумовлено старінням і вибуттям рухомого складу і частково у зв'язку з їх реалізацією стороннім організаціям і фізичним особам. Ця тенденція підсилюється неможливістю оновлення парку вантажних автомобілів через брак внутрішніх інвестиційних джерел та обмеженою можливістю залучення зовнішніх джерел надходжень капітальних вкладень. Означені структурні зрушення характеризують якісні зміни, які відбуваються в складі основних фондів і відображають процес реформування речового складу основних засобів вантажного автомобільного транспорту.

## 1.3. Відтворення основних виробничих фондів – важлива передумова підвищення ефективності роботи вантажних автотранспортних підприємств

В умовах переходу до ринкових відносин кількість промислових, торгових, побутових і інших підприємств, особливо малих за останні роки різко зросла, зросли також обсяги будівництва приватних житлових і дачних будинків. Помітно зросла потреба в перевезеннях вантажів. Однак потужності більшості автотранспортних підприємств Івано-Франківської, Львівської та Тернопільської областей за останні роки швидко знижувались, старіли парки автомобілів.

Положенням про технічне обслуговування і ремонт автомобілів передбачається проведення, переважно, одного капітального ремонту, що відповідає пробігу 250 – 300 тис. км для І-ої категорії умов експлуатації (при середньорічному пробігу 30 – 35 тис. км; строк служби повинен складати 8 – 9 років), після чого автомобіль вибраковується; однак в силу згаданих вище причин фактичний пробіг автомобілів значно перевищує нормативний. Так, при вивченні фактичних строків експлуатації вантажних автомобілів автотранспортних підприємств Івано-Франківської, Львівської і Тернопільської областей встановлено, що більше третини автомобілів експлуатується понад установлені строки служби (табл.1.3.1).

Рухомий склад автотранспортних підприємств дуже зношений. Якщо в Україні частка рухомого складу вантажного автотранспорту, що відслужив понад 8 років, складає 62,4 % і з них більше 10 років – 32,1 % (див. рис. 1.3.1), то у досліджуваних

Таблиця 1.3.1

Розподіл вантажних автомобілів АТП за тривалістю експлуатації і пробігу

(в % до загальної чисельності) станом на 1 січня 1997 року

| Тривалість експлуатації автомобілів | Питома вага автомобілів, % | | | | | |
|---|---|---|---|---|---|---|
| Загальний пробіг автомобілів | До 3 років | Від 3,1 до 5 років | Від 5,1 до 8 років | Від 8,1 до 10 років | більше 10 років | Всього |
| До 100 тис. км | 0,6 | 0,2 | 0,5 | – | – | 1,3 |
| Від 101 до 300 тис. км | 0,1 | 4,3 | 19,3 | 14,3 | – | 38,0 |
| Від 301 до 400 тис. км | – | 2,4 | 6,4 | 7,6 | 11,7 | 28,1 |
| Понад 400 тис. км | – | 0,3 | 2,4 | 5,7 | 24,2 | 32,6 |
| Всього | 0,7 | 7,2 | 28,6 | 27,6 | 35,9 | 100,0 |

АТП Івано-Франківської, Львівської та Тернопільської областей питома вага автомобілів, що прослужили більше 8 років, складає 63,5 %, з яких 35,9 % від загальної кількості припадає на автомобілі, що експлуатувались більше 10 років.

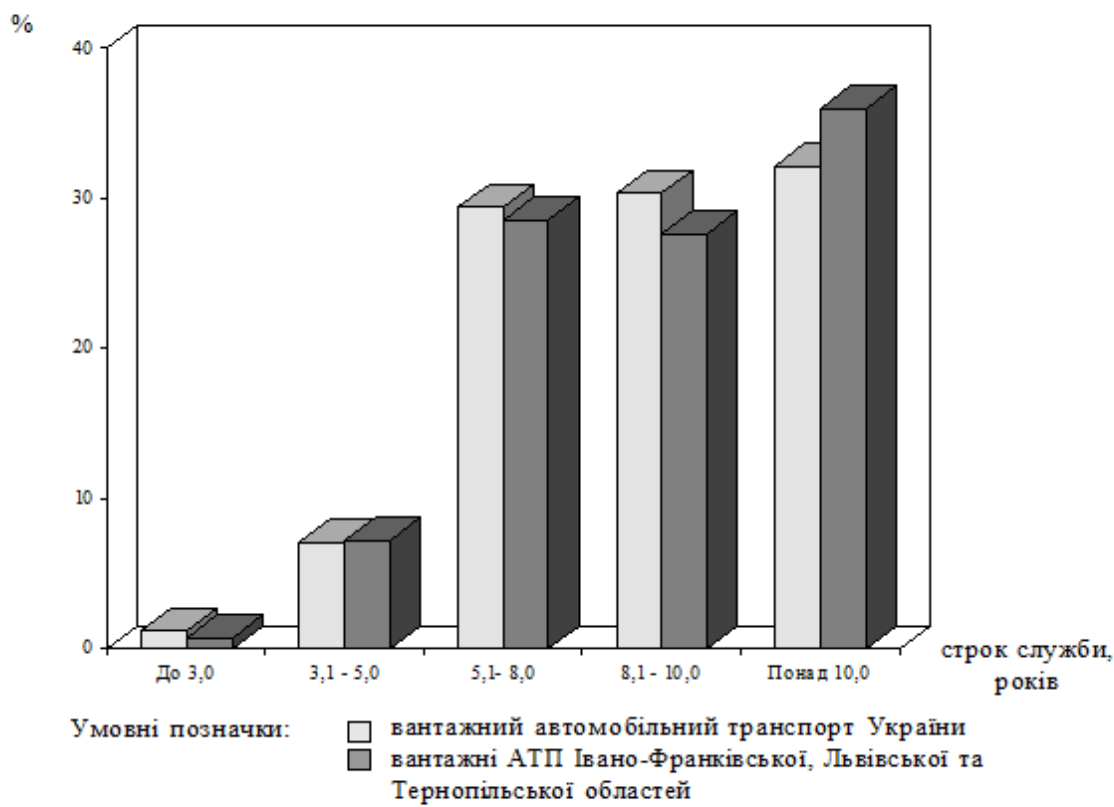

Рис. 1.3.1. Вікова структура вантажного автотранспорту в 1996 році, %

Частка автомобілів, які пройшли понад 300 тис. км і підлягає за нормативами списанню, сягає понад 60 %. Практика роботи автотранспортних підприємств свідчить про те, що експлуатація автомобілів, строк служби яких понад 10 років і загальний пробіг більше 300 тис. км, технічно невиправдана та економічно недоцільна, оскільки продуктивність у них надзвичайно низька, а затрати на перевезення значно перевищують затрати при перевезеннях вантажів автомобілями, що відносяться до двох перших груп. Це наочно підтверджується даними табл. 1.3.2.

**Таблиця 1.3.2**

Залежність технічної готовності і продуктивності автомобілів автотранспортних підприємств Івано-Франківської, Львівської та Тернопільської областей від загального їх пробігу за 1996 рік

| Групи автомобілів залежно від їх загального пробігу | Кількість автомобілів в % до підсумку | Коефіцієнт технічної готовності | Продуктивність на одну автомобіле-тонну | |
|---|---|---|---|---|
| | | | тис. км. | в % до першої групи |
| До 100 тис. км | 1,3 | 0,981 | 37,1 | 100,0 |
| Від 101 до 300 тис. км | 38,0 | 0,863 | 36,7 | 98,9 |
| Від 301 до 400 тис. км | 28,1 | 0,735 | 25,9 | 69,8 |
| Понад 400 тис. км | 32,6 | 0,497 | 17,2 | 46,4 |
| Всього | 100,0 | X | X | X |

Наведені в табл. 1.3.1 дані свідчать про те, що автопарк Івано-Франківської, Львівської і Тернопільської областей оновлюється надто повільно, слабо охоплюється і капітальним ремонтом. За 1996 рік частка автомобілів, що не пройшли капітального ремонту, становила 21,7 %. Відносно низький рівень технічної готовності автомобілів в транспортних господарствах Львівщини, де він складав 0,737. А у таких автотранспортних підприємствах, як Бориславське № 24659, Рава-Руське № 24668, коефіцієнт технічної готовності відповідно 0,607 і 0,617, що обумовлено відносно високою у них питомою вагою автомобілів з тривалим строком експлуатації. Більшість автотранспортних підприємств з метою забезпечення виконання своєї транспортно-виробничої програми змушені використовувати для роботи автомобілі з понад амортизаційним строком служби. Природньо, що для підтримування цих автомобілів у технічно справному стані автотранспортні підприємстванамагаються розширювати виробничі потужності ремонтних майстерень, профілакторіїв та інших приміщень, підвищувати їх технічну оснащеність.

Як засвідчують розрахунки українських вчених П.П. Нєдова та С.А. Прусенка, «... питомі матеріальні витрати на один кілометр пробігу в інтервалі 250 – 300 тис. км з початку експлуатації на 28 % вищі, ніж в інтервалі 50 – 100 тис. км пробігу» [143, с. 3]. Аналіз структури вантажних автомобілів за віком, здійснений українськими вченими УТУ, показує, що витрати запасних частин суттєво залежать від пробігу з початку експлуатації транспортного засобу. Так, для автомобілів марок «КрАЗ – 256», «КамАЗ – 5320», «ЗІЛ – 130» залежність між сумарними витратами запасних частин і пробігом в інтервалі до 50 тис. км відповідно становить: 2,8, 2,0, 1,0 %; в інтервалі до 150 тис. км – 31,8, 26,0 і 12,5 %; в інтервалі до 300 тис. км – 213,4, 190,0, 106,0 % [див.: 105, с. 8]. Фактично експлуатація вантажних автомобілів після 300 тис. км пробігу потрапляє в зону збитковості. Ці збитки на автопідприємстві не завжди одразу помітні, оскільки покриваються за рахунок прибутку, отриманого від експлуатації більш нових автомобілів, якщо такі є в АТП. «Морально і фізично застаріла структура парку транспортних засобів, непристосована до існуючих і прогнозованих вантажопотоків спеціалізація парку транспортних засобів призвели транспортні підприємства до стану банкрутства» [8, с. 2].

Важкий фінансовий стан автотранспортних підприємств негативно позначається на оновленні основних фондів. Слід зауважити, що нині діюча методологія побудови показників оновлення і вибуття основних виробничих фондів є небездоганною.

По-перше, у чисельнику і знаменнику цих показників порівнюються не-співставні величини ( у чисельнику використовуються інтервальні, а у знамен-нику – моментні дані).

По-друге, вони між собою неспівставні, оскільки ці показники розраховуються відносно різних баз. За їх величиною важко судити про те, що переважало у досліджуваному періоді: оновлення чи вибуття основних фондів.

У недосконалості цих показників неважко переконатися, якщо розглянути приклад про рух основних фондів у ВАТ «Львіввантажавтотранс» за 1996 рік.

Таблиця 1.3.3

Баланс основних фондів ВАТ «Львіввантажавтотранс» за 1996 рік, тис. грн.

| Наявність на початок року | Надійшло у звітному році | | Вибуло у звітному році | | Наявність на кінець року |
|---|---|---|---|---|---|
| | всього | в т. ч. введено в дію нових основних фондів | всього | в т. ч. ліквідовано основних фондів | |
| 16919 | 509 | 274 | 271 | 87 | 17157 |

І отже, за нині діючою методологією обчислення коефіцієнт оновлення основних фондів за рік складав:

$$K_{оновл.} = \frac{274}{17157} \cdot 100 = 1{,}597\ \%.$$

Коефіцієнт вибуття основних фондів:

$$K_{вибут.} = \frac{271}{16919} \cdot 100 = 1{,}602\ \%.$$

Проведені обчислення показують, що рівень вибуття основних фондів дещо перевищує рівень їх оновлення, з чим, звичайно, не можна погодитись. Суперечливість розрахованих показників не викликає сумніву.

У практиці роботи підприємств з метою одержання порівняльних показників обидва коефіцієнти часто розраховують відносно однієї бази – вартості основних виробничих фондів на початок року, що теоретично також неправильно.

На нашу думку, коефіцієнти оновлення і вибуття основних виробничих фондів повинні, без сумніву, обчислюватися відносно однієї бази. При цьому з тією порівняльною базою повинні бути пов'язані однаковою мірою обидва показники: і коефіцієнт оновлення, і коефіцієнт вибуття. Такою, на нашу думку, може бути середньорічна вартість основних виробничих фондів. Вона, по-перше, є найбільш типовою величиною основних фондів в даному році, по-друге, забезпечить найнижчу похибку обчислених показників.

Для підтвердження обчислимо за запропонованим нами способом ці ж показники, використовуючи дані табл. 1.3.3.

– коефіцієнт оновлення:

$$K_{оновл.} = \frac{274}{17038} \cdot 100 = 1,61\ \%;$$

– коефіцієнт вибуття всіх основних фондів:

$$K_{вибут.} = \frac{271}{17308} \cdot 100 = 1,59\ \%;$$

– коефіцієнт вибуття основних фондів внаслідок їх зношення:

$$K_{виб.зн.} = \frac{87}{17308} \cdot 100 = 0,5\ \%.$$

Таким чином, зіставляючи обчислені показники з вихідними даними та обчисленими за нині діючою методологією, неважко переконатися у правильності відображення суті останніми.

Обчислені запропонованим нами способом показники оновлення і вибуття основних виробничих фондів забезпечують: по-перше, порівняльність між цими показниками; по-друге, зручність у користуванні ними при економічному аналізі руху основних фондів; по-третє, побудовані таким способом коефіцієнти не порушують математичних вимог.

Проведені вище обчислення свідчать про надзвичайно низькі темпи оновлення основних виробничих фондів на ВАТ «Львіввантажавтотранс», хоча за своєю величиною дещо і переважають темпи оновлення в цілому по Західному регіону. Так, якщо у 1992 р. коефіцієнт оновлення в досліджуваних АТП Івано-Франківської, Львівської і Тернопільської областей в цілому становив 8,22 %, то у 1996 р. – лише 0,28 % (див. табл. 1.3.4).

Слід зазначити, що із загального обсягу засобів праці, які надійшли в автотранспортні підприємства, лише частина з них були новими. Значна частина засобів праці передається на баланс автотранспортних підприємств таких, що уже

**Таблиця 1.3.4**

Динаміка вартості основних виробничих фондів АТП Івано-Франківської, Львівської та Тернопільської областей за 1996 році

| Показники | Області | | | Загалом по західному регіону |
|---|---|---|---|---|
| | Івано-Франківська | Львівська | Тернопільська | |
| Вартість основних фондів на початок року, з урахуванням індексації 1996 р., тис. грн. | 46085 | 131749 | 64942 | 242776 |
| Надійшло у звітному році, всього, тис. грн. | 489 | 767 | 447 | 1703 |
| в т. ч. введено в дію нових основних фондів, тис. грн. | 22 | 338 | 293 | 653 |
| Коефіцієнт оновлення, % | 0,05 | 0,27 | 0,46 | 0,28 |
| Вибуло у звітному році, всього, тис. грн. | 2232 | 6804 | 1952 | 10988 |
| Коефіцієнт вибуття, % | 4,84 | 5,16 | 3,0 | 4,53 |
| в т. ч. ліквідовано основних фондів, всього, тис. грн. | 920 | 5697 | 1775 | 8392 |
| Коефіцієнт ліквідації, % | 2,0 | 4,32 | 2,73 | 3,46 |
| Вартість основних фондів на кінець року, тис. грн. | 44341 | 126890 | 63450 | 234681 |
| Середньорічна вартість виробничих основних фондів, тис. грн. | 45313 | 126962 | 64379 | 236654 |

перебували в експлуатації. Про це свідчить зниження частки нових основних засобів у загальному надходженні з 81,3 % у 1992 р. до 38,3 % у 1996 р. При цьому оновлення парку рухомого складу практично не спостерігалося. І навіть те, що питома вага нових основних активів автопідприємств Тернопільської області у 1996 р. була найвищою серед областей досліджуваного регіону, сягаючи 65,5 %, жодним чином не позначилось на оновленні рухомого складу.

Подібні факти стосуються і вибуття основних виробничих фондів. Лише частина їх вибуває через списання, внаслідок зношення чи старіння, решта передається на баланс інших організацій та підприємств. Протягом п'яти років – з 1992 по 1996 – коефіцієнт вибуття знизився з 8,71 % до 4,53 %, в тому числі коефіцієнт ліквідації фондів з 4,5 % до 3,46 %.

Якщо простежити характер оновлення основних виробничих фондів і їх вибуття протягом означених п'яти років, то можна зауважити, що темп надходження нових засобів праці знизився на 96,59 % і становив лише 3,41 % від рівня 1992 р. Темп вибуття основних фондів від ліквідації знизився лише на 23,1 % і склав 76,9 % до рівня 1992 р. Інтенсивність оновлення основних виробничих фондів у 1992 р. у 1,83 раза перевищувала темпи їх ліквідації, проте на 0,49 % відставала від темпів вибуття основних засобів. Капітальні активи у 1996 р. оновлювалися в 16,2 раза повільніше, ніж їх ліквідація від старіння та непридатності для використання.

Разюча різниця, що склалася між темпами оновлення і вибуття основних виробничих фондів в галузі протягом п'яти років, відображає вкрай критичну ситуацію, яка все більше поглинає автопідприємства транспорту, розмиваючи їх основний капітал.

Темпи оновлення рухомого складу ще стриманіші. Протягом трьох років – з 1994 по 1996 – нові автомобілі придбало лише вісім із ста досліджуваних АТП Західного регіону, з них шість – автогосподарства Львівщини. Щодо автогосподарств Івано-Франківщини і Тернопільщини, то їхній парк рухомого складу протягом цього ж періоду оновився 8 вантажними автомобілями, що становить менше 0,14 автомобіля в розрахунку на одне АТП.

Аналогічний стан оновлення та вибуття вантажних автомобілів в цілому по автотранспорту України. За даними Державного департаменту автомобільного транспорту питома вага нових автомобілів скоротилася з 5,42 % у 1990 р. до 0,12 % у 1996 р. (див. табл. 1.3.5).

**Таблиця 1.3.5**

Динаміка оновлення та списання рухомого складу вантажного автомобільного транспорту України протягом 1990 – 1996 років

| Показники | Одиниця виміру | 1990 | 1992 | 1993 | 1994 | 1995 | 1996 | 1996 в % до 1990 |
|---|---|---|---|---|---|---|---|---|
| Поповнення | од. | 5435,0 | 2747,0 | 794,0 | 281,0 | 128,0 | 62,0 | 1,14 |
| питома вага | % | 5,41 | 3,04 | 0,98 | 0,40 | 0,23 | 0,12 | X |
| Списання | од. | 7630,0 | 8205,0 | 7303,0 | 8350,0 | 5281,0 | 4443 | 58,2 |
| питома вага | % | 7,60 | 9,08 | 8,97 | 12,02 | 9,31 | 8,71 | X |
| Процент відтворення рухомого складу | % | 71,2 | 33,5 | 10,9 | 3,4 | 2,4 | 1,4 | X |

Поруч з показниками оновлення та вибуття основних фондів в умовах ринкових відносин з'являється потреба в обчисленні показника, за допомогою якого можна було б оцінити рух основних фондів, визначити його напрям. Таким показником, на нашу думку, може бути відношення нових основних фондів, введених в дію за період, до повної вартості вибулих основних фондів внаслідок їх зношення за цей же період. Впровадження такого показника в практику роботи дозволить встановити: по-перше, ступінь випередження нових основних фондів, що поступили за досліджуваний період у порівнянні з вибулими основними фондами за цей ж період; по-друге, напрям руху основних фондів. На нашу думку, цей показник можна назвати коефіцієнтом відтворення основних засобів. Згідно даних табл. 1.3.4 коефіцієнт відтворення основних фондів досліджуваних автопідприємств у 1996 р. становив лише 7,78 % (653/8392 · 100 %), що означає перевищення їх вибуття у зв'язку із зношенням над оновленням на 92,22 %. При цьому складніша ситуація спостерігається на АТП Івано-Франківщини, де значення цього коефіцієнта становить 2,39 % і свідчить про майже повне призупинення відтворення основних виробничих фондів.

Низькі темпи відтворення основних виробничих фондів супроводжуються ще нижчими темпами відтворення рухомого складу вантажних автопідприємств. Ще у 1990 р. списання рухомого складу вантажного автомобільного транспорту України на 28,8 % перевищувало його оновлення (див. табл. 1.3.5). До 1993 р. цей розрив склав 89,1 %, а у 1996 р. – 98,6 %.

Відставання темпів оновлення основних виробничих фондів від темпів його вибуття, як зазначалося, є негативним і небезпечним явищем, що відображає процес «проїдання» основного капіталу, наслідки якого відбиваються в усіх сферах суспільного життя. Зокрема, в найболючішій сфері – соціальній – це виявляється у заборгованості по заробітній платі, вимушених відпустках, безробітті та ін.

Брак власних коштів АТП, відсутність зовнішнього кредитування, висока вартість нових автомобілів – ось основні причини, через які у автомобільному транспорті не відбувається навіть просте відтворення. У підприємств не вистачає коштів на те, щоб підтримувати старий парк автомобілів у технічно справному стані, не говорячи вже про забезпечення належного технічного рівня рухомого складу.

Поняття технічного рівня автомобіля в широкому розумінні передбачає наявність таких головних чинників споживчих властивостей, як ефект застосування, матеріальні витрати в експлуатації та сервісі, ступінь задоволення вимог споживача, наявність у моделі прогресивних конструкторських, технологічних та експлуатаційних рішень.

Згідно з підрахунками вітчизняних вчених В.М. Реви та В.В. Рудзинського підтримання автомобіля у працездатному стані за весь строк його роботи вимагає витрат, що в 9 разів перевищують первісну вартість автомобіля. Отже, якщо прийняти усі витрати, пов'язані з придбанням, утриманням і обслуговуванням автомобіля протягом строку служби за 100 %, то вартість його придбання (ціна) складатиме 10 %, щодобовий догляд – 17 %, технічне обслуговування – 30 %, поточний та капітальний ремонти – 43 % [201, с. 3]. Отже, належний технічний рівень в такому розумінні є суттєвим чинником формування економічної стабільності автотранспортних господарств. Як наслідок цього – високі темпи списання з балансу витратного застарілого рухомого складу АТП. «Настільки значні зміни режиму вибуття,– на думку певної частини економістів,– передбачають використання таких радикальних заходів, як ліквідація без відшкодування певної кількості найбільш застарілих виробничих потужностей, експлуатація яких вимагає підвищених витрат всіх видів ресурсів» [252, с. 71].

Якщо простежити динаміку основних фондів за ряд років, то неважко помітити щорічне їх скорочення (табл. 1.3.6).

На основі даних табл. 1.3.6. можна прийти до висновку, що в цілому по регіону вартість транспортних засобів скоротилася на 80920 млн. грн. при середньорічних темпах зниження – 9,5 %. Основні фонди зменшилися на 36,8 %. Отже, зменшення вартості основних фондів проходило в основному за рахунок швидких темпів зниження їх активної частини.

Для виявлення характеру зміни вартості транспортних засобів необхідно зупинитися на темпах їх щорічного поповнення і вибуття. Дані табл. 1.3.7 ха актеризують динаміку транспортних засобів в автотранспортних підприємствах Івано-Франківської, Львівської і Тернопільської областей.

Таблиця 1.3.6

Динаміка середньорічної вартості основних фондів вантажних АТП
Івано-Франківської, Львівської та Тернопільської областей
протягом 1992 – 1996 рр. (в порівняльних цінах)

| Показники | Роки | | | | | 1996 р. в % до 1992 р. |
|---|---|---|---|---|---|---|
| | 1992 | 1993 | 1994 | 1995 | 1996 | |
| Основні виробничі фонди, всього, тис. грн. | 371637 | 349843 | 304598 | 277922 | 234681 | 63,2 |
| в % до попереднього року | – | 94,2 | 87,1 | 91,2 | 84,4 | X |
| Вартість транспортних засобів, тис. грн. | 212948 | 196125 | 172403 | 156702 | 132028 | 62,0 |
| в % до попереднього року | – | 92,1 | 87,9 | 90,8 | 84,2 | X |

Таблиця 1.3.7

Оновлення та вибуття рухомого складу вантажних АТП
Івано-Франківської, Львівської та Тернопільської областей за 1996 рік

| Показники | По регіону | в т. ч. по областях | | |
|---|---|---|---|---|
| | | Івано-Франківська | Львівська | Тернопільська |
| Процент поповнення вантажних автомобілів | 0,24 | 0,08 | 0,34 | – |
| в т. ч. новими | 0,16 | – | 0,29 | – |
| Процент вибуття вантажних автомобілів | 9,54 | 9,77 | 8,95 | 10,53 |
| в т. ч. у зв'язку із списанням | 8,65 | 9,16 | 8,07 | 9,39 |

Як бачимо, в практиці виробничо-фінансової діяльності автотранспортних підприємств у нових умовах господарювання основною проблемою на рівні господарства є проблема фінансування процесу оновлення основних фондів, особливо рухомого складу. «Раніш відпрацьована і централізовано координована стратегія простого чи екстенсивного розширеного оновлення парку, що базувалось на стабілізації і рості його перевізної спроможності, уже неприйнятна» [240, с. 8]. У більшості автогосподарств не вистачає фінансових ресурсів навіть на покриття нестачі обігових коштів, не говорячи вже про довгострокові вкладення в основні виробничі фонди.

Некерованість процесу оновлення парку рухомого складу з боку автогосподарств, а також високий рівень його фізичного спрацювання зумовлюють різке зниження потенційної конкурентоспроможності кожного автопідприємства України. Господарства вантажного автотранспорту опинилися в своєрідному завороженому колі, коли значна частина коштів, як зазначалося вище, спрямовується на ремонт і підтримання рухомого складу в робочому стані, а, з другого боку, через нестачу грошових засобів автопідприємства позбавлені можливості оновлювати свій парк високоефективними зразками автомобілів.

Вирішення цієї виробничо-фінансової проблеми бачиться як наслідок досконалого вивчення, аналізу та поступового залучення керівництвом автопідприємств усіх можливих сучасних джерел фінансування, що сприятиме нагромадженню достатніх засобів для придбання рухомого складу.

Виділяють дві групи джерел фінансування відтворення основних фондів та парку автомобілів зокрема – внутрішні і зовнішні (див. рис. 1.3.2).

Специфічною особливістю наведеної класифікації є її відповідність принципам системного підходу, що виявляється у максимальному наближенні до сучасних економіко-правових умов функціонування акціонерних товариств, питома вага яких серед досліджуваних автопідприємств є найвищою. Переваги розробленої класифікації в порівнянні з іншими подібними класифікаціями джерел фінансування відтворення основних фондів [див., напр., 79, с. 158; 95, с. 28] полягають у тому, що до її складу внесені нетрадиційні джерела, такі як позики кредитних спілок, засоби приватних інвесторів та лізингових компаній, інвестиції, захищені державними зобов'язаннями. Фігурування держави як зовнішнього джерела фінансування може виявитися у субсидуванні окремих програм акціонерних товариств, що стосуються інтересів ряду суміжних виробництв чи галузей.

Становлення ринкових відносин суттєво вплинуло на структуру джерел фінансування капітальних вкладень, підвищивши за останні роки частку власних коштів автопідприємств. Однак слід врахувати, що в умовах високих темпів інфляції, платіжної кризи, зниження обсягів виробництва, а відповідно і перевезень, необґрунтованого зростання цін і тарифів, реальні можливості реалізації коштів на інвестиційні цілі надзвичайно обмежені. До того ж недоліки в оцінці вартості основних фондів сприяли зниженню розміру амортизаційного

фонду – одного з найбільш важливих власних джерел фінансування. Так, на початок 1992 р., як зазначається у «Концепції переходу автотранспорту загального користування до ринкових відносин» заступника начальника економічного управління корпорації «Укравто транс» А.П.Скорика «стан справ в автотранспортних

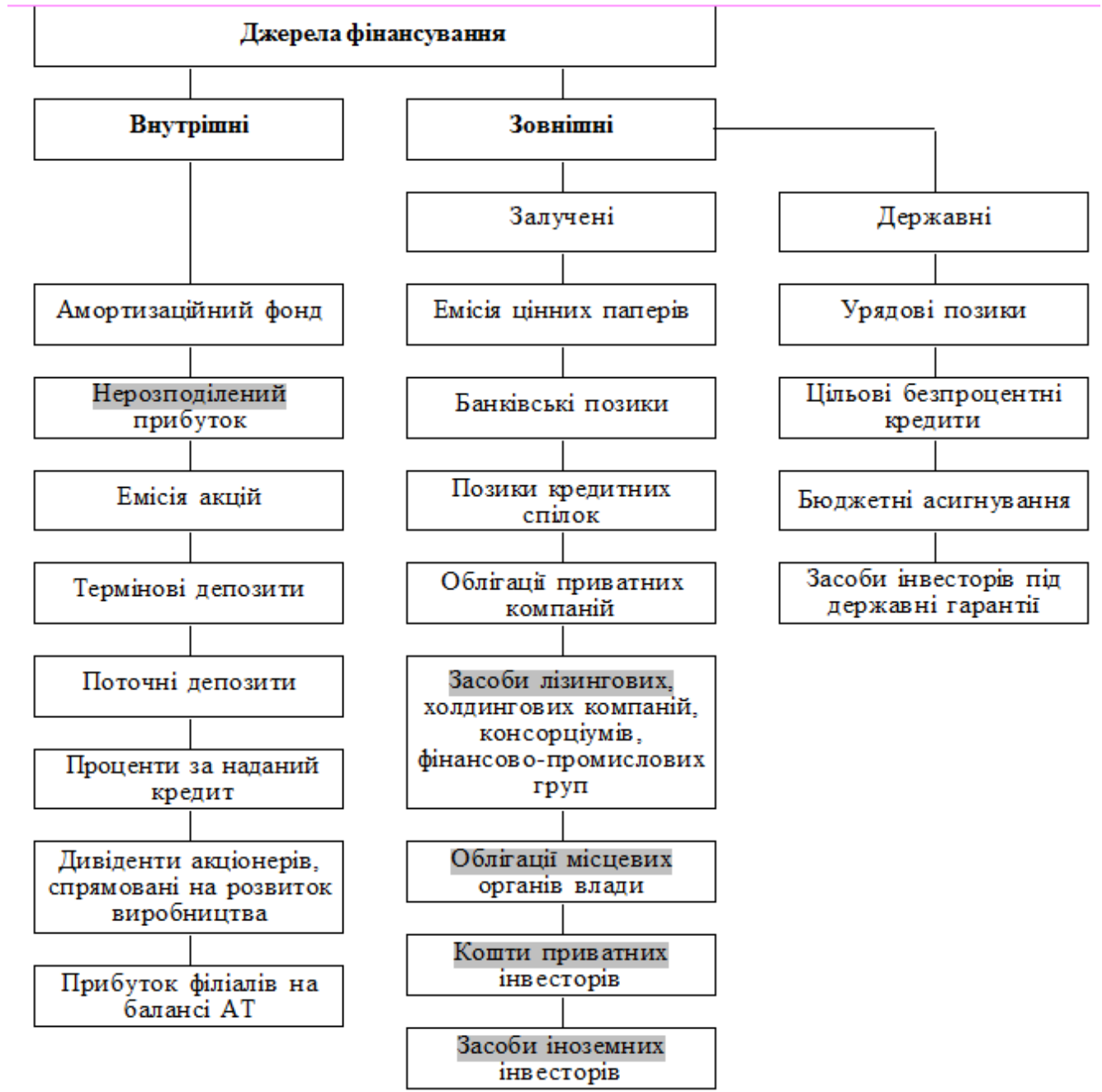

Рис. 1.3.2. Джерела фінансування відтворення основних фондів акціонерних товариств підприємствах по нарахуванню амортизаційних засобів на придбання рухомого складу засвідчує, що ці засоби незначні і їх достатньо (в автотранспортному підприємстві, що володіє 100 – 200 автомобілями) для придбання 2 – 3 нових автомобілів при нормативному оновленні за рік 12 – 25 автомобілями» [212, с. 2]. Свідченням цього є також тенденція зниження питомої ваги амортизаційних відрахувань в структурі витрат вантажних автотранспортних підприємств Івано-Франківської, Львівської та Тернопільської областей з 14,08 % у 1992 р. до 6,14 % в 1996 р.

Як відомо, найважливішою умовою ефективного функціонування підприємств у ринковому середовищі є забезпечення принципу самофінансування, який передбачає здійснення господарської та інвестиційної діяльності за рахунок власних джерел (амортизації, що покликана забезпечити просте відтворення попередньо інвестованого капіталу, та нерозподіленого прибутку як засобу розширеного відтворення капіталу), а також залучених кредитних ресурсів. Ефективність фінансування відтворення основних фондів і рівень його інтенсивності залежить від питомої ваги власних коштів підприємств. Серед джерел самофінансування амортизаційні відрахування посідають головне місце, оскільки завдяки синхронності між кругообігом основних фондів і формуванням амортизаційного фонду, вони в меншій мірі схильні до кон'юнктурних коливань, зумовлених результатами господарської діяльності. У цьому зв'язку амортизаційні відрахування більшою мірою, ніж прибуток, можуть бути фінансовою базою самостійності автотранспортних підприємств у відтворенні основних засобів. Ось чому невипадково в економіці розвинутих країн самофінансування відіграє визначальну роль. Так, наприклад, «в Японії у 1985 р. частка власних джерел фінансування в основний капітал становила 68,9 %, а залучених – 31,1 %, у США, відповідно, – 63,5 і 36,5 %, Великобританії – 62,1 і 37,9 %, ФРН – 55,3 і 44,7 %, Франції – 52,1 і 47,9 %» [206, с. 54]. Отже, практично в усіх розвинених країнах більша частина потреб у фінансуванні капітальних вкладень забезпечується на основі власних коштів.

Зупинимося на амортизаційній політиці як одному з основних елементів у регулюванні та управлінні процесом відтворення основних фондів. Відомо, що амортизаційна політика – це сукупність заходів, які здійснює держава з метою забезпечення нормального режиму відтворення основних фондів.

Дієздатність і ефективність амортизаційної політики значною мірою визначається обґрунтованістю строків служби засобів праці і нормами амортизації.

У господарській практиці розрізняють фактичні, нормативні та економічні строки служби основних фондів. Останні характеризують період продуктивного функціонування основних засобів, протягом якого за їх допомогою отримують економічний ефект.

До недавнього часу амортизаційний період обчислювався переважно за технічним строком служби основних фондів і передбачав значну його нормативну тривалість. Як наслідок амортизаційні відрахування «не давали власникові можливості компенсувати затрати на їх відшкодування в умовах постійно зростаючих цін. І підприємства «зациклювалися» на старому устаткуванні, не маючи можливості придбати нове» [28, с. 89]. У зв'язку з незначними накопиченнями сум амортизації аналогічні явища і до сьогодні спостерігаються на підприємствах автотранспорту.

Останнім часом в амортизаційній політиці України намітились певні позитивні тенденції, що виявляються у її загальній орієнтації, власне, на економічні строки служби основних засобів. Згідно з проектом закону України «Про амортизацію», прийнятого Верховною Радою України в першому читанні 2 грудня 1997 р., та Законом України «Про оподаткування прибутку підприємств», який набув чинності з 1 липня 1997 р., основні фонди підлягають розподілу за трьома амортизаційними групами [85, ст. 8]. Для основних фондів, до яких відноситься автомобільний транспорт та вузли (запасні частини) до нього, встановлена річна норма амортизації в розмірі 25%. Принагідно слід зауважити, що вказана норма амортизації на чверть перевищує аналогічні амортизаційні норми у США, встановлені відповідно до податкової реформи 1986 р. Щодо норм амортизації по інших групах основних фондів, то вони теж є дещо вищими, ніж у США [206, с. 54]. На нашу думку, такий амортизаційний «радикалізм» вже у найближчі роки позитивно позначиться на відтворенні основних засобів.

Серед переваг сучасної амортизаційної політики в Україні слід виділити й те, що підприємства «мають право протягом звітного року віднести до валових витрат будь-які витрати, пов'язані з поліпшенням основних фондів, у сумі, що не перевищує п'ять відсотків сукупної балансової вартості груп основних фондів на початок звітного року» [85, п. 8.7.1].

Важливим кроком у напрямку ринкових перетворень амортизаційної політики України, початок якому поклав відповідний Указ Президента України, виданий у 1995 р. [див.: 196], є закріплення на законодавчому рівні ліквідації колишньої радянської практики спрямовування амортизаційних відрахувань до централізованих державних та міністерських фондів. Саме такою слід вважати практику позаекономічного перерозподілу амортизаційного фонду підприємств, згідно з положеннями нормативних документів якої з 1993 р. 25% суми амортвідрахувань підприємств перераховувалось в бюджет [див.: 111, с. 88]. Відтепер амортизаційні відрахування належать підприємствам і не підлягають вилученню [192, с. 14].

Нові позитивні аспекти в реформуванні амортизаційної політики проявляються і в особливостях індексації балансової вартості основних фондів, за індексом інфляції в разі, коли останній перевищуватиме 10% за рік. «За змістом цей індекс має відображати зміни цін на капітальні вкладення, а за формою – результат зчеплення помісячних ланцюгових індексів» [135, с. 37]. Останнє положення знайшло своє відображення спочатку в Указі Президента України «Про заходи щодо реформування інвестиційної політики в Україні» [див.: 233], зокрема у «Положенні про порядок визначення амортизації та віднесення амортизаційних відрахувань на витрати виробництва (обігу)» [див.: 171; 187], а далі продубльовано в Законах «Про амортизацію» та «Про оподаткування прибутку підприємств». «Це сприятиме наближенню фінансової амортизації до критеріїв споживання основного капіталу» [135, с. 37]. Однак, протягом 1991 – 1996 рр., як свідчить аналіз темпів інфляції та темпів індексації основного капіталу (див. табл. 1.3.8), спостерігалося «заниження відновної вартості виробничих фондів, що зумовило різке погіршення структури джерел фінансового забезпечення» [18, с. 6]. Коефіцієнти індексації «не були прив'язані до ринкових цін, внаслідок чого вартість окремих об'єктів основних фондів було спотворено» [13, с. 26]. Непродуманість індексацій основних фондів протягом вказаного періоду фактично спричинилася до їх знецінення. Так, загальна вартість основних фондів в Україні на кінець 1990 р. становила за комерційним курсом 180,5 млрд. американських доларів, що приблизно відповідало їх реальній вартості. Після проведених перших двох переоцінок вартість основних фондів на кінець 1993 р. становила в перерахунку за комерційним курсом 10,9 млрд. американських доларів, тобто скоротилася майже в 11 разів [див. 25, с. 48].

**Таблиця 1.3.8**

Динаміка споживчих цін та балансової вартості
основних фондів автотранспорту України

| Показники | РОКИ | | | | | |
|---|---|---|---|---|---|---|
| | 1991 | 1992 | 1993 | 1994 | 1995 | 1996 |
| Індекс зростання цін | 390,0 | 2100,0 | 10256,0 | 501,0 | 281,7 | 139,7 |
| Індекс зростання цін наростаючим підсумком | 390,0 | 8190,0 | 839966,4 | 4208231,6 | 11854588,0 | 16560859,0 |
| Індекс зростання балансової вартості основних фондів автотранспорту | – | 1570,0 | 2680,0 | – | 4880,0 | 830,0 |
| в т. ч.: | | | | | | |
| – будівель, споруд, передавальних пристроїв | – | 1100,0 | 3760,0 | – | 4400,0 | 750,0 |
| – транспортних засобів, машин та устаткування | – | 1800,0 | 2080,0 | – | 5200,0 | 870,0 |
| – інших основних фондів | – | 1800,0 | 2440,0 | – | 1800,0 | 740,0 |
| Індекс споживчих цін на момент індексації балансової вартості основних фондів | – | 2390,1 травень | 101160,7 серпень | – | 6023544,7 лютий | 14793627,0 травень |
| Індекс зростання балансової вартості основних фондів автотранспорту наростаючим підсумком | – | 1570,0 | 42076,0 | 42076,0 | 2053308,8 | 17042463,0 |
| в т. ч.: | | | | | | |
| – будівель, споруд, передавальних пристроїв | – | 1100,0 | 41360,0 | 41360,0 | 1819840,0 | 13648800.0 |
| – транспортних засобів, машин та устаткування | – | 1800,0 | 37440,0 | 37440,0 | 1946880,0 | 16937856,0 |
| – інших основних фондів | – | 1800,0 | 43920,0 | 43920,0 | 790560,0 | 5850144,0 |

Відставання нормативних розмірів індексації балансової вартості основних фондів автотранспорту від динаміки споживчих цін спостерігалося до індексації, що відбулася у травні 1996 р. Як видно з даних табл. 1.3.8 та графіку (рис.1.3.3), коефіцієнти індексації лише частково ліквідовували розрив, що виникав між темпами інфляції та динамікою відновної вартості основних фондів автотранспорту. Так, до моменту першої індексації, що здійснювалася у травні 1992 р., балансова вартість основних фондів автотранспорту України в 23,9 раза відставала від існуючого рівня цін; після її проведення цей розрив скоротився і становив 1,52 раза. До наступної індексації 1993 р. інфляція випереджала відновну

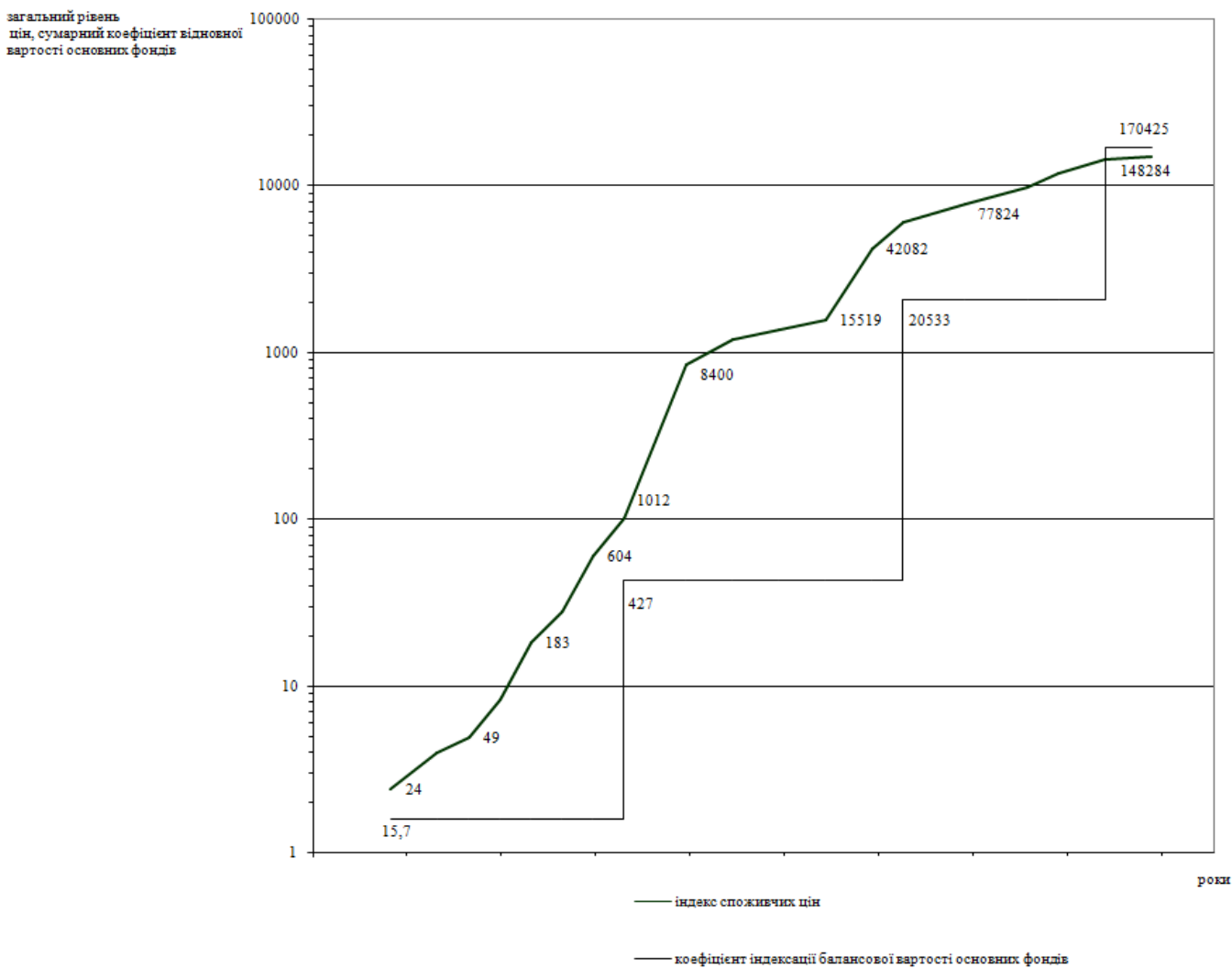

Рис. 1.3.3. Динаміка споживчих цін та коефіцієнтів індексації балансової вартості основних фондів автотранспорту України з 1992 р. по червень 1996 р.

вартість основних фондів у 64,4 раза, а після її завершення – в 2,4 раза. Але вже через півтора року загальний рівень цін у 143,2 раза випереджав сумарний коефіцієнт відновної вартості основних фондів і в результаті індексації, що здійснювалася протягом лютого 1995 р., скоротився до 2,9 раза. Однак, до травня 1996 р., часу проведення наступної індексації, він сягнув 7,2 раза. Зрозуміло, що аналогічні тенденції спостерігалися і в темпах зростання амортизаційного фонду, що нараховувався, виходячи з нової вартості основних фондів за діючими нормами амортизації. Проте, слід зауважити, що затримка у темпах зростання останніх простежувалась у порівнянні з темпами підвищення відновної вартості основних фондів. В основному це зумовлено застосуванням понижуючих коефіцієнтів щодо амортизаційних норм протягом певного періоду після здійснення індексації балансової вартості основних фондів [див.: 177; 178; 183; 186], вимогами інструкцій починати нараховувати амортизацію з наступного місяця після проведення індексацій [див.: 178] та ін.

У зв'язку з інфляцією суми амортизаційних відрахувань на повне оновлення основного капіталу передбачалось щоквартально коригувати відповідно до індексів, встановлених Міністерством фінансів України. «Проте в 1993 і 1994 рр. таке коригування амортизаційних відрахувань не проводилось. Це призвело до того, що зростаюча інфляція знецінила власні кошти підприємств, нагромаджені за рахунок амортизаційних відрахувань, і фактично девальвувала це джерело капіталовкладень» [237, с. 48].

Як справедливо зауважив український економіст А. Фукс, «ми почали «про- їдати» реноваційний фонд. В останні роки практично зупинився процес нагромадження та оновлення основних фондів» [238, с. 92]. Усе це – наслідки неадекватного відображення зростання цін у відтворювальних процесах основних фондів. Проте в економічній літературі знаходимо і протилежні міркування. Так, на думку вітчизняних вчених О. Рудченка, Н. Омельянчика, М. Тютюна, відповідність темпів індексації основного капіталу динаміці цін «призведе до ще більшої активізації інфляційних процесів, що створить своєрідне «замкнуте» коло; тобто практично немає можливості при значних річних темпах інфляції, які мають місце в Україні, здійснювати адекватну індексацію основних засобів» [206, с. 57]. Але при такому підході фактично заперечується економічний зміст амортизації, порушується безперервність процесу оновлення виробничого потенціалу підприємств. Крім того, беручи до уваги значне зниження питомої ваги амортизаційних відрахувань у собівартості продукції і послуг, що спостерігалося протягом останніх років, важко погодитись, що амортизація є одним з основних чинників, котрі зумовлюють інфляцію. Її роль у цьому процесі цілком другорядна. А тому підвищення питомої ваги амортизації у собівартості на кілька пунктів не спричинить того провокуючого впливу на інфляційні процеси, які зумовлюються, наприклад, невиправдано високими нарахуваннями на заробітну плату, неефективною податковою та кредитною політикою держави та ін.

Непродуманість і необґрунтованість фіскального законодавства України, яке до того ж не враховує особливостей інфляційної економіки, є те, що у більшості автопідприємств витрати виробництва перевищують ту суму коштів, яка залишається після сплати всіх обов'язкових платежів. Виникнення цієї диспропорції зумовлене тим, що згідно з підрахунками не менше 65 % всього заново створеного продукту у сфері вантажних автоперевезень витрачається на сплату податків. Так, наприклад, річна сума податків в розрахунку на один великотоннажний автомобіль, що здійснює міжнародні перевезення вантажів, більше, ніж у 10 разів переважає відповідний середньоєвропейський рівень (див. рис. 1.3.4). «Українські підприємці, що займаються міжнародними перевезеннями, змушені віддавати 80 – 85 % доходу, виплачуючи різноманітні податки і збори» [50, с. 72]. І це при тому, що рівень технічного стану рухомого складу, а відповідно і рівень якості транспортних послуг автопідприємств України в цілому знаходиться поза межами конкуренції.

Повертаючись до аналізу сучасної амортизаційної політики в Україні, серед її переваг слід відзначити й новизну в питаннях оцінки основних засобів. Згідно з Законом України «Про амортизацію» та «Про оподаткування прибутку підприємств» облік основних фондів здійснюється за їх залишковою вартістю. В результаті цього нововведення полегшується аналіз технічного стану основних засобів, їх зношеності та планування відтворення.

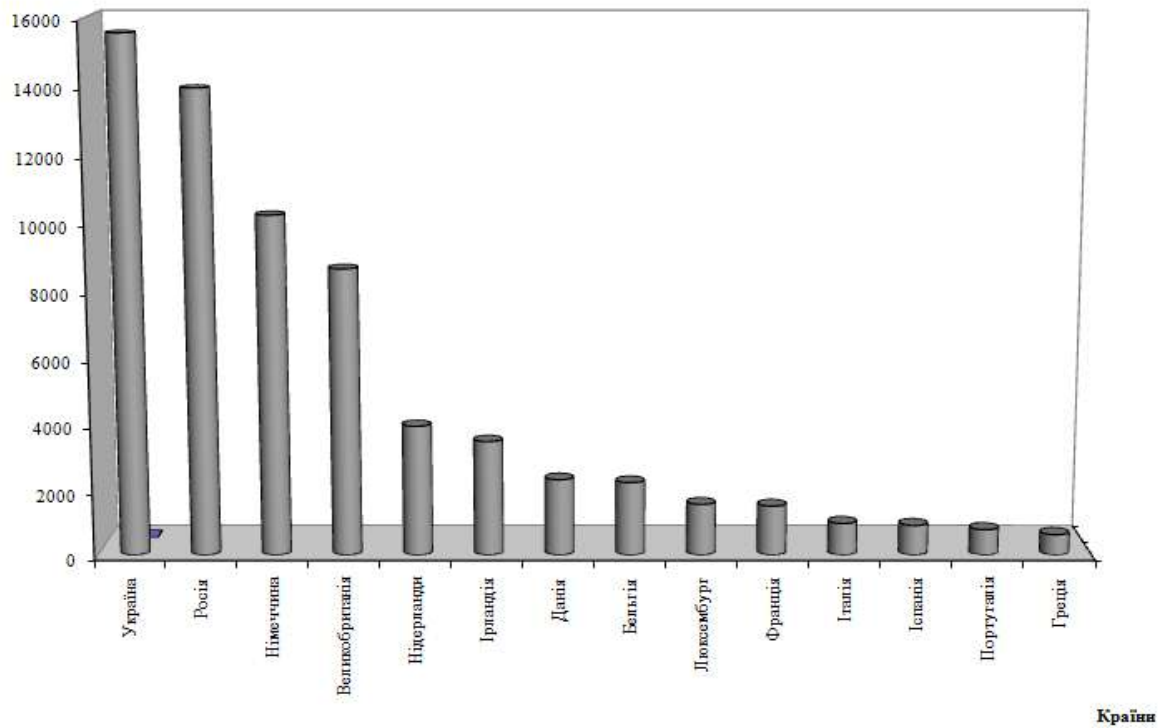

Рис. 1.3.4. Річна сума податків в розрахунку на один великотоннажний автомобіль станом на кінець 1996 року [див. також 119, с. 57]

Відповідно до чинної інструкції з бухобліку балансової вартості груп основних фондів відтепер в бухгалтерській звітності «відображається сума зносу придбаних основних засобів, що були в експлуатації» [102, п. 13]. Попередня методика оцінки основних фондів в бухгалтерській звітності «спотворювала бачення ступеня придатності (спрацьованості) основних засобів, що були в експлуатації, придбаних або отриманих з будь-яких джерел, і створювала викривлене уявлення про те, що підприємство володіє цілковито новими основними засобами» [126, с. 51]. Крім цього, усувається позаекономічна практика переамортизації основних фондів, що «створювало нічим не обґрунтовану ситуацію, коли сума нарахованої за весь період експлуатації об'єкта амортизації могла значно перевищувати його початкову вартість. Таким чином певна частина суми нарахованої амортизації наче по інерції включалася у собівартість продукції, не маючи при цьому ніякого відношення до відшкодування здійснених на підприємстві капітальних затрат.

І коли вже сам порядок нарахування амортизації допускав подібні відступи від економічного призначення даного процесу, то якоюсь мірою можна вважати виправданим механізм позаекономічного перерозподілу засобів амортизаційного фонду підприємств, який діяв у ті роки» [111, с. 88].

Серед причин недугів, якими вражена українська економіка, слід виділити недостатню теоретичну обґрунтованість нормативних документів, які регламентують економічні процеси, відсутність регулярної оцінки дієвості регламентованих положень. Такі міркування певною мірою зумовлені недосконалістю і мінливістю засад щодо застосування прискореної амортизації основних фондів. Згідно з положеннями останніх законодавчих актів України значно розширилися можливості в застосуванні прискореної амортизації основних фондів підприємств всіх форм власності. Залишилася в минулому амортизаційна політика, що включала необхідність отримання бюрократичних дозволів центральних органів на застосування прискореної амортизації, оскільки, відповідно до положень, лише незначна частка суб'єктів господарювання могла скористатися її перевагами.

Недоліком активної амортизаційної політики України, що переходить з одних положень в інші, є її однобокий підхід до встановлення способу нарахування прискореної амортизації та спрямованість лише на активну частину основних фондів. У світовій практиці серед різноманітних методів нарахування прискореної амортизації виділяють декілька загальноприйнятих способів її обчислення, що можуть використовуватись на вибір суб'єкта господарювання [див., напр., 34, с. 34 – 43; 72, с. 30 – 32; 95, с. 26; 115, с. 6 – 8; 124, с. 29; 146, с. 26 – 31; 158, с. 52 – 54; 223, с. 100 – 102;   238, с. 90 – 93;   243, с. 16 – 22 ]. Починаючи з 1991 р. прискорену амортизацію дозволялося обчислювати лише за методом зменшуваного залишку, при якому базову норму амортизації, що відповідала рівномірному способові її нарахування, дозволялося підвищувати не більше, як у два рази. Згідно з сучасними вимогами щодо нарахування прискореної амортизації виділяється не тільки єдиний спосіб її нарахування за законодавчо закріпленими диференційованими річними нормами, але й вузьку спрямованість на окремі групи активної частини основних фондів. Пришвидшення процесу амортизації пасивної частини основних фондів в Україні не практикувалось. В результаті цього автотранспортні підприємства опинилися в дискримінаційних умовах, оскільки частка основних фондів, до якої може бути застосована прискорена амортизація, не перевищує 10% як в досліджуваних вантажних автопідприємствах Івано-Франківської, Львівської, Тернопільської областей, так і по вантажному автотранспорту України в цілому. Тобто останні законодавчі акти України з питань амортизації припинили практику застосування прискореної амортизації транспортних засобів, що існувала певний час до 1997 р. [див., напр., 136, с. 50].

Звичайно, можна заперечувати доцільність прискореної амортизації рухомого складу вантажного автотранспорту, зважаючи на законодачо закріплені високі річні норми його амортизації (17,5% у 1997- 98 рр. і 25% починаючи з 1999 р.). Однак, не можна не визнавати того, що мова йде, по-перше, не просто про оновлення рухомого складу, а про необхідність докорінного реформування його застарілої структури, що не відповідає сучасним ринковим умовам з притаманною їм жорсткою конкуренцією; по-друге, «в умовах будь-якого виробництва, будь-якої соціально-економічної формації на кожній стадії розвитку людської спільноти, транспортні системи повинні розвиватися більш прискореними темпами порівняно з іншими господарськими об'єктами, співмірно як з ресурсами і можливостями держави, так і з її перспективними потребами з метою досягнення збалансованого прогресу економіки країни» [231, с. 26].

Зважаючи на сказане, стає очевидною необхідність з'ясування ролі власне прискореної амортизації рухомого складу автопідприємств України, яке буде неповним без бодай стислого аналізу реального стану нарахування амортизації та його впливу на процеси оновлення основних фондів.

За останні роки на підприємствах вантажного автотранспорту спостерігається тенденція скорочення реальних амортизаційних сум, а відповідно і їх частки у відтворювальних процесах. Виникнення цієї тенденції є цілком закономірним явищем з огляду на вікову структуру активної частини основних фондів, рухомого складу зокрема. У цьому зв'язку уже в найближчі роки за відсутності кардинальних змін в процесах оновлення парку вантажних автомобілів питома вага автотранспорту із строком служби понад 10 років, тобто повністю замортизованих (див. табл. 1.3.1), перевищить 60 %. І це при тому, що на кінець 1996 р. в цілому зношеність основних фондів досліджуваних автопідприємств становила 60,22 % (у 1992 р. – 37,03 %), коливаючись від 58,80 % в Тернопільській області до 67,74 % у Львівській області. Ось чому вже зараз заявила про себе й інша тенденція – зростання витрат на ремонт основних фондів, до того ж в обсягах, що у все більшій мірі переважають суми нарахованої амортизації. За підсумками 1994 р. співвідношення між витратами на капітальний ремонт основних фондів і річними амортвідрахуваннями досліджуваних вантажних АТП Івано-Франківської, Львівської та Тернопільської областей становило 1,08, а за результатами обчислень 1996 р. – 1,48 (див. табл. 1.3.9).

**Таблиця 1.3.9**

Порівняльна характеристика показників оновлення і вибуття основних фондів вантажних АТП Івано-Франківської, Львівської і Тернопільської областей

| Показники | Роки | |
|---|---|---|
| | 1994 | 1996 |
| Коефіцієнт оновлення основних фондів, % | 0,83 | 0,28 |
| Коефіцієнт ліквідації основних фондів, % | 3,91 | 3,46 |
| Відношення коефіцієнта ліквідації до коефіцієнта оновлення основних фондів, разів | 4,71 | 12,35 |
| Співвідношення вартості ліквідованих основних фондів і річної суми амортизації, разів | 2,05 | 3,50 |
| Співвідношення між вартістю основних фондів, що поступили, і річною сумою амортизації | 0,67 | 0,27 |
| Відношення витрат на капітальний ремонт до амортизації, разів | 1,08 | 1,48 |
| Відношення нарахованої амортизації до залишкової вартості основних фондів на початок року, % | 4,41 | 2,89 |

Критичність ситуації виявляється і в тому, що співвідношення між вартістю ліквідованих основних фондів і нарахованою амортизацією зросло протягом цього ж періоду з 2,05 раза до 3,5 раза.

Все це свідчення того, що в сучасних умовах амортизація перестає виконувати роль основного джерела відтворення основних фондів автотранспорту. Недосконалість формування амортизаційного фонду призводить у кінцевому підсумку до значних втрат купівельної спроможності накопичених фінансових ресурсів і не дозволяє підприємствам здійснювати серйозні фінансові інвестиції.

На нашу думку, амортизаційні відрахування зможуть виконувати свою економічну функцію в повному обсязі лише за умови, якщо методика їх нарахування базуватиметься на таких економічних засадах:

1) врахування закону спадної ефективності основного капіталу, що виявляється у поступовому зменшенні амортизаційного навантаження відповідно до його фізичного спрацювання та морального старіння;

2) врахування фактора часу, у зв'язку з яким критерієві забезпечення простого відтворення основних засобів через амортизацію відповідає не просто однакова величина сум нарахованої амортизації і початкової вартості капітальних активів, а приблизна рівність початкової відновної вартості основних фондів та поточної дисконтованої вартості суми амортизаційних відрахувань.

Виходячи із зазначених засад, слід оцінювати відповідність чинної методики нарахування амортизації на відновлення рухомого складу вимогам сучасного фінансового менеджменту в умовах ринку [див., напр., 27, с. 393 – 399; 36, с. 261 – 269; 115, с. 9 – 11; 131, с. 288 – 293; 146, с. 114 – 116; 163, с. 493 – 501; 166, с.82 – 83; 242, т. 2, с. 203 – 211; 256, s. 394 – 395; 258, p. 266 – 269; 263, p. 499 – 503].

«Зміни, що відбуваються в народногосподарському комплексі країни в пере-хідний період, вимагають врахування фактора постійного знецінювання грошей при проведенні техніко-економічних розрахунків» [118, с. 94]. У зв'язку з цим для визначення ефективності будь-яких фінансових інвестицій, віддача від яких розподілена в часі, використовується критерій чистої дисконтованої вартості (ЧДВ). Якщо за цим критерієм досягається додатне або нульове значення, то інвестиції вважаються ефективними. Оскільки амортизаційний фонд, процес накопичення якого відбувається в часі, покликаний забезпечити просте відтворення основного капіталу, то критерієм ефективності його формування можна вважати нульове значення чистої дисконтованої вартості. При цьому величина чистої дисконтованої вартості визначається за формулою:

$$ЧДВ = \sum \left(\left(\frac{A_t}{(1+r)^t} + \frac{E_t}{(1+r)^t}\right) - \left(Ф_в + \frac{Л_т}{(1+r)^T}\right)\right), \quad (1.3.1)$$

де $A_t$ – сума амортизації, що нарахована за період $t$, грн.;

$Ф_в$ – початкова відновна вартість основних засобів, грн.;

$E_t$ – ефект від використання амортизаційних нарахувань в інших видах діяльності підприємства до закінчення амортизаційного періоду, грн.;

$r$ – норма дисконта;

$Л_т$ – ліквідаційна вартість основних виробничих фондів на момент їх списання, грн.;

$T$ – тривалість амортизаційного періоду, роки [див., напр., 143].

Пасивне накопичення амортизації ніколи не забезпечить навіть простого відтворення основних фондів, оскільки фактор часу призводить до фінансових втрат. Часткова або повна компенсація цих втрат може бути досягнута тільки шляхом активного використання накопичуваних фінансових ресурсів в інших видах діяльності, що ґарантують швидку віддачу. Цим пояснюється наявність в наведеній формулі величини ефекту від використання накопиченої амортизації в інших видах діяльності до закінчення амортизаційного періоду і настання часу цільової реінвестиції.

Отже, враховуючи закон спадної віддачі капіталу і зміну вартості грошей в часі, проаналізуємо на відповідність ефективному відтворенню рухомого складу методику нарахування амортизації, що діяла протягом аналізованого періоду часу. Недоліком цієї методики є те, що вона передбачала однакове фінансове навантаження на один кілометр пробігу протягом всього строку експлуатації автомобіля. Сума амортизаційних відрахувань при цьому визначалася за формулою:

$$A_t = \frac{Ф_в \cdot H_a}{100} \cdot \frac{L_t}{1000}, \quad (1.3.2)$$

де $H_a$ – норма амортвідрахувань на 1000 км пробігу, %;

$L_t$ – пробіг автомобіля протягом розрахункового періоду, км.

Такий підхід в нарахуванні амортизації абсолютно не враховував об'єктивні структурні зміни в собівартості експлуатації вантажного автомобіля в процесі його зношення.

Аналіз динаміки формування амортизаційного фонду вантажного автомобіля початковою вартістю 25000 грн. засвідчив, що нарахування коштів на реновацію забезпечує до кінця строку експлуатації повернення 20700 грн. або 82,8 % початкової вартості (див. табл. 1.3.10). Навіть при мінімальній ставці дисконту в 15 %, що і так у кілька разів занижена з огляду на сучасний економічний клімат, поточна дисконтована вартість суми амортизаційних відрахувань складає лише 13557,6 грн., що становить 54,2 % від початкової вартості.

**Таблиця 1.3.10**

Динаміка надходжень амортизації при різних способах її нарахування

| Роки експлуа-тації, t | Річний пробіг автомобіля, $L_t$, тис. км | Коефіцієнт дисконтува-ння при $r = 0,15$ | Методи нарахування амортизації | | | |
|---|---|---|---|---|---|---|
| | | | рівномірний | | дегресивний | |
| | | | Річна сума амортизаційних відрахувань, грн. | Поточна дис-контована вар-тість аморти-заційних відра-хувань, грн. | Річна сума амортизаційних відрахувань, грн. | Поточна дис-контована вар-тість аморти-заційних відра-хувань, грн. |
| 1 | 70 | 0,870 | 4200,0 | 3654,0 | 8400,0 | 7308,0 |
| 2 | 66 | 0,756 | 3960,0 | 2993,76 | 5258,88 | 3975,71 |
| 3 | 61 | 0,658 | 3660,0 | 2408,28 | 3320,68 | 2185,01 |
| 4 | 56 | 0,572 | 3360,0 | 1921,92 | 2155,89 | 1233,17 |
| 5 | 50 | 0,497 | 3000,0 | 1491,0 | 1407,49 | 699,52 |
| 6 | 42 | 0,432 | 2520,0 | 1088,64 | 898,54 | 388,17 |
| Всього | 345 | X | 20700,0 | 13557,60 | 21441,48 | 15789,58 |

Отже, мінімальна сума реальних фінансових втрат автопідприємства при нарахуванні амортизації на одиницю рухомого складу становить 11442,4 грн. Крім того, у зв'язку із зростанням собівартості експлуатації автомобіля отримання останніх 2520,0 грн. досить проблематичне, оскільки в цей період експлуатація вантажівки, про що зазначалося вище, перебуває в зоні збитковості.

Як бачимо, методика нарахування амортизації, якою користувалися автопід-приємства до 1997 р., в силу притаманних їй недоліків (ігнорування закону спадної віддачі та дії фактора часу), а також неґативний вплив макроекономічних чинників як наслідок непродуманої амортизаційної політики уряду України (недоліки у встановленні відновної вартості основних фондів та її відставання від загального рівня цін і ін., що розглядалися вище) не дозволяла забезпечити ефективне оновлення рухомого складу.

Спрощена методика нарахування амортизації на відновлення рухомого складу (без врахування пробігу автомобіля), що запроваджена в Україні у 1997 р. згідно з Законом «Про оподаткування прибутку підприємств», хоча і передбачає високі амортизаційні норми, та все ж не ліквідовує вказаних внутрішніх її недоліків. А тому на кінець амортизаційного періоду автомобіля поточна дисконтована вартість амортвідрахувань при попередній ставці дисконту не перевищить 70 % його початкової вартості. Звести до мінімуму фінансові втрати в процесі формування амортизаційного фонду дозволило б застосування прискореної амортизації. Використовуючи дегресивний метод, який до згаданого закону практикувався підприємствами України, втрати в результаті знецінення вартості грошей вдалось би зменшити за рахунок різного розподілу сум амортизації протягом реноваційного періоду. При цьому, за нашими підрахунками, поточна дисконтована вартість амортизаційних відрахувань становила б 74 % від початкової вартості вантажівки. Якщо ж для нарахування амортизації скористатися кумулятивним методом, то остання величина сягнула б 76 %. Недосконалість чинного законодавства щодо застосування прискореної амортизації господарствами вантажного автотранспорту позбавляє їх переваг, що досягаються засобами активної амортизаційної політики.

На нашу думку, звести до мінімуму фінансові втрати в процесі формування реноваційного фонду на відновлення рухомого складу дозволить поєднання прискореної амортизації з методом, що врахує фактичний пробіг автомобіля. З точки зору економіки доцільність такого своєрідного тандему зумовлена, по-перше, відображенням взаємозв'язку між величиною суми нарахованої амортизації та фактичним пробігом автомобіля, тобто з його реальною спрацьованістю; по-друге, врахуванням фактора часу, що виявляється в захищеності амортизаційних сум від знецінення.

Використання методів прискореної амортизації обмежується специфічними особливостями, притаманними автотранспорту. У зв'язку з цим формування амортизаційного фонду на відновлення рухомого складу, на нашу думку, слід здійснювати дегресивним методом. При цьому сума амортизаційних відрахувань у будь-який момент часу визначатиметься за формулою:

$$A_t = Ф_з \cdot \frac{n \cdot H_a}{100} \cdot \frac{L_t}{1000}, \qquad (1.3.3)$$

де $Ф_з$ – залишкова вартість автомобіля у діючих цінах, грн.;

$n$ – коефіцієнт збільшення норм амортизації при рівномірному методі її нарахування.

З наведених в табл. 1.3.10 даних розрахунків видно, що 76 % суми амортизації поступає до 250 тис. км пробігу, а інші 24 % припадають приблизно на 100 тис. км пробігу, що залишились. Крім цього, поточна дисконтована вартість суми нарахувань при дегресивному методі на 2232 грн. перевищує аналогічну суму, нараховану рівномірним способом, а отже, і більш точно відображає реальну різницю у розмірах сум нарахованої амортизації. До речі, різниця у величині сум нарахованої амортизації обома методами без врахування фактора часу не настільки значна і становить лише 741,5 грн.

Компенсувати втрати від знецінення амортизаційних сум можна шляхом ефективного вкладання засобів, отриманих протягом перших 2 – 3-х років амортизаційного циклу. Використання наступних надходжень не тільки забезпечить реальне оновлення рухомого складу автопідприємств, але й дозволить отримати прибуток від фінансових операцій.

Використання запропонованої методики нарахування амортизації дозволить значно прискорити повернення попередньо вкладених засобів в рухомий склад, не збільшуючи собівартості перевезень і зменшуючи амортизаційне навантаження на автомобіль в міру об'єктивного зниження ефективності його експлуатації. Для визначення впливу амортвідрахувань на динаміку собівартості одного кілометра пробігу скористаємося формулою, отриманою в результаті досліджень кафедрою економіки і організації виробництва ХДАДТУ. Виведена функціональна залежність відображає кількісний взаємозв'язок між матеріальними витратами на один кілометр пробігу та загальним пробігом автомобіля з початку експлуатації. За незмінних інших умов ця залежність описується фукцією:

$$B_{м} = B_{мн} \cdot e^{\frac{m \cdot L}{100000}}, \qquad (1.3.4)$$

де $B_{м}$ – матеріальні витрати на один кілометр пробігу;

$B_{мн}$ – питомі матеріальні витрати на один кілометр пробігу для нового автомобіля;

$m$ – коефіцієнт, що враховує категорію умов експлуатації (набуває значень від 0,08 до 0,14);

$L$ – загальний пробіг автомобіля з початку експлуатації [див.: 143, с. 3].

Результати відповідних розрахунків наведено в табл. 1.3.11. Аналізуючи їх, неважко зауважити, що використання рівномірного нарахування амортизації (яке не враховує об'єктивних структурних змін у собівартості експлуатації автомобіля в процесі його спрацювання) спричинює значно вищі коливання собівартості одного кілометра пробігу, аніж використання дегресивного методу прискореної амортизації.

**Таблиця 1.3.11**

Вплив амортизаційних відрахувань на собівартість
одного кілометра пробігу вантажного автомобіля
($m = 0{,}1$; $B_{мн} = 26{,}92$ коп.; $B_{пост} = 20{,}83$ коп.; $n = 2$)

| Роки експлуа-тації, t | Річний пробіг автомобіля, $L_t$, тис. км | Питомі матеріальні втрати на 1 км пробігу, коп. | Амортизаційне навантаження на 1 км пробігу, грн. | | Собівартість 1 км пробігу, грн. | |
|---|---|---|---|---|---|---|
| | | | рівномірне нарахування амортизації, коп. | дегресивне нарахування амортизації, коп. | при рівномірному нарахуванні амортизації, коп. | при дегресивному нарахуванні амортизації, коп. |
| 1 | 70 | 27,49 | 6,0 | 12,0 | 54,32 | 60,32 |
| 2 | 66 | 29,52 | 6,0 | 7,97 | 56,35 | 58,32 |
| 3 | 61 | 31,49 | 6,0 | 5,44 | 58,32 | 57,76 |
| 4 | 56 | 33,35 | 6,0 | 3,85 | 60,18 | 58,03 |
| 5 | 50 | 35,11 | 6,0 | 2,81 | 61,94 | 58,75 |
| 6 | 42 | 36,63 | 6,0 | 2,14 | 63,46 | 59,60 |

Як видно з рис.1.3.5., амортизаційні нарахування в розрахунку на один кілометр пробігу автомобіля знижуються синхронно до зростання матеріальних витрат на забезпечення одного кілометра пробігу.

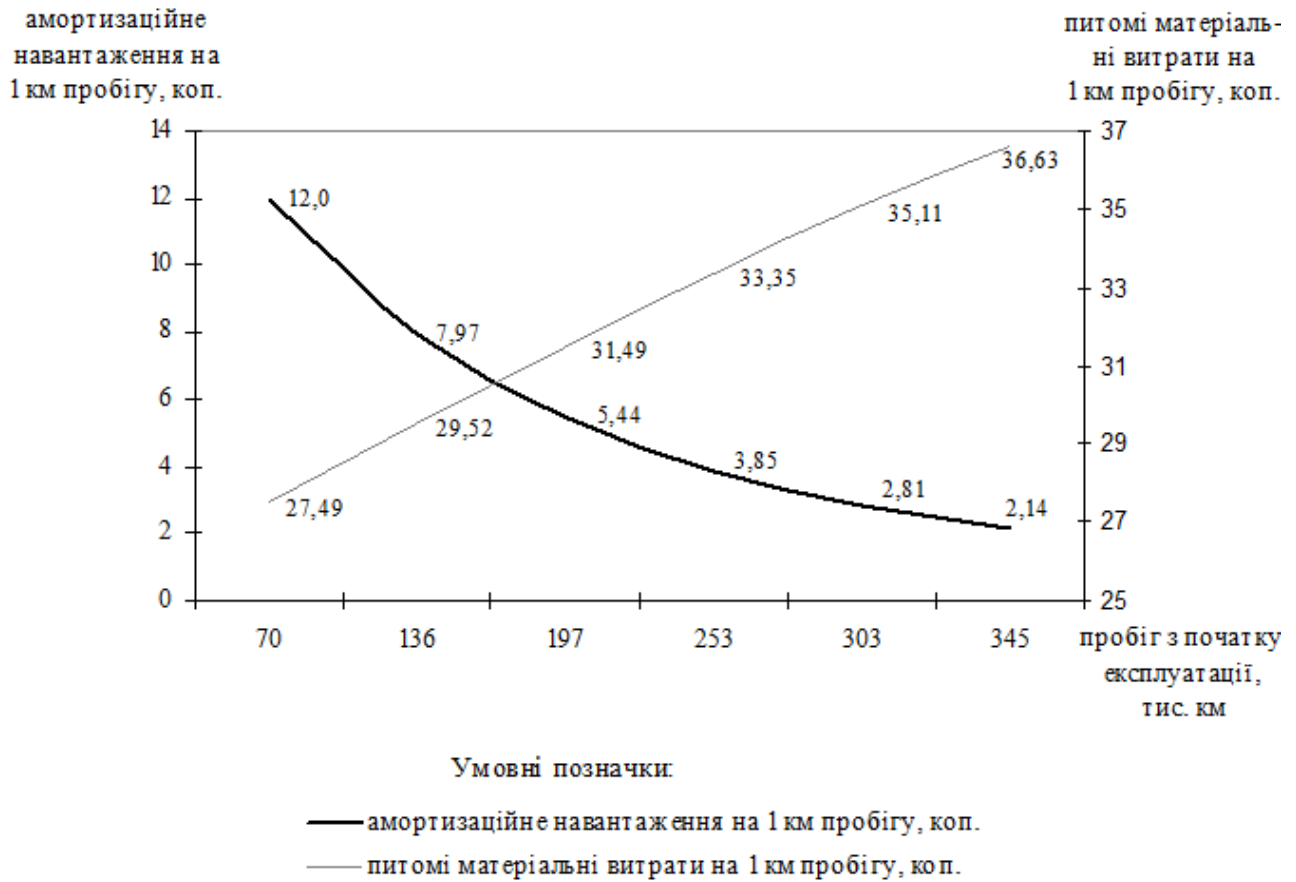

Рис. 1.3.5. Динаміка питомих матеріальних витрат і амортизаційного навантаження на один кілометр пробігу при дегресивному методі нарахування амортизації

Управління процесом формування амортизаційного фонду пропонованим методом забезпечує відповідні фінансові умови для підтримання раціонального рівня рентабельності експлуатації автомобіля протягом всього амортизаційного періоду.

Коефіцієнт зростання рівномірної норми амортизації, від якого залежать темпи нагромадження реноваційного фонду, кожне автопідприємство повинне визначати самостійно, керуючись при цьому співвідношенням собівартості перевезень і тарифом, що склався на ринку транспортних послуг. А тому побоювання, пов'язані з ймовірним різким зростанням тарифів на автоперевезення вантажів при наданні самостійності підприємствам у виборі способу нарахування амортизації, не зовсім обґрунтовані. Основним реґулятором в цій ситуації виступає ринкова вартість транспортних послуг, рівень якої зумовлюється не бажанням окремого автогосподарства, а співвідношенням попиту і пропозиції на автомобільні перевезення вантажів.

Практика запровадження прискореної амортизації пов'язана не тільки з реґулюванням відтворювальних процесів. Її вплив виявляється в усіх сферах економічної та виробничої діяльності як окремого суб'єкта господарювання, так і економічної системи в цілому. Підсумовуючи, окреслимо головні наслідки та форми вияву позитивної дії активної амортизаційної політики:

1) прискорюється обіг як основного, так і оборотного капіталу. Амортизуючи протягом першої половини строку служби основних фондів до 65 – 80 % їх вартості, підприємства отримують можливість не тільки швидше накопичити кошти для оновлення виробничого потенціалу, але й реалізувати недоамортизовані основні засоби, замінивши їх принципово новими. Такі дії підприємств значно підвищують ефективність виробництва, знижуючи ресурсомісткість продукції та послуг, підвищуючи їх конкурентоздатність і прискорюючи обіг основного капіталу.

За підрахунками англійської комісії з трудових ресурсів зростання загальної ефективності американської та японської промисловості на 25 % є результатом змін у характері використання капіталовкладень і на 60 % – за рахунок змін у технології.

До того ж додатковий приплив коштів в результаті прискореної амортизації основних фондів розширює можливості підприємств, дозволяючи оперувати значно більшими сумами вільних коштів та вкладати їх у інші сфери підприємницької діяльності, прискорюючи цим обіг оборотного капіталу.

Надаючи підприємствам можливості швидкого відшкодування витрат, які вкладаються в основні фонди, держава має право розраховувати на те, що кошти, які побували в обігу, знову використовуватимуться для фінансування капітальних вкладень. Однак у певних умовах і в конкретні проміжки часу для підприємства більш важливим може стати не інвестиційна, а яка-небудь інша діяльність, що вимагає затрат коштів. У цих випадках у підприємства не повинно бути ніяких перешкод на шляху вибору сфери застосування власних фінансових ресурсів. «Економічний сенс такого підходу полягає у створенні підприємствам можливостей формувати фінансові фонди для технологічного оновлення виробництва, не чекаючи вироблення технічного ресурсу об'єкта» [111, с. 89];

2) таке постійне маніпулювання тимчасово вільними амортизаційними коштами частково знижує згубний вплив інфляційних процесів на фінансовий стан підприємств;

3) прискорена амортизація внаслідок пришвидшення обігу основного капіталу знижує його втрати від морального старіння. «Проведені дослідження стверджують, що структура наявного парку автомобілів в Україні не відповідає попиту на транспортні послуги як за вантажністю, так і за продуктивністю, прибутковістю, паливною економічністю, технічними та експлуатаційними якостями» [160, с. 8]. Розгляд у площині морального старіння невідповідності наявного рухомого складу сучасним техніко-економічним вимогам виявляє себе, у першу чергу, в послабленні позицій як на внутрішньому, так і на зовнішньому ринках транспортних послуг. Так, зокрема, уже сьогодні це пов'язано з обмеженим допуском вантажних автомобілів вітчизняних автопідприємств до країн ЄС через їх невідповідність екологічним нормам міжнародного стандарту EURO – 1, запровадженого з січня 1995 р. та стандарту EURO – 2, запровадженого у 1996 р. [див.: 138, с. 12; 249, с. 84]. «У першу чергу така заборона стосується українських КрАЗів, російських КамАЗів, ЗІЛів, ГАЗів та білоруських МАЗів, які не відповідають жодній екологічній вимозі ЄЕС» [138, с. 12] в основному через укомплектованість двигунами, що «мають незадовільні показники по задимленню і малу потужність, не призначену для далеких перевезень» [14, с. 10]. Саме такими автомобілями укомплектовані вітчизняні автогосподарства. Ось чому багато українських автопідприємств, які займаються вантажними перевезеннями в Європі, змушені перекомплектовувати свої автопарки. За словами заступника міністра транспорту України В.Реви, щоб стати конкурентоспроможними, автопідприємства змушені купувати вантажні іномарки «Volvo», «Mercedes», «MAN», «Scania», «Iveco» [див.: 65, с. 11]. У зв'язку з цим використання прискореної амортизації сприятиме не тільки швидкій перекомплектації, але й оновленню рухомого сладу відповідно до вимог міжнародних нормативних документів, яке дозволить формувати зустрічні вантажопотоки, що є неодмінною умовою рівноцінного взаємообміну відкритих економічних систем;

4) прискорена амортизація як дієвий засіб оновлення виробничого апарату забезпечує на цій основі підвищення конкурентоспроможності підприємств на внутрішньому та зовнішньому ринках, а щодо автогосподарств виявляється, в першу чергу, у формуванні ефективної структури рухомого складу, що незаперечно позначиться на рівневі якості надання транспортних послуг.

До цього слід додати, що просте відтворення передбачає таку заміну засобів праці, при якій кількість і якість виконуваної роботи не змінюється. Проте вартість їх заміщення може бути вищою від вартості купівлі. Ось чому «амортизаційні нарахування слід обчислювати не на базі ціни купівлі, а саме на основі ціни заміщення» [12, с. 106]. Роль активної амортизаційної політики у цьому процесі незаперечна;

5) застосування підприємствами активної амортизаційної політики при відповідному фіскальному законодавстві призводить до різного розподілу прибутку в часі, а отже, й різної в часі структури податку, що стягується з підприємств. Тобто, господарства, використовуючи прискорену амортизацію, зменшують величину прибутку, що підлягає оподаткуванню, а отже, й самого податку;

6) використовуючи прискорену амортизацію, підприємства отримують певні «податкові пільги». Занижені суми податку протягом першої половини амортизаційного періоду в такій же мірі будуть завищені протягом другої, і в цілому підприємство сплатить таку ж суму податку, як і при рівномірному нарахуванні амортизації. Але, нараховуючи її прискореними методами, підприємство фактично відстрочує сплату певної частини податку, залишаючи її в своєму розпорядженні, отримуючи, таким чином, від держави своєрідну «безпроцентну позику», а сам факт такої відстрочки у сплаті частини податку збільшує прибутковість капітальних вкладень. Крім того, відстрочуючи день сплати податку, підприємці сподіваються, що й величина самого податку з часом зменшиться.

В американській літературі наводяться факти, згідно з якими деяким фірмам вдалося досягнути значних успіхів у своїй діяльності завдяки застосуванню прискореної амортизації основного капіталу, що дозволило їм отримувати високі прибутки і виплачувати, відповідно, високі дивіденди по акціях, сплачуючи при цьому низькі податки;

7) широке застосування прискореної амортизації на підприємствах України за нормальних умов оподаткування призвело б до зменшення потреби підприємств в коштах на реновацію. Отже, прискорена амортизація – це спосіб певною мірою послабити залежність підприємств від зовнішнього кредитування і тим самим скоротити грошову емісію. Це тим більше актуально за умов, коли частка амортизаційних відрахувань у сукупному самофінансуванні підприємств України, за підрахунками вітчизняних вчених, складає сьогодні від 1 до 7 % (у США, наприклад, – 70 – 90 %) [див.: 156, с. 31];

8) активна реноваційна політика підприємств забезпечує раціональне поєднання амортизації із зростанням витрат на капітальний ремонт. «Практика показує, що спрацювання машин збільшується у перший період їх експлуатації, період їх освоєння» [238, с. 90];

9) впровадження прискореної амортизації основних фондів дозволяє «об'єктивно оцінити амортизаційну складову в структурі ринкової ціни і відповідно забезпечити (при незмінних інших умовах) адекватність обміну між суб'єктами господарювання різних форм власності» [10, с. 98];

10) активна амортизаційна політика дозволяє якомога швидше повернути початкові вкладення з метою зниження наступного ризику [див., напр., 250, с. 48];

11) активна амортизаційна політика як один із засобів покращення фінансових показників підприємства в кінцевому підсумку слугує орієнтиром для потенційного інвестора і таким чином дозволяє формувати ринковий інвестиційний клімат. Адже, як відомо, амортизаційні відрахування – це база для визначення ефективності інвестицій. Раніше ці ключові економічні поняття були відірваними одне від одного: амортизація визначалася, виходячи з технічних факторів в основному – фізичної спрацьованості основних фондів; ефективність капіталовкладень – через приведені витрати чи строк окупності;

12) в сучасних умовах ризик завищення амортизаційних відрахувань менш шкідливий, ніж їх недобір, оскільки останній пов'язаний з необхідністю тривалої експлуатації застарілих основних виробничих фондів та здійсненням ремонтів, що дорого коштують. А прискорене оновлення основного капіталу в межах економічної системи сприяє формуванню ринку засобів виробництва та розвитку гуртової торгівлі ними.

Таким чином, система прискореної амортизації виступає не тільки як завуальована форма державного фінансування капітальних вкладень, але й при відповідному законодавчому забезпеченні може стати дієвим чинником нормалізації процесу відтворення основних виробничих фондів народного господарства України, а також підвищення готовності підприємств до інновацій та пришвидшення їх адаптації до найновіших науково-технічних результатів. Враховуючи те, що 30 % потенціалу конкретного інноваційного процесу припадає на підприємство, яке його реалізує, і 70 % – опосередковано підвищує ефективність інших підприємств, роль прискореної амортизації в цьому процесі заслуговує на значно більшу увагу з боку конкретних суб'єктів підприємницької діяльності та державної інноваційної політики в цілому.

Підсумовуючи сказане, слід зазначити, що відтворювальний процес є системою, якій притаманні внутрішні пропорції, методи їх регулювання і балансування. З точки зору системного підходу особливо важливим елементом є виробничі пропорції. Зазвичай, увага надається їх окремим елементам, а не розгляду у взаємозв'язку. Пояснення цьому, на нашу думку, криється в нерозробленості чіткої критеріальної основи, що визначає, по-перше, види пропорцій, їх місце і роль як окремих елементів системи; по-друге, внутрішню структуру пропорцій відтворення, її спрямованість на перехід до інтенсивного типу оновлення основних фондів.

Реалізація цього вимагає вдосконалення відтворювальних пропорцій перш за все співвідношенням між новими засобами праці, спрямованими, з одного боку, на розширення основних фондів, з другого – на заміну їх застарілої частини. Це першопочаткова відтворювальна пропорція, що визначає найважливіший орієнтир інтенсивного оновлення фондів – економічний строк служби. Різновидом цього співвідношення є пропорція в капітальних вкладеннях, що використовуються для розширення і заміни діючих основних фондів новими, які поліпшують технологічність виробничого процесу, а також на формування активної і пасивної частин цих фондів. Таким чином, перша пропорція відображає єдність руху натуральної та вартісної форм функціонування основних фондів.

Похідною від неї є пропорція, що відображає методи оновлення основних фондів у співвідношенні капіталовкладень на переозброєння і реконструкцію виробничого потенціалу, з одного боку, та розширення виробництва і нове

будівництво – з другого. Ця пропорція, незважаючи на її похідний характер від першої, і на те, що вона виступає конкретною формою її реалізації, має цілком самостійну значущість. Внаслідок цього, її слід вважати другою основною пропорцією з тих, які розглядаються в цьому дослідженні.

Суть третьої відтворювальної пропорції становлять співвідношення між різними джерелами капіталовкладень. В економічній літературі з цього приводу ведеться мова, як правило, лише про два джерела капіталовкладень – амортизаційний фонд і частину нерозподіленого прибутку, які є далеко не єдиними серед переліку власних інвестиційних засобів підприємств. І зовсім мало уваги надається зовнішнім, як залученим, так і державним, джерелам фінансування. В сучасних умовах такий підхід слід вважати несправедливо завуженим.

Особливо важливу роль в реалізації інтенсивного типу оновлення основних фондів відіграє співвідношення між елементами активної і пасивної частин основних засобів, з одного боку, і ремонтним господарством – з другого. Серед досліджуваних ця пропорція є четвертою. Вона теж рідко розглядалася та аналізувалася в економічній літературі, хоча у суттєвій значущості співвідношення між обсягами засобів, скерованих на реновацію і ремонт, сумнівів не викликає. Конкретний вислід цієї пропорції змінюється при переглядах строків і норм амортизації. Згідно з чинними положеннями власник всього амортизаційного фонду і основних засобів виступає в одній особі. Проте встановлення величини норм амортизації є прерогативою уряду. Підтримання четвертої пропорції на економічно обґрунтованому, раціональному рівні стає для власників основних фондів одним із головних завдань, від вирішення яких залежить і метод нарахування амортизації, і питома вага накопичень за рахунок засобів амортизаційного фонду, які виділяються на модернізацію виробництва (а отже, і співвідношення між власними засобами фінансування відтворення та залученими чи позиченими). Таким чином, від того, як визначається ця пропорція, в значній мірі залежить перспективна програма інвестування.

Інтенсивний тип оновлення основних фондів припускає, що вказані вище відтворювальні пропорції забезпечують своєчасність такого оновлення згідно з економічно обґрунтованими строками (при рівномірному нарахуванні амортизації) або строками, що визначаються кон'юнктурними міркуваннями власників основних фондів (при використанні методів прискореної амортизації чи інших способів).

Аналіз процесу відтворення виробничого апарату України засвідчив, що в період економічної кризи саме ця сфера виявилася найбільш вразливою. Це зумовлено тим, що, по-перше, процес відтворення основних фондів був деформований і в періоди позитивних темпів зростання показників соціально-економічного розвитку. По-друге, в умовах високих темпів інфляції, а тим більше гіперінфляції, об'єктивно скорочується інтерес до нагромадження та інвестицій. По-третє, в останні роки проблемам реґулювання відтворення виробничого апарату та підтримки інвестицій надавалася недостатня увага. Державні програми економічних реформ були спрямовані, головним чином, на впровадження жорстких монетаристських методів реґулювання економіки. Тим часом досвід країн ринкової економіки, які пройшли кризові періоди, переконує в необхідності структурної перебудови, активізації виробництва на основі стимулювання відтворення виробничого апарату, його оновлення і технічного переозброєння. Формування ринку капіталу та процес відтворення основних фондів потребують цілеспрямованого реґулювання, оскільки вони в значній мірі визначають ефективність суспільного виробництва, темпи впровадження науково-технічного проґресу, зростання національного доходу і добробуту населення.

# РОЗДІЛ 2
# ЕФЕКТИВНІСТЬ ВИКОРИСТАННЯ ОСНОВНИХ ВИРОБНИЧИХ ФОНДІВ: ФАКТОРИ ВПЛИВУ, РЕЗЕРВИ ЗРОСТАННЯ

## 2.1. Показники ефективності використання основних виробничих фондів вантажних автотранспортних підприємств

Серед показників оцінки ефективності використання основних виробничих фондів центральне місце займає фондовіддача. В основі її обчислення лежить кількісне співвідношення ефекту і витрат, необхідних для його отримання. Звернемо увагу на ефект, тобто чисельник показника фондовіддачі. З цього приводу серед економістів є різні точки зору. Одні вважають, що для обчислення показника фондовіддачі слід використовувати обсяг валової, товарної або реалізованої продукції. Інші дотримуються думки, що лише чиста продукція і прибуток найбільш повно і точно відображають віддачу основних виробничих фондів, і що саме вона може бути вимірником фондовіддачі.

Не можна не погодитися, що запропонована рядом економістів [3; 19; 72;73] в якості вимірника фондовіддачі валова продукція повніше, ніж інші показники, відображає загальний обсяг виробництва. Проте вона не позбавлена і певних недоліків. На величину валової продукції істотно впливають зрушення в асортименті продукції, зміна цін на сировину, матеріали, паливо, енергію, зміна матеріаломісткості продукції, зміна галузевої структури народного господарства та масштаби спеціалізації і кооперування.

Небездоганний у цьому розумінні і показник товарної продукції, оскільки не враховує затрат праці у напівфабрикатах і незавершеному виробництві.

Рекомендований деякими авторами [4; 19] показник обсягу реалізованої продукції найбільш повно відображає результативність діяльності підприємства в умовах ринкової економіки. Проте в сучасних економічних умовах, що склалися в Україні, показник реалізованої продукції, як і попередні, хибує деякою неточністю. Його величина значною мірою залежить від фінансового стану підприємств-споживачів, а отже, він не може точно відображати ефективність використання основних фондів.

Значна частина економістів [162; 220] перевагу віддають показнику чистої продукції. Однак, не зважаючи на його істотні переваги порівняно із розглянутими показниками, йому притаманні і певні недоліки: по-перше, чиста продукція може змінюватися за рахунок зміни чисельності працівників і фонду заробітної плати; по-друге, як і попередні показники, він чутливий до зміни цін і асортиментних зрушень. Тому і цей показник не завжди відображає справжні витрати праці на підприємствах.

«В якості показника продукції для автотранспортних підприємств використовується загальний обсяг доходів» [162, с. 142], отриманий за перевезення вантажів та надання інших послуг. Цей показник дозволяє більш чітко пов'язувати роботу автопідприємств із споживачами їхніх послуг.

Ряд науковців [6; 107] твердять, що показником, який «найбільш повно виражає критерій економічної ефективності використання основних виробничих фондів діючих автотранспортних підприємств, слід вважати величину прибутку на одиницю основних фондів – рентабельність основних виробничих фондів. Головна перевага, що дозволяє вважати рентабельність основних фондів синтетичним показником ефективності їх використання, полягає у тому, що цей показник відображає як кількісний, так і якісний бік використання основних фондів» [107, с. 30 – 31].

Вартісні показники, що використовуються для обчислення фондовіддачі автотранспортних підприємств, відображають загальні обсяги виконаних робіт (перевезення вантажів, ремонт і технічний огляд рухомого складу інших організацій, реалізація продукції ремонтних майстерень і профілакторіїв, експедиційні операції тощо). За даними роботи вантажних автотранспортних підприємств Івано-Франківської, Львівської і Тернопільської областей за 1992 і 1996 рр. їх доходи за перевезення вантажів складали відповідно 98,0 і 97,6 % від загальної суми доходів. Отже, найбільш повним показником використання основних виробничих фондів автотранспортних підприємств є випуск продукції (надання транспортних послуг) у грошовому виразі в розрахунку на одиницю вартості основних виробничих фондів.

Вартісний показник фондовіддачі має ряд важливих переваг: він найбільш простий і наочний, може обчислюватись як по окремих цехах, дільницях, так і в цілому по автотранспортному підприємству. Він забезпечує можливість охопити діяльність всього комплексу різноіменних засобів праці. Тобто цей показник дозволяє відобразити у величині фондовіддачі автотранспортного підприємства фондовіддачу окремих робочих і силових машин та устаткування, а також інших основних фондів, за допомогою яких виробляються і надаються послуги іншим підприємствам та організаціям.

Однак, не дивлячись на велике значення вартісного показника фондовіддачі, йому притаманні певні недоліки, головним з яких є його знеособленість. Вартісний показник фондовіддачі не дає конкретного уявлення про те, які засоби виробництва і в якій кількості включає в себе той чи інший обсяг фондів, виражених у грошовій формі.

На вартісний показник фондовіддачі основних фондів автотранспортних підприємств значний вплив має величина середньої дохідної ставки, яка визначається рівнем тарифів і умовами перевезення вантажів, а також величина цін на продукцію допоміжних виробництв, що реалізується автотранспортним підприємством іншим організаціям і підприємствам.

Таким чином, зміна рівнів тарифів на вантажні перевезення і відпускних цін на продукцію окремих цехів і дільниць автотранспорних підприємств при незмінних об'ємних показниках продукції викликає відповідну зміну вартісного показника фондовіддачі основних фондів, що є його недоліком, внаслідок чого аналіз величини вартісного показника фондовіддачі повинен проводитись лише у порівняльних тарифах і цінах.

На грошову оцінку обсягу продукції автомобільного транспорту виявляє певний вплив і спеціалізація рухомого складу. Використовуючи для перевезень вантажів спеціалізовані автомобілі, автотранспортні підприємства відповідно до прейскуранту на перевезення вантажів можуть підвищувати тарифи до 25 %, а при перевезенні крупногабаритних вантажів – до 60 %, внаслідок чого змінюється і величина їх доходів за перевезення вантажів, при цьому підвищення доходів часто непропорційне підвищенню вартості спеціалізованого рухомого складу і його продуктивності. «Оскільки величина фондовіддачі значною мірою залежить від рівня цін і тарифів на транспортну продукцію, то розрахунок її у вартісній формі дещо умовний. Різні тарифи за транспортну роботу можуть призвести до підвищення або зниження фондовіддачі проти справжньої» [19, с. 34]. Тому обчислення вартісного показника фондовіддачі можна здійснювати після попереднього аналізу вихідних даних, виявлення конкретних умов, встановлення порівняльності обсягів транспортної роботи, цін і тарифів, величини основних фондів.

У сучасних умовах, що склалися в Україні, використання останніх є проблематичним з огляду на їх неспівставність. Значна дебіторська заборгованість, що у структурі обігових коштів транспортних підприємств складає майже 50 % [див. 228], спотворює уявлення про обсяги виконаної транспортної роботи, а, відповідно, і величину показника фондовіддачі. У зв'язку з цим на сучасному етапі економічних перетворень доцільнішим є використання натуральних показників фондовіддачі, аналіз яких слід доповнювати комплексом споріднених показників (наприклад, продуктивністю спискової автомобіле-тонни, рентабельністю основних виробничих фондів та ін.), що дозволяє створити цілісну характеристику ефективності використання основних засобів автогосподарств.

Показник фондовіддачі основних виробничих фондів у натуральній формі, який використовується сьогодні на автомобільному транспорті (величина транспортної роботи, що припадає на одиницю вартості основних фондів), характеризує використання основних виробничих фондів вантажних автотранспортних підприємств в цілому, з точки зору його транспортної роботи.

Зупинимо свою увагу на показнику обсягу роботи автомобільного транспорту. Відомо, що продукцією вантажного автомобільного транспорту є перевезення вантажів. Обсяг роботи автомобілів, які працюють за відрядною формою, характеризується натуральними показниками – обсягом перевезеного вантажу – в тоннах або вантажообігом – в тонно-кілометрах, а для автомобілів, що працюють почасово – кількістю автомобіле-годин роботи на лінії.

Однак, застосування натуральних показників для характеристики транспортної роботи ускладнюється специфічними для автомобільного транспорту особливостями: залежністю їх величини від відстані перевезення вантажів для автомобілів, що працюють відрядно, і неможливістю відобразити обсяг роботи автотранспортного підприємства в цілому одним показником.

Що стосується транспортної роботи, то і тут серед економістів немає єдиної думки. Одні з них вважають, що в якості вимірника фондовіддачі повинен виступати вантажообіг (тонно-кілометри), інші – віддають перевагу тоннам перевезеного вантажу, а треті – дотримуються думки, що таким показником може виступати «тран», який обчислюється шляхом добутку «тонно-кілометрів» на «квадрат швидкості» [див., напр., 21].

Отже, всі ці показники певною мірою збагачують методику визначення обсягу виконаної транспортної роботи, хоча кожен з них несе своє відповідне економічне навантаження. В кожному з цих показників по-своєму відображається обсяг роботи, виконаної транспортом. Разом з тим всім цим показникам притаманні певні недоліки. Якщо перший (вантажообіг) орієнтує транспортників на далекі перевезення, то другий (тонни перевезених вантажів) стимулює перевезення вантажів на близькі відстані. Цих недоліків певною мірою позбавлений показник «тран». Однак методика його розрахунку (добуток «тонно-кілометрів» на «квадрат швидкості доставки») знижує його зрозумілість і зручність користування, завищує обсяг транспортної роботи.

На нашу думку, в основі обчислення головного оціночного показника роботи транспорту визначальними повинні бути: **кількісна, просторова і часова характеристики** (виділ. наше. – О.В.). Це випливає з того, що обсяг і якість роботи транспорту можуть бути виражені трьома параметрами: кількістю тонн перевезених вантажів, відстанню перевезення та витратами часу на перевезення. Будувати цей показник необхідно таким чином: спочатку слід перемножити кількість тонно-кілометрів на затрати часу за нормою (виходячи з експлуатаційної швидкості), після чого одержаний результат розділити на фактичні затрати часу, тобто як добуток обсягу фактично виконаного вантажообігу на коефіцієнт виконання норм часу, пов'язаного із перевезеннями вантажів. Його можна виразити у вигляді такої формули:

$$P = \Sigma m \cdot L \cdot \frac{t_\text{н}}{t_\text{ф}}, \qquad (2.1.1)$$

де $P$ – вантажообіг, скоректований на коефіцієнт виконання норм часу;

$m$ – кількість тонн перевезених вантажів;

$L$ – відстань перевезення, км;

$t_\text{н}$, $t_\text{ф}$ – затрати часу на перевезення вантажів та вантажно-розвантажувальні роботи відповідно за нормами і фактично.

Розглянемо методику розрахунку показника на прикладі автомобільного транспорту (дані умовні).

**Таблиця 2.1.1**

Вихідні дані для розрахунку транспортної роботи

(за запропонованим нами методом)

| Варіант | Перевезено вантажів, тонн | Відстань перевезення, км | Затрати часу, годин | | Коефіцієнт виконання норм часу гр.4 : гр.3 | Вантажообіг з врахуванням затрат часу на перевезення гр.1 х гр.2 х гр.5 |
| | | | за нормою | фактично | | |
|---|---|---|---|---|---|---|
| А | 1 | 2 | 3 | 4 | 5 | 6 |
| І | 10 | 50 | 15 | 14 | 1,071 | 535,5 |
| ІІ | 10 | 50 | 15 | 15 | 1,000 | 500,0 |
| ІІІ | 10 | 50 | 15 | 16 | 0,9375 | 468,75 |

Отже, загальний обсяг вантажообігу з врахуванням затрат часу на перевезення вантажів становить: для першого варіанту 10 · 50 · 1,071 = 535,5 ткм; для другого варіанту 10 · 50 · 1,0 = 500,0 ткм; для третього варіанту 10 · 50 · 0,9375 = 468,75 ткм.

Таким чином, зниження затрат часу на перевезення вантажів і вантажно-розвантажувальні роботи порівняно із встановленим нормативом веде до збільшення обсягу вантажообігу, і навпаки, збільшення затрат часу на виконання цих робіт – до зменшення обсягу транспортної роботи.

Так, у першому варіанті зниження затрат часу на перевезення вантажів порівняно із нормативом на одну годину привело до збільшення обсягу вантажо-обігу на 35,5 ткм (535,5 – 500). У третьому варіанті збільшення затрат часу на перевезення вантажів порівняно із нормативом на одну годину привело до зменшення обсягу вантажообігу на 31,25 ткм (468,75 – 500).

Пропонуючи цей показник, ми не претендуємо на його бездоганність. Він, як і вантажообіг та «тран», стимулює дальні перевезення вантажів. Однак, на відміну від них, водночас орієнтує транспортників на пошуки найкоротших маршрутів доставки вантажів, зниження затрат часу на їх доставку.

Впровадження цього показника в практику роботи транспортних підприємств дозволить стимулювати зниження затрат часу на оформлення документів, вантажно-розвантажувальні роботи, часу перебування вантажів в дорозі, часу на простоювання їх на станціях, у портах і аеропортах, на здійснення транзитних операцій.

Це забезпечить підвищення ефективності роботи не лише транспортних підприємств, ступінь використання їх основних виробничих фондів, а й підприємств і організацій, які обслуговуються ними. В свою чергу, скорочення часу транспортування вантажів дозволяє знизити витрати матеріальних, трудових і фінансових ресурсів на перевезення і вантажно-розвантажувальні роботи, підвищити продуктивність рухомого складу.

До цього часу ми розглядали лише чисельник показника фондовіддачі. Правильне визначення рівня використання засобів праці вимагає розгляду й другого його елемента – знаменника.

Щодо знаменника показника фондовіддачі серед економістів немає єдиної точки зору. Одні вважають, що для побудови показника використання засобів праці в якості знаменника слід брати середньорічну вартість основних фондів; інші рекомендують основні фонди брати за залишковою вартістю. Ця точка зору ґрунтується на тому, що основні виробничі фонди в процесі використання втрачають свою продуктивність пропорційно до рівня їх зношення. До того ж такий підхід відповідає сучасним вимогам бухгалтерського обліку основних засобів.

Побудова показників використання основних фондів на базі залишкової вартості, на наш погляд, викривлювала б справжній стан їх використання, штучно занижуючи темпи зміни фондовіддачі в періоди оновлення основних виробничих фондів, а впровадження такого показника в практику не стимулювало б заміни старих основних засобів новими.

Багато науковців в якості знаменника показника фондовіддачі пропонують брати не вартість використаних основних виробничих фондів, а їх спожиту частину, тобто амортизацію. Такий підхід найбільшою мірою відповідає співвідношенню ефекту (результату діяльності) і витрат.

Недоліком такого підходу є його обмежений характер, оскільки доцільність використання нарахованих сум амортизації в якості знаменника показника фондовіддачі не поширюється на амортизаційні суми, нараховані прискореними методами.

Аналіз різних підходів в обчисленні показника фондовіддачі дозволяє зробити висновок про те, що всі вони збагачують методологію економічного дослідження ефективності використання основних виробничих фондів, оскільки кожен із них несе відповідне економічне і смислове навантаження.

Фондовіддача основних виробничих фондів автотранспортних підприємств формується під впливом великої кількості факторів, особливості взаємодії яких досить різнобічні. За видами впливу їх можна розподілити на прямі і непрямі, а за характером і умовами впливу – на виробничі, організаційні, природньо-кліматичні.

Наявність великої кількості факторів, що впливають на рівень фондовіддачі вантажних АТП, вимагає розробки їх обґрунтованої класифікації, яка давала б можливість одержання об'єктивної оцінки сили впливу кожного з них на його рівень і динаміку.

Визначення впливу різних факторів на фондовіддачу здійснюється шляхом поділу їх на дві групи: фактори, що впливають на формування обсягів вантажообігу, і фактори, що впливають на величину основних виробничих фондів (рис. 2.1.1). В процесі аналізу факторів фондовіддачі вивчається динаміка обсягів виконаної транспортної роботи і вартості основних виробничих фондів.

Одним із найбільш важливих факторів, що впливає на обсяг основних виробничих фондів, є їх оцінка. «Оцінка основного капіталу відіграє велику роль для правильного визначення його загальної величини, складу і структури, а також

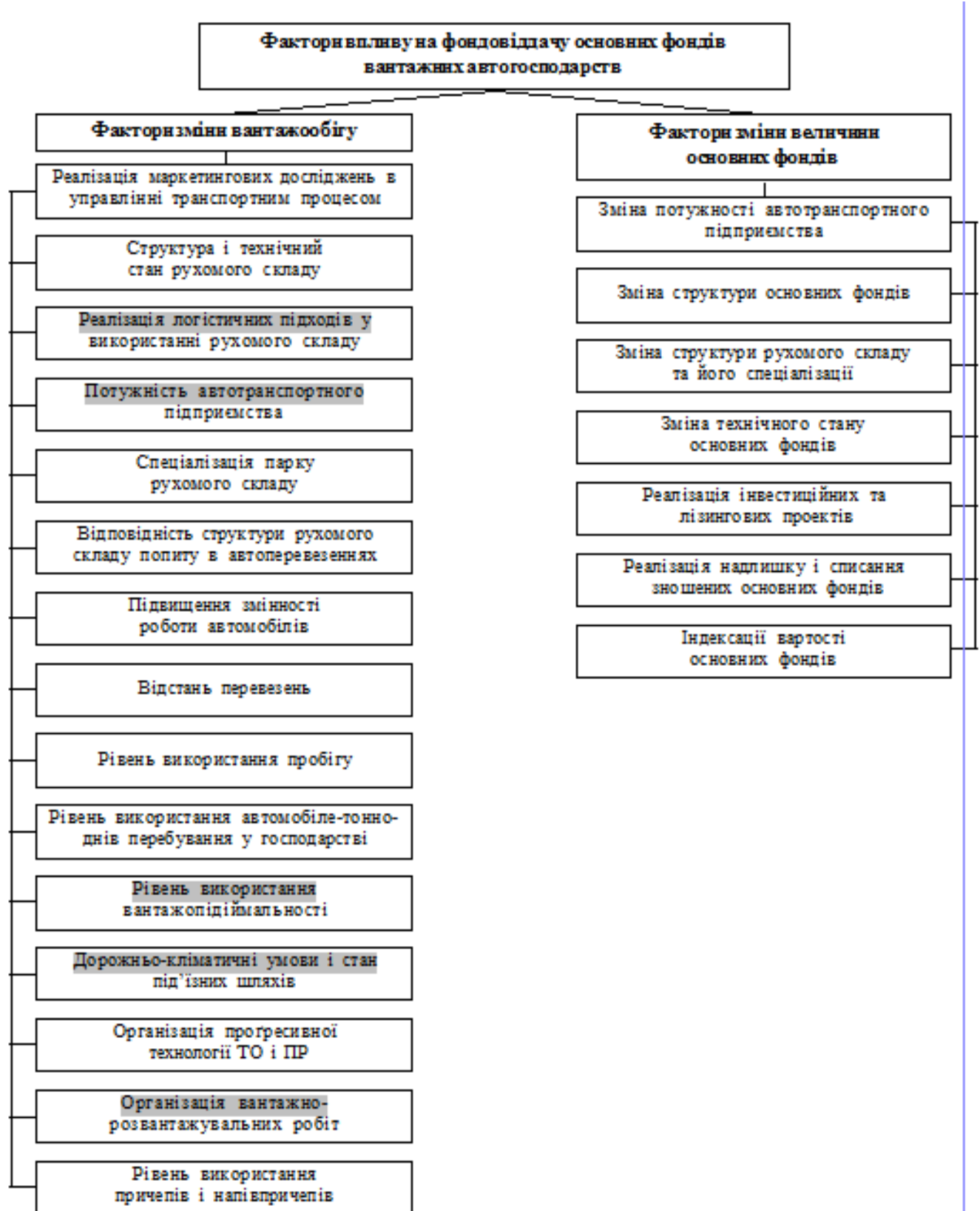

Рис. 2.1.1. Класифікація факторів, що впливають на фондовіддачу вантажних автотранспортних підприємств амортизаційних відрахувань, вартості продукції, прибутку, рентабельності та інших показників роботи підприємств» [237, с. 46].

До 1997 р. облік основних фондів здійснювався за їх повною первісною вартістю в цінах періоду створення чи придбання з урахуванням індексації. Щорічний приріст основних фондів також обліковується в цінах тих років, коли вони вводились в експлуатацію. Внаслідок цього одержуємо нашарування різних цін різних років, які викривлюють дійсний обсяг основних фондів.

Деякою мірою спотворення первісної структури основних фондів зумовлене їх індексацією, що здійснювалася згідно з постановами Кабінету Міністрів України [див.: 180; 181; 183; 186] з метою заміни традиційної трудомісткої переоцінки основних фондів. Результати таких переоцінок в умовах зростання цін потребували б постійних коригувань.

Починаючи з 1992 р. в Україні індексація балансової вартості основних фондів здійснювалася чотири рази. В інструктивних матеріалах щодо її проведення щоразу передбачалось використання різних за величиною коефіцієнтів індексації як для окремих груп основних засобів, так і з врахуванням часу введення їх у дію [див.: 100; 176; 177; 178; 222]. Так, згідно з постановою Кабінету Міністрів України від 7 травня 1992 р. «Про проведення загальної переоцінки основних фондів, залік взаємної заборгованості підприємств і поповнення оборотних засобів державних підприємств та організацій» [180], а також інструкцією на її виконання від 21.05.1992 р. «Про індексацію балансової вартості основних фондів підприємств, організацій та установ у зв'язку з підвищенням цін» [100] підприємства всіх галузей і форм власності проіндексували свої основні засоби, введені в дію до 1 січня цього ж року, на коефіцієнти, встановлені для кожної галузі окремо по пасивній та активній частинах засобів праці. Зокрема, для автогосподарств коефіцієнти, за якими індексувалися будівлі, споруди, передавальні пристрої, передбачали зростання їх вартості в 11 разів, а для транспортних засобів, машин і устаткування – у 18 разів. Розрив між коефіцієнтами індексації для вказаних груп основних фондів, а відповідно, і їх вартості в 1,64 раза перекривався наступною постановою Кабінету Міністрів України від 3 серпня 1993 р. «Про проведення індексації основних фондів у народному господарстві України» [181]. Інструкцією до цієї постанови [176] вартість попередньо індексованих будівель, споруд та передавальних пристроїв збільшувалася в 37,6 раза, а транспортних засобів, машин і устаткування – у 20,8 раза. Індексація 1993 р. за рахунок вищих коефіцієнтів для пасивної частини основних засобів в основному ліквідовувала її відставання в структурі основних виробничих фондів, зумовлене попередньою індексацією, а для автотранспортних підприємств навіть її випередження приблизно на 10 %.

Структурні зрушення, що відбувалися в складі основних виробничих фондів автогосподарств України в результаті наступних індексацій їх вартості у 1995 і 1996 рр., позначилися тенденцією зростання питомої ваги активної частини. В загальному підсумку результатом вказаних індексацій для вітчизняних автогосподарств стало зростання питомої ваги активної частини основних виробничих фондів. Коефіцієнти, за якими індексувалася вартість останніх в цілому, в 1,24 раза переважали сумарний коефіцієнт індексації будівель, споруд та передавальних пристроїв. А це, у свою чергу, відбилося на показниках ефективності використання основних фондів, спотворюючи їх економічне навантаження. В результаті чого при визначенні вказаних показників важливою умовою їх достовірності є досягнення порівняльності вихідних даних. «Відсутність реальної оцінки основного капіталу викликає необхідність дальшого вивчення цього питання» [237, с. 50], вдосконалення варіантів і методів її здійснення.

Експерти ООН з міжнародних стандартів обліку та звітності пропонують використовувати кілька видів оцінки вартості основних засобів, не даючи при цьому конкретних рекомендацій щодо застосування того чи іншого її способу. В багатьох країнах з розвинутою ринковою економікою допускається облік основних засобів за їх ринковою вартістю [див.: 141, с. 57].

На сьогодні в Україні найприйнятнішим варіантом переоцінки основних фондів, що дозволяє власнику засобів виробництва мати дані про їх ринкову оцінку, є варіант, запропонований В.Г. Гетьманом [див.: 56, с. 92]. Викладені ним положення частково знайшли своє відображення в новій редакції Закону України «Про оподаткування прибутку підприємств», що набув чинності у 1997 р. Згідно з цим законом підприємства мають право застосовувати щорічну індексацію балансової вартості груп основних фондів при умові, що рівень інфляції року, за підсумками якого проводиться індексація, перевищує 110 % [див.: 240, п. 8.3.3].

Окрім вказаних чинників, на обсяг основних фондів впливає також спосіб оцінки будівель і споруд. Їх приріст відображається у балансі по-різному: якщо будівництво велось господарським способом, то воно враховувалось по собівартості, а в тих випадках, коли будівництво велось підрядним способом, то воно оцінювалося за договірними цінами, які значно перевищують їх собівартість.

Таким чином, однакові обсяги основних виробничих фондів часто одержують різний грошовий вираз. Внаслідок цього показник використання основних фондів включає в себе ряд умовностей. Незначні спотворення можливі вже при оцінці використання основних фондів на одному і тому ж автотранспортному підприємстві, їх значно більше при порівнянні роботи окремих автотранспортних підприємств. Тому при розгляді величини основних виробничих фондів необхідно особливу увагу звернути на формування їх вартості та враховувати дані індексації основних фондів.

Для оцінки економічної ефективності використання основних виробничих фондів можна використовувати показники повної фондовіддачі ($f$) і граничної фондовіддачі ($\Delta f$).

Гранична фондовіддача дозволяє виявити тенденції зміни у використанні засобів праці, які у подальшому приведуть до зміни величини повної фондовіддачі, розкриваючи при цьому внутрішній характер її формування.

Перший показник може бути поданий як відношення загального обсягу транспортної роботи автотранспортних підприємств ($P$) до величини середньорічної вартості основних виробничих фондів ($Ф$), другий – відношенням приросту транспортної продукції за звітний період у порівнянні з базисним періодом до приросту середньорічної вартості основних виробничих фондів за цей же період, тобто:

$$f = \frac{P}{Ф}; \quad \Delta f = \frac{\Delta P}{\Delta Ф}, \qquad (2.1.2)$$

де $\delta P = P_1 - P_0$; $\delta Ф = Ф_1 - Ф_0$; $\delta f = f_1 - f_0$;

$\delta f$ – гранична фондовіддача, ткм/грн.;

$P_0$, $P_1$ – обсяг транспортної роботи відповідно за базисний і звітний періоди, ткм;

$\delta P$ – приріст транспортної роботи, ткм;

$\delta \Phi$ – приріст середньорічної вартості основних фондів, грн.

Відношення граничної фондовіддачі до повної фондовіддачі показує, як змінилась ефективність використання основних виробничих фондів у даному періоді порівняно з базисним внаслідок виробничої діяльності автотранспортного підприємства.

$$\frac{\Delta f}{f} = K, \qquad (2.1.3)$$

де $K$ – коефіцієнт оцінки ефективності використання основних виробничих фондів.

$$K = \frac{\Delta P \cdot \Phi}{\Delta \Phi \cdot P} = \frac{\Delta P}{P} : \frac{\Delta \Phi}{\Phi} = \frac{I_р - 1}{I_р} : \frac{I_ф - 1}{I_ф} = \frac{I_р - 1}{I_ф - 1} \ ; \quad (2.1.4)$$

де $I_ф$ – індекс динаміки вартості основних фондів;

$I_р$ – індекс динаміки обсягу транспортної роботи.

У тих випадках, коли $K = 1$, то це означає, що ефективність використання основних виробничих фондів у звітному періоді порівняно із базисним періодом залишилася без змін.

У процесі виробничої діяльності автотранспортного підприємства вартість основних виробничих фондів і обсяги транспортної роботи можуть змінюватися як в бік їх збільшення, так і в бік зменшення, при цьому з точки зору оцінки ефективності використання основних виробничих фондів важливі не тільки темпи їх зміни, але і їх співвідношення. Для оцінки економічної ефективності використання основних виробничих фондів застосуємо графічний метод.

Для побудови графіка використаємо формулу коефіцієнта оцінки ефективності використання основних фондів у розгорнутому вигляді:

$$K = \frac{\Delta P \cdot \Phi}{\Delta \Phi \cdot P} ; \qquad (2.1.5)$$

Місце знаходження точки «*К*» в системі координат $x = \delta P \cdot \Phi$; $y = \delta \Phi \cdot P$ дозволяє оцінити ефективність використання основних фондів (рис. 2.1.2).

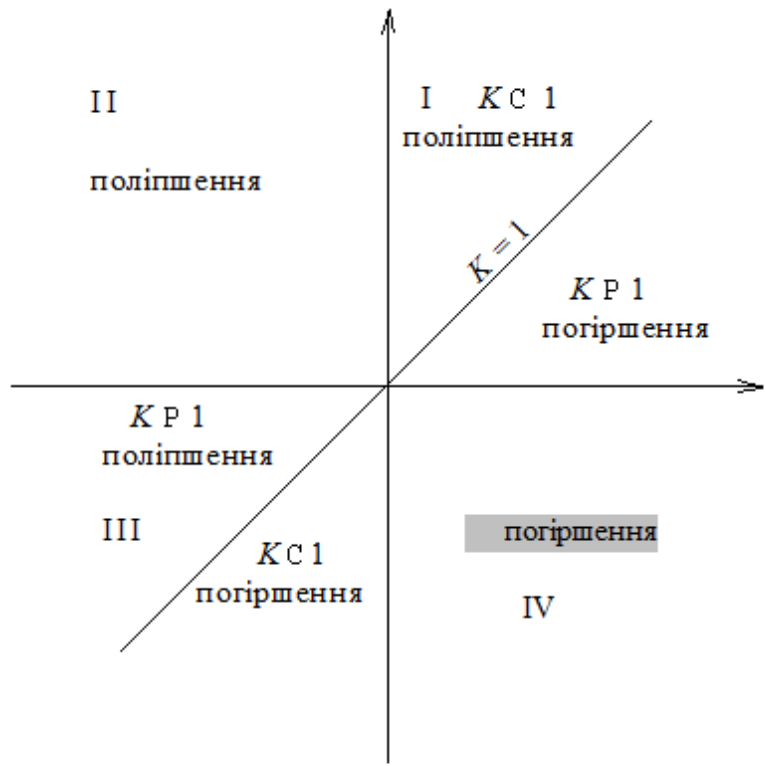

Рис. 2.1.2. Графік оцінки ефективності використання основних виробничих фондів

Так, в I чверті при одночасному зростанні вартості основних виробничих фондів і обсягів перевезень може бути і поліпшення (при *К* > 1), і погіршення (при *К* < 1) використання основних фондів залежно від співвідношення темпів росту основних фондів і обсягів транспортної роботи.

В III чверті у випадку зменшення як величини вартості основних фондів, так і обсягів перевезень також може спостерігатися як поліпшення (при *К* < 1), так і погіршення (при *К* > 1) використання основних фондів, залежно від темпів їх зниження. При одночасному зростанні величини вартості основних фондів і зменшенні обсягів транспортної роботи (IV чверть) спостерігається погіршення використання основних фондів і, відповідно, при зменшенні величини вартості основних фондів та збільшенні обсягів транспортної роботи (II чверть) використання основних фондів покращується.

Користуючись наведеним графіком оцінки ефективності використання основних виробничих фондів, можна дати оцінку їх використання в автотранспортних підприємствах Івано-Франківської, Львівської та Тернопільської областей (табл. 2.1.2).

**Таблиця 2.1.2**

Оцінка ефективності використання основних виробничих фондів автотранспортних підприємств

| Показники | Роки | | | | |
|---|---|---|---|---|---|
| | 1992 | 1993 | 1994 | 1995 | 1996 |
| Повна фондовіддача в тонно-кілометрах | 4,113 | 3,238 | 2,216 | 1,510 | 0,632 |
| Гранична фондовіддача в тонно-кілометрах | X | 18,159 | 10,118 | 9,571 | 6,275 |
| Коефіцієнт оцінки ефективності використання основних виробничих фондів | X | 5,608 | 4,566 | 6,339 | 9,929 |
| Оцінка ефективності використання* | – | – | – | – | – |

\* Знаки "+" чи "–" показують відповідно поліпшення, чи погіршення використання основних виробничих фондів.

Так, у 1996 р. порівняно з 1995 р. в цілому по АТП Західного регіону коефіцієнт оцінки ефективності використання основних фондів $K = 9,929$ (ІІІ чверть).

Враховуючи, що протягом 1992 – 1996 рр. у зв'язку з різким скороченням вантажних автоперевезень відбувалося зниження обсягів транспортної роботи, яке спостерігалося на тлі зменшення реальної вартості їх основних виробничих фондів, то для розгляду слід брати ІІІ чверть наведеного графіка (див. рис. 2.1.2). Обчислені величини коефіцієнта оцінки ефективності використання основних виробничих фондів дають підстави стверджувати, що темпи зниження обсягів транспортної роботи значно перевищували темпи зменшення загальної вартості основних виробничих фондів автогосподарств регіону. Свідченням цього є й дані табл. 2.1.3.

Так, на кінець 1996 р. порівняно з 1992 р. вартість основних виробничих фондів у порівняльних цінах зменшилася на 36,9 %, а обсяги перевезень – на 88,6 %.

Динаміка повної і граничної фондовіддачі вантажних автотранспортних підприємств за 1992 – 1996 рр. (див. табл. 2.1.2 і рис. 2.1.3) відображають досить сталу тенденцію зниження рівня фондовіддачі.

**Таблиця 2.1.3**

Динаміка вартості основних виробничих фондів і обсягів транспортної роботи АТП (в порівняльних цінах у відсотках до 1992 р.)

| Показники | Роки | | | | |
|---|---|---|---|---|---|
| | 1992 | 1993 | 1994 | 1995 | 1996 |
| Вартість основних виробничих фондів | 100,0 | 94,1 | 82,0 | 74,8 | 63,1 |
| Вартість транспортних засобів | 100,0 | 92,1 | 80,9 | 73,6 | 62,0 |
| Обсяг транспортної роботи | 100,0 | 77,2 | 35,9 | 20,7 | 11,4 |

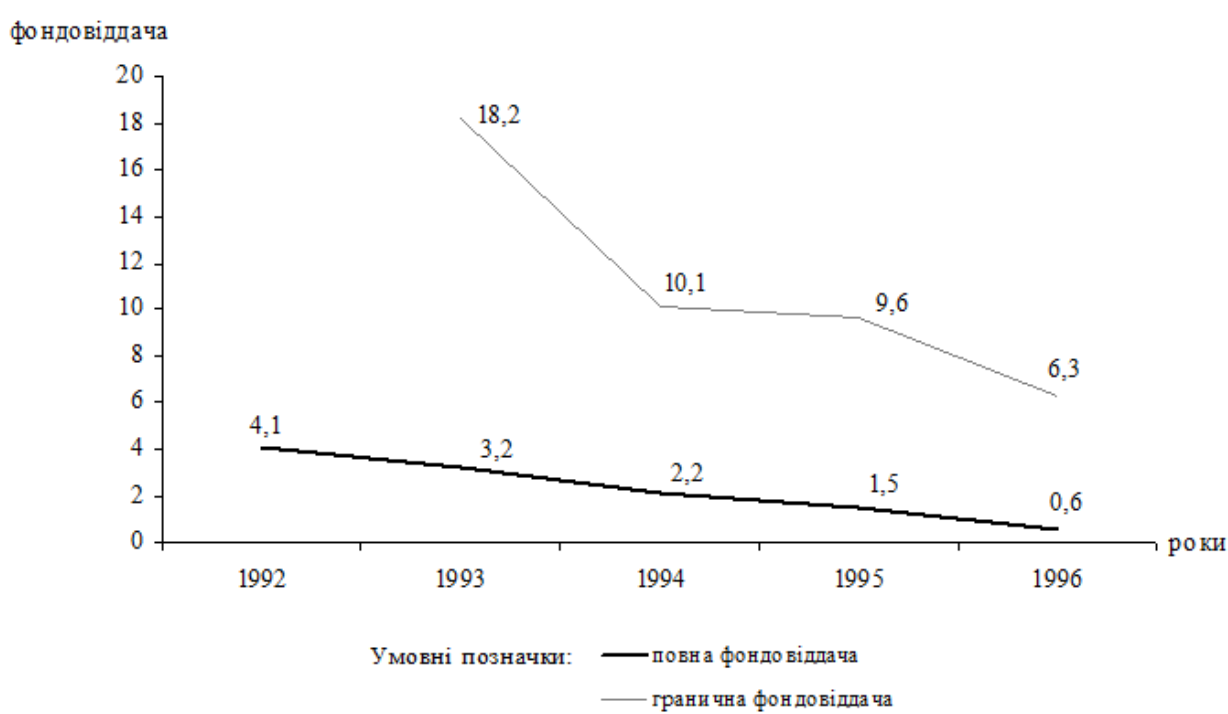

Рис. 2.1.3. Динаміка фондовіддачі АТП Івано-Франківської, Львівської і Тернопільської областей

Прикметно, що темпи зниження фондовіддачі основних виробничих фондів та фондовіддачі транспортних засобів суттєво не різняться між собою (див. табл. 2.1.4). Проте різниця між натуральним і вартісним показниками фондовіддачі характеризується значно більшою розбіжністю їх значень (див. рис. 2.1.4). В основному це зумовлено коливаннями реальної вартості автоперевезень.

Підсумовуючи вищесказане, слід зазначити, що протягом 1992 – 1996 рр. на

Таблиця 2.1.4

Динаміка фондовіддачі автотранспортних підприємств
Івано-Франківської, Львівської та Тернопільської областей

| Показники | Роки | | | | |
|---|---|---|---|---|---|
| | 1992 | 1993 | 1994 | 1995 | 1996 |
| Фондовіддача основних виробничих фондів: | | | | | |
| у тонно-кілометрах; | 4,113 | 3,238 | 2,216 | 1,510 | 0,632 |
| те саме у відсотках до 1992 р.; | 100,0 | 78,7 | 53,9 | 36,7 | 15,4 |
| у гривнях доходу; | 1,777 | 1,457 | 0,778 | 0,492 | 0,213 |
| те саме у відсотках до 1992 р. | 100,0 | 82,0 | 43,8 | 27,7 | 18,1 |
| Фондовіддача транспортних засобів: | | | | | |
| у тонно-кілометрах; | 7,179 | 5,776 | 3,915 | 2,678 | 1,535 |
| те саме у відсотках до 1992 р.; | 100,0 | 80,5 | 54,5 | 37,3 | 21,4 |
| у гривнях доходу; | 3,101 | 2,602 | 1,377 | 0,871 | 0,571 |
| те саме у відсотках до 1992 р. | 100,0 | 83,9 | 44,4 | 28,1 | 18,4 |

вантажних автотранспортних підприємствах Івано-Франківської, Львівської та Тернопільської областей поряд із значним зменшенням вартості основних виробничих фондів (майже на 40 %) спостерігалося зниження обсягів виконуваної транспортної роботи. Стрімкі темпи падіння останньої – наслідок спаду промислового виробництва в Україні. За період з 1991 до 1996 рр. обсяги промислового виробництва скоротились в Україні на 50,3 %, в тому числі в Івано-Франківській області на 51,0 %, Львівській – на 68,9 %, Тернопільській – на 52,1 % [див.: 234, 235].

Скорочення вантажоперевезень негативно позначилося на загальному фінансовому стані автогосподарств. Постійний брак коштів, заборгованість змусили автопідприємства відмовитися від утримання надлишкових виробничих потужностей шляхом їх списання чи реалізації. Все це не могло не позначитися на показниках ефективності використання основних виробничих фондів. Як наслідок – постійне зниження фондовіддачі, що протягом означеного періоду зменшилася понад 80 % до рівня 1992 р.

Зміна величини показника фондовіддачі залежить від ряду факторів як макроекономічного, так і мікроекономічного характеру. Вивчення впливу останніх на

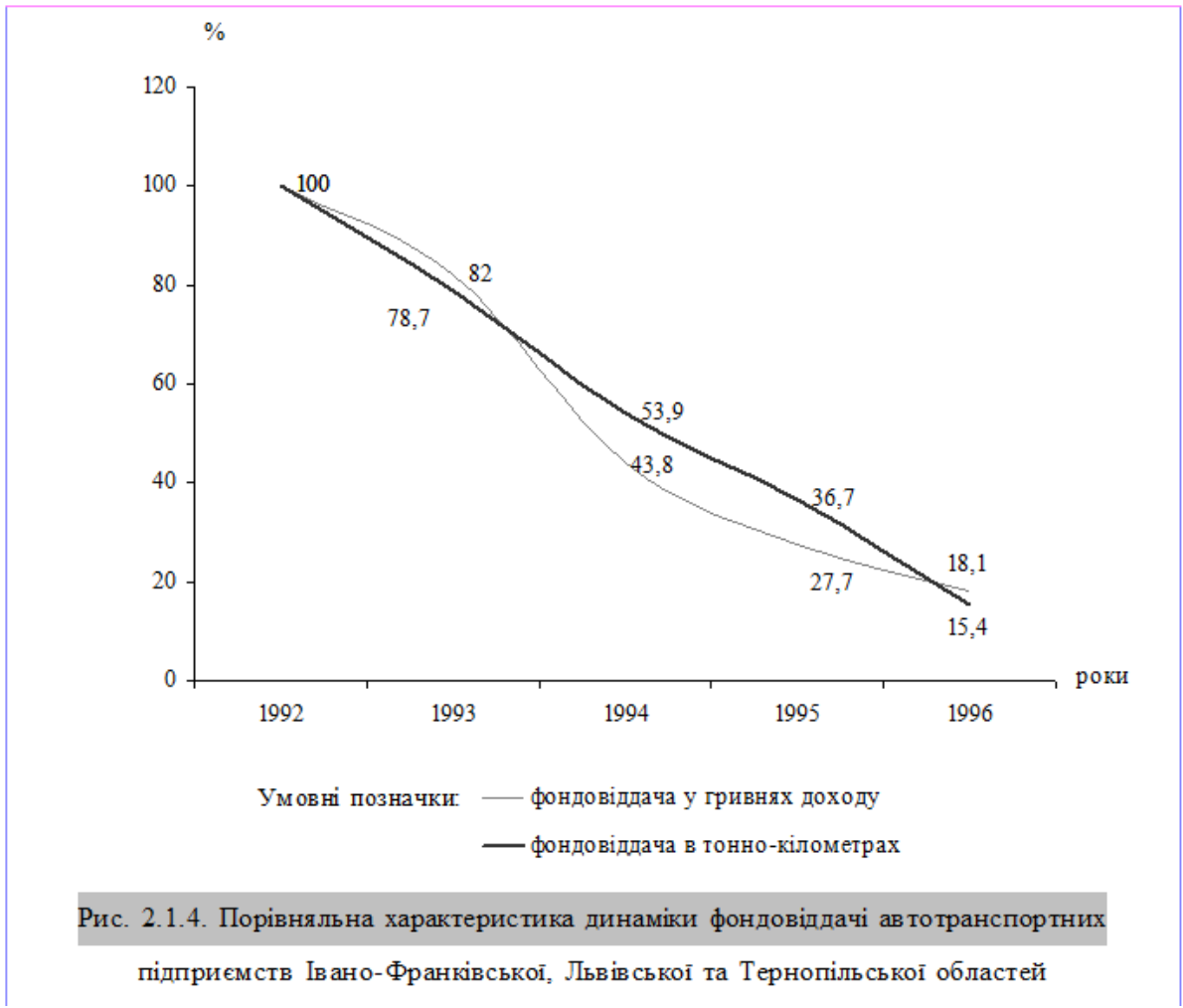

Рис. 2.1.4. Порівняльна характеристика динаміки фондовіддачі автотранспортних підприємств Івано-Франківської, Львівської та Тернопільської областей фондовіддачу вантажних автотранспортних підприємств Івано-Франківської, Львівської та Тернопільської областей здійснюватиметься нижче.

## 2.2. Аналіз факторів впливу на рівень і динаміку фондовіддачі основних виробничих фондів

Підвищення ефективності роботи вантажних автотранспортних підприємств в умовах економічної нестабільності є одним із найбільш важливих завдань. Його реалізація залежить від ряду факторів. Серед них три вважаємо найсуттєвішими: ступінь використання рухомого складу, рівень використання пробігу, виробіток рухомого складу.

## 2.2.1. Залежність фондовіддачі основних виробничих фондів від рівня використання автомобіле-тонно-днів перебування у господарстві

Поліпшення використання парку вантажних автомобілів – важливий чинник зростання обсягів транспортної роботи та підвищення рівня фондовіддачі основних виробничих фондів. У зв'язку з цим для характеристики екстенсивного завантаження рухомого складу скористаємося коефіцієнтом використання автомобіле-тонно-днів перебування у господарстві, що обчислюється за формулою:

$$K_в = \frac{\Sigma A_р}{\Sigma A_п} \qquad (2.2.1)$$

де $K_в$ – коефіцієнт використання автомобіле-тонно-днів перебування у господарстві;

$A_р$ – автомобіле-тонно-дні в роботі;

$A_п$ – автомобіле-тонно-дні перебування у господарстві.

Перевага цього коефіцієнта у порівнянні з іншими показниками екстенсивного використання полягає у тому, що методика його обчислення враховує не тільки тривалість використання рухомого складу, але і його вантажопідіймальність.

Недоліком коефіцієнта є те, що він не враховує режиму роботи автотранспортних підприємств. Однак в умовах низької завантаженості роботою рухомого складу, коли на один автомобіле-день перебування у господарстві в середньому припадає 2,1 год. часу перебування в наряді (в т. ч. у Івано-Франківській області 2,9 год., Львівській – 2,0 год., Тернопільській – 1,4 год.) доводиться говорити про те, що цей недолік не є суттєвим.

Протягом останніх років у міру зниження обсягів промислового виробництва намітилася стала тенденція зниження коефіцієнта використання автомобіле-тонно-днів перебування у господарстві. Як результат сьогодні лише у 5 % автогосподарств його значення перевищує 0,4. У переважної більшості вантажних АТП, що складають 82,5 %, коефіцієнт використання автомобіле-тонно-днів перебування у господарстві не перевищує 0,3. Але і при цьому цей чинник характеризується значним впливом на рівень фондовіддачі автогосподарств як досліджуваного регіону, так і кожної з його областей зокрема.

Суттєвий вплив рівня використання автомобіле-тонно-днів перебування у господарстві на ефективність використання основних виробничих фондів підтверджується практикою роботи досліджуваних вантажних АТП. У таблицях 2.2.1 та 2.2.2 проведено групування 97 вантажних автотранспортних підприємств регіону залежно від рівня використання автомобіле-тонно-днів перебування у господарстві, з якого видно, що із збільшенням величини даного коефіцієнта зростає і рівень фондовіддачі. Так, в середньому по регіону рівень фондовіддачі АТП з найвищими коефіцієнтами використання автомобіле-тонно-днів перебування у господарстві у 5,2 раза перевищує рівень фондовіддачі АТП, що мають найнижчі значення цього коефіцієнта. Зокрема, в Івано-Франківській області це співвідношення становить 3,1 раза, Львівській – 2,6 раза, Тернопільській – 3,6 раза. При цьому по регіону

**Таблиця 2.2.1**

Вплив рівня використання автомобіле-тонно-днів перебування у господарстві на фондовіддачу основних виробничих фондів вантажних АТП Івано-Франківської, Львівської та Тернопільської областей за 1996 р.

| Показники | Групи автопідприємств за величиною коефіцієнта використання автомобіле-тонно-днів перебування у господарстві | | | | | В середньому по автопід-приємствах регіону |
|---|---|---|---|---|---|---|
| | до 0,10 | 0,11 – 0,20 | 0,21 – 0,30 | 0,31 – 0,40 | понад 0,41 | |
| Кількість виконаних тонно-кілометрів у розрахунку на 1 грн. вартості основних виробничих фондів | 0,209 | 0,505 | 0,817 | 0,821 | 1,084 | 0,632 |
| Кількість перевезених тонн вантажів у розрахунку на 1 грн. вартості основних виробничих фондів | 0,012 | 0,027 | 0,043 | 0,044 | 0,148 | 0,038 |
| Дохід у розрахунку на 1 грн. вартості основних виробничих фондів, тис. грн. | 0,051 | 0,123 | 0,289 | 0,698 | 0,394 | 0,213 |
| Рентабельність основних виробничих фондів, ткм/грн. | 0,004 | 0,005 | 0,005 | 0,085 | 0,089 | 0,029 |
| Річна продуктивність однієї спискової автомобіле-тонни: | | | | | | |
| а) тис. тонн; | 0,098 | 0,144 | 0,303 | 0,331 | 0,836 | 0,212 |
| б) тис. тонно-кілометрів | 1,589 | 2,740 | 5,337 | 5,793 | 5,807 | 4,127 |

Таблиця 2.2.2

Залежність фондовіддачі вантажних автотранспортних підприємств від величини коефіцієнта використання автомобіле-тонно-днів перебування у господарстві за 1996 р.

| Групи автопідприємств за величиною коефіцієнта використання автомобілем тонно-днів перебування у господарстві | Область | | | | | | | | | Регіон | | |
|---|---|---|---|---|---|---|---|---|---|---|---|---|
| | Івано-Франківська | | | Львівська | | | Тернопільська | | | | | |
| | кількість підприємств у % до підсумку | фондовіддача | | кількість підприємств у % до підсумку | фондовіддача | | кількість підприємств у % до підсумку | фондовіддача | | кількість підприємств у % до підсумку | фондовіддача | |
| | | ткм/грн. | у % до середнього значення | | ткм/грн. | у % до середнього значення | | ткм/грн. | у % до середнього значення | | ткм/грн. | у % до середнього значення |
| до 0,10 | 20,8 | 0,083 | 13,8 | 7,7 | 0,442 | 56,8 | 23,5 | 0,185 | 46,0 | 16,5 | 0,209 | 33,1 |
| 0,11 – 0,20 | 25,0 | 0,557 | 92,8 | 38,5 | 0,531 | 68,3 | 52,9 | 0,435 | 108,2 | 40,2 | 0,505 | 79,9 |
| 0,21 – 0,30 | 37,5 | 0,737 | 122,8 | 25,6 | 1,249 | 160,5 | 17,6 | 0,471 | 117,2 | 25,8 | 0,817 | 129,3 |
| 0,31 – 0,40 | 12,5 | 0,590 | 98,3 | 20,5 | 0,844 | 108,5 | 3,0 | 0,780 | 194,0 | 12,4 | 0,821 | 130,0 |
| понад 0,41 | 4,2 | 0,259 | 43,2 | 7,7 | 1,168 | 150,1 | 3,0 | 0,666 | 165,7 | 5,1 | 1,084 | 171,5 |
| Разом: | 100,0 | – | – | 100,0 | – | – | 100,0 | – | – | 100,0 | – | – |
| Середня фондовіддача | – | 0,600 | – | – | 0,778 | – | – | 0,402 | – | – | 0,632 | – |

фондовіддача у гривнях доходу (див. табл. 2.2.1) зростає у 7,7 раза, кількість перевезених тонн вантажів у розрахунку на 1 грн. вартості основних виробничих фондів – у 12,3 раза, виробіток на 1 автомобіле-тонну в тонно-кілометрах – в 3,7 раза, рентабельність основних виробничих фондів – у 22,3 раза.

У процесі дослідження ефективності використання основних виробничих фондів вантажних автотранспортних підприємств за формами власності, встановлено тенденцію деякого відставання рівня фондовіддачі у державних автогосподарствах порівняно з автогосподарствами, що перебувають у колективній власності. Так, якщо середнє значення фондовіддачі досліджуваних автопідприємств регіону у 1996 р. склало 0,632 ткм/грн., то в автогосподарствах з державною формою власності – 0,624, колективною – 0,709 ткм/грн. (табл. 2.2.3). Ця тенденція простежується і для кожної групи підприємств, згрупованих за величиною коефіцієнта використання автомобіле-тонно-днів перебування у господарстві. Загалом максимальна різниця у рівнях фондовіддачі автопідприємств цих форм власності знаходиться в межах 20 %. Прикметно, що й

**Таблиця 2.2.3**

Характеристика рівня фондовіддачі автотранспортних підприємств за формами власності Івано-Франківської, Львівської та Тернопільської областей у 1996 р.

| Показники | Групи автопідприємств за величиною коефіцієнта використання автомобіле-тонно-днів перебування у господарстві | | | | | | | | | | В середньому по автогосподарствах регіону | |
|---|---|---|---|---|---|---|---|---|---|---|---|---|
| | до 0,10 | | 0,11 – 0,20 | | 0,21 – 0,30 | | 0,31 – 0,40 | | понад 0,41 | | | |
| | державна | колективна | державна | колективна | державна | колективна | державна | колективна | державна | колективна | державна | колективна |
| Кількість виконаних тонно-кілометрів у розрахунку на 1 грн. основних виробничих фондів | 0,175 | 0,218 | 0,477 | 0,583 | 0,763 | 0,859 | 0,771 | 0,883 | – | 1,084 | 0,624 | 0,709 |
| Кількість перевезених тонн вантажів у розрахунку на 1 грн. основних виробничих фондів | 0,020 | 0,011 | 0,023 | 0,034 | 0,042 | 0,043 | 0,043 | 0,046 | – | 0,148 | 0,027 | 0,043 |
| Річна продуктивність однієї спискової автомобіле-тонни: а) тис. тонн; | 0,109 | 0,095 | 0,114 | 0,190 | 0,208 | 0,334 | 0,305 | 0,390 | – | 0,836 | 0,138 | 0,251 |
| б) тис. тонно-кілометрів | 0,954 | 1,870 | 2,517 | 3,242 | 4,658 | 6,399 | 4,638 | 6,670 | – | 5,807 | 3,187 | 4,127 |

продуктивність однієї спискової автомобіле-тонни в автогосподарствах з колективною формою власності в основному є вищою, ніж у державних підприємствах.

Аналізуючи залежність між коефіцієнтом використання автомобіле-тонно-днів перебування у господарстві та фондовіддачею основних виробничих фондів, установлено, що із збільшенням величини першого зростають: коефіцієнт випуску автомобілів на лінію, коефіцієнт використання пробігу, продуктивність автомобіля, що у свою чергу дозволяє збільшити обсяг транспортної роботи в розрахунку на списковий автомобіль і, отже, фондовіддачу.

Зіставлення коефіцієнтів використання автомобіле-тонно-днів перебування у господарстві і фондовіддачі основних виробничих фондів, а також дослідження характеру і форми зв'язку між цими показниками переконують в його існуванні. Отримані емпіричні лінії регресії (див. рис. 2.2.1) відображають характер цього зв'язку. Кількісно його можна виразити, застосувавши кореляційний метод – рівняння параболи другого порядку:

$$y_х = a + bx + cx^2 \qquad (2.2.2)$$

де $y_х$ – середня фондовіддача основних виробничих фондів, ткм/грн.;

$x$ – коефіцієнт використання автомобіле-тонно-днів перебування у господарстві;

*a, b, c* – параметри рівняння.

Наведені моделі рівняння (табл. 2.2.4) виражають зв'язок між коефіцієнтом використання автомобіле-тонно-днів перебування у господарстві та рівнем використання основних виробничих фондів вантажних АТП у розрізі досліджуваних областей.

На рис. 2.2.1 наведений графік кореляційної залежності рівня фондовіддачі основних фондів від величини коефіцієнта використання автомобіле-тонно-днів перебування у господарстві.

Отримані регресійні рівняння зв'язку показують, що із збільшенням коефіцієнта використання автомобіле-тонно-днів перебування у господарстві на 0,1 фондовіддача основних виробничих фондів в цілому по вантажних автопідприємствах регіону зростає на 4,517 ткм/грн. (в т. ч.: в Івано-Франківській області – на 5,445, Львівській – на 5,309, Тернопільській – на 2,541).

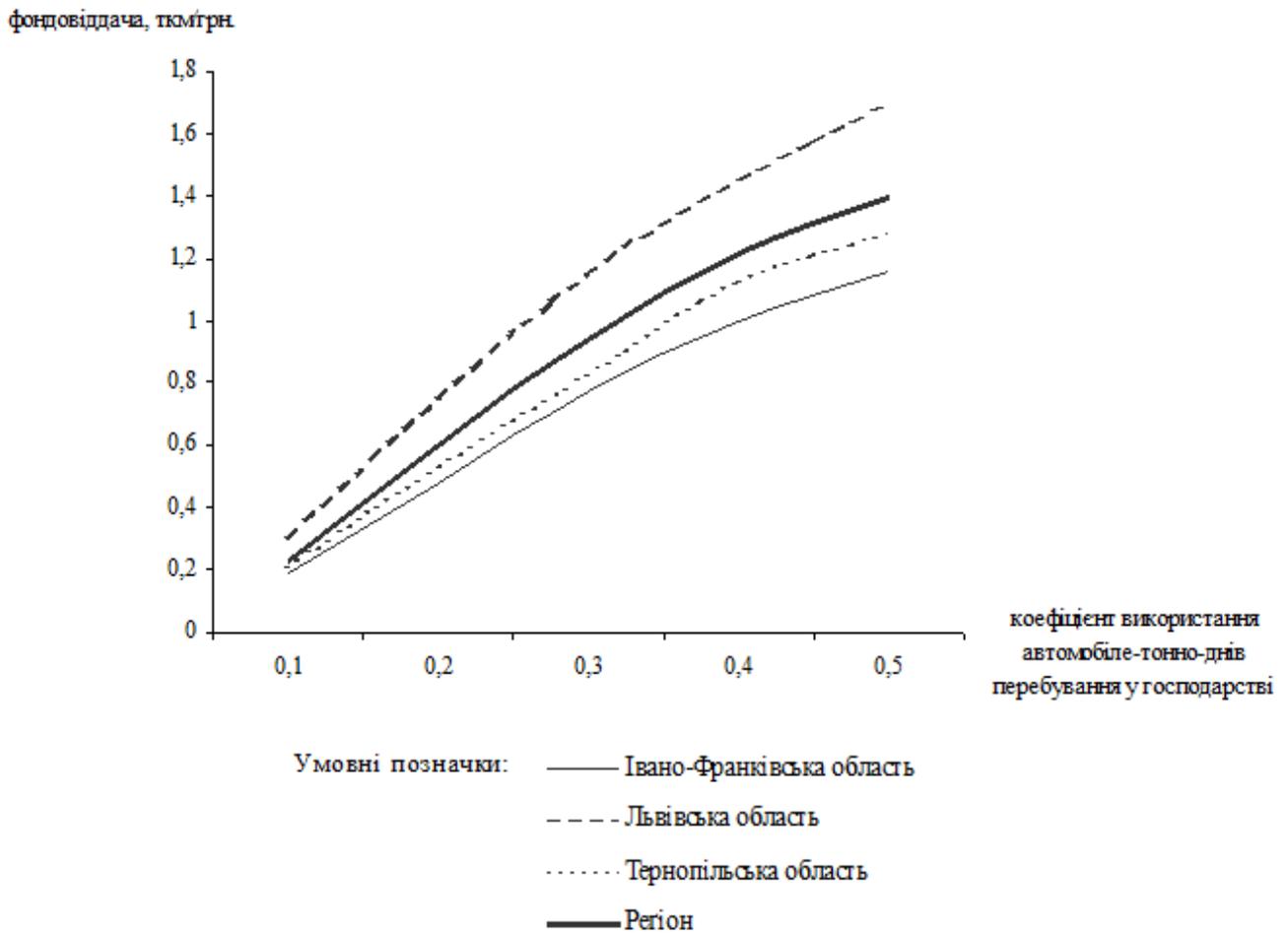

Рис. 2.2.1. Залежність фондовіддачі від коефіцієнта використання автомобіле-тонно-днів перебування у господарстві на рівень фондовіддачі

Однак у невеликих щодо обсягу сукупностях коефіцієнт регресії схильний до випадкових коливань. Межі цих коливань обчислюються з допомогою граничної помилки, визначеної на основі середньої помилки коефіцієнта регресії, скоригованої на величину коефіцієнта довіри [див., напр., 220, с. 91]. При величині граничної помилки 0,395, довірчі межі коефіцієнта регресії становитимуть від 4,122 до 4,912. Отже, при збільшенні коефіцієнта використання автомобіле-тонно-днів перебування у господарстві на 0,1 фондовіддача основних виробничих фондів автопідприємств регіону підвищиться не менше, як на 4,122 ткм/грн., і не більше як на 4,912 ткм/грн. Для автогосподарств Івано-Франківської області довірчі межі коефіцієнта регресії становлять від 4,901 до 5,988, Львівської – від 2,493 до 8,126, Тернопільської – від 2,248 до 2,834.

**Таблиця 2.2.4**

Показники кореляційної залежності рівня фондовіддачі від коефіцієнта використання автомобіле-тонно-днів перебування у господарстві за 1996 р.

|  | Область | | | По регіону |
|---|---|---|---|---|
|  | Івано-Франківська | Львівська | Тернопільська | |
| Рівняння зв'язку | $y_x = -0,2344 + 5,4446x - 9,7343x^2$ | $y_x = 0,10596 + 5,3093x - 6,4688x^2$ | $y_x = 0,0690 + 2,5408x - 3,9647x^2$ | $y_x = -0,04321 + 4,5168x - 5,5561x^2$ |
| Середня помилка коефіцієнта регресії | 0,3724 | 1,2353 | 0,1665 | 0,1865 |
| Гранична помилка коефіцієнта регресії | 0,5437 | 2,8164 | 0,2930 | 0,3953 |
| Довірчі межі коефіцієнта регресії | $4,9009 \leq b \leq 5,9883$ | $2,4929 \leq b \leq 8,1257$ | $2,2478 \leq b \leq 2,8338$ | $4,1215 \leq b \leq 4,9121$ |
| Коефіцієнт детермінації | 0,7321 | 0,9557 | 0,8490 | 0,9332 |
| Теоретичне кореляційне відношення | 0,8556 | 0,9776 | 0,9214 | 0,9660 |
| Фактичне значення F-критерію | 28,7 | 388,2 | 84,4 | 656,6 |
| Критичне значення F-критерію | 3,49 | 3,23 | 3,32 | 3,07 |
| Коефіцієнт еластичності | 2,990 | 1,412 | 1,244 | 1,623 |

Найвища залежність між фондовіддачею основних виробничих фондів і коефіцієнтом використання автомобіле-тонно-днів перебування у господарстві спостерігається у Львівській області. Так, якщо в цілому по регіону 93,32 % варіації фондовіддачі пояснюється різним рівнем коефіцієнта використання автомобіле-тонно-днів перебування у господарстві, то у Львівській області – 95,57 %. Дещо нижчі значення варіації фондовіддачі у Тернопільській та Івано-Франківській областях, де вони становлять відповідно 84,9 % і 73,21 %.

Обчислені значення теоретичного кореляційного відношення, що сягають від 0,856 в Івано-Франківській області до 0,978 у Львівській, свідчать про наявність високої щільності зв'язку між факторною та результативною ознаками. Як бачимо, найвище значення теоретичного кореляційного відношення спостерігається у Львівській області. Це пояснюється значно меншими коливаннями величини факторної ознаки, ніж у інших областях.

Розрахований F-критерій, фактичне значення якого становить по регіону 656,6, що значно перевищує критичне – 3,07 (для рівня істотності $a = 0,05$), свідчить про те, що зв'язок між фондовіддачею і коефіцієнтом використання автомобіле-тонно-днів перебування у господарстві носить не випадковий, а закономірний характер.

Досліджуючи рівняння регресії на екстремум, можна помітити, що максимальна фондовіддача спостерігається при збільшенні коефіцієнта використання автомобіле-тонно-днів перебування у господарстві в АТП регіону до 0,41 (в т. ч. в Івано-Франківській області – до 0,28, Львівській – до 0,41, Тернопільській – до 0,32).

Обчислені коефіцієнти еластичності в розрізі областей свідчать, що при збільшенні коефіцієнта використання автомобіле-тонно-днів перебування у господарстві на 1% від його середнього значення фондовіддача в автогосподарствах Івано-Франківської області підвищується на 2,99%, Львівської – на 1,41 %, Тернопільської – на 1,24 %.

### 2.2.2. Залежність фондовіддачі від рівня використання пробігу

Важливим показником інтенсивності використання вантажних автомобілів є коефіцієнт використання пробігу. Він характеризує ефективність використання рухомого складу і визначають його за формулою:

$$\beta = \frac{\Sigma L_{в}}{\Sigma L_{з}}, \qquad (2.2.3)$$

де $\beta$ – коефіцієнт використання пробігу;

$L_{в}$ – пробіг з вантажем, км;   $L_{з}$ – загальний пробіг, км.

Коефіцієнт використання пробігу дозволяє виявити співвідношення продуктивного (з вантажем) і непродуктивного (порожнього і нульового) пробігу рухомого складу. В умовах загального спаду виробництва, зниження обсягів перевезень випуск автомобілів на лінію складає лише 20 – 30 % наявного парку автомобілів. Величина цього показника значною мірою зазнає впливу зовнішніх факторів та умов організації транспортного процесу. В кінцевому підсумку коефіцієнт використання пробігу відображає рівень раціоналізації транспортних зв'язків і визначає рівень продуктивності рухомого складу – найбільш активної частини основних виробничих фондів вантажного автомобільного транспорту.

У табл. 2.2.5 проведено групування 97 автотранспортних підприємств Івано-Франківської, Львівської та Тернопільської областей залежно від величини коефіцієнта використання пробігу. Воно доводить, що із збільшенням коефіцієнта використання пробігу продуктивність рухомого складу зростає. Так, рівень фондовіддачі підприємств з найнижчими значеннями цього коефіцієнта у 4,1 раза менший від рівня фондовіддачі автогосподарств з найвищими значеннями, а рівень продуктивності спискової автомобіле-тонни у тонно-кілометрах – у 3,3 раза.

**Таблиця 2.2.5**

Вплив коефіцієнта використання пробігу на фондовіддачу основних виробничих фондів вантажних автогосподарств Івано-Франківської, Львівської і Тернопільської областей за 1996 р.

| Показники | Групи автопідприємств за величиною коефіцієнта використання пробігу | | | | | В середньому по автопід-приємствах регіону |
|---|---|---|---|---|---|---|
| | до 0,15 | 0,16 – 0,30 | 0,31 – 0,45 | 0,46 – 0,60 | понад 0,60 | |
| Кількість виконаних тонно-кілометрів у розрахунку на 1 грн. вартості основних виробничих фондів | 0,270 | 0,601 | 0,645 | 0,900 | 1,108 | 0,632 |
| Кількість перевезених тонн вантажів у розрахунку на 1 грн. вартості основних виробничих фондів | 0,012 | 0,030 | 0,045 | 0,110 | 0,117 | 0,038 |
| Дохід у розрахунку на 1 грн. вартості основних виробничих фондів, тис. грн. | 0,061 | 0,123 | 0,261 | 0,378 | 0,948 | 0,213 |
| Рентабельність основних виробничих фондів, ткм/грн. | 0,001 | 0,008 | 0,042 | 0,084 | 0,051 | 0,029 |
| Річна продуктивність однієї спискової автомобіле-тонни: | | | | | | |
| а) тис. тонн; | 0,075 | 0,166 | 0,307 | 0,511 | 0,615 | 0,212 |
| б) тис. тонно-кілометрів | 1,668 | 3,280 | 4,259 | 5,430 | 5,512 | 4,127 |

Чітка залежність рівня фондовіддачі від величини коефіцієнта використання пробігу простежується і в розрізі окремих областей (табл. 2.2.6). Зберігається вона

Таблиця 2.2.6

Залежність фондовіддачі вантажних АТП Івано-Франківської, Львівської та Тернопільської областей від величини коефіцієнта використання пробігу за 1996 р.

| Групи автопідприємств за величиною коефіцієнта використання пробігу | Область | | | | | | | | | Регіон | | |
|---|---|---|---|---|---|---|---|---|---|---|---|---|
| | Івано-Франківська | | | Львівська | | | Тернопільська | | | | | |
| | кількість підприємств у % до підсумку | фондовіддача | | кількість підприємств у % до підсумку | фондовіддача | | кількість підприємств у % до підсумку | фондовіддача | | кількість підприємств у % до підсумку | фондовіддача | |
| | | ткм/грн. | у % до середнього значення | | ткм/грн. | у % до середнього значення | | ткм/грн. | у % до середнього значення | | ткм/грн. | у % до середнього значення |
| до 0,15 | 8,3 | 0,041 | 6,8 | 5,1 | 0,389 | 50,0 | 8,8 | 0,096 | 23,9 | 7,2 | 0,270 | 42,7 |
| 0,16 – 0,30 | 29,2 | 0,458 | 76,3 | 23,1 | 0,771 | 91,4 | 20,6 | 0,430 | 107,0 | 23,7 | 0,601 | 95,1 |
| 0,31 – 0,45 | 16,7 | 0,661 | 110,2 | 41,0 | 0,836 | 107,5 | 44,1 | 0,468 | 116,4 | 36,1 | 0,645 | 102,1 |
| 0,46 – 0,60 | 37,5 | 0,694 | 115,7 | 20,5 | 1,154 | 148,3 | 14,7 | 0,539 | 134,1 | 22,7 | 0,900 | 142,4 |
| понад 0,61 | 8,3 | 1,059 | 176,5 | 10,3 | 1,251 | 160,8 | 11,8 | 0,681 | 169,4 | 10,3 | 1,108 | 175,3 |
| Разом: | 100,0 | – | – | 100,0 | – | – | 100,0 | – | – | 100,0 | – | – |
| Середня фондовіддача | – | 0,600 | – | – | 0,778 | – | – | 0,402 | – | – | 0,632 | – |

і при зіставленні їх середніх значень. Так, у Тернопільській області середнє значення коефіцієнта використання пробігу – 0,36. При цьому середній рівень фондовіддачі автопідприємств становить 0,402 ткм/грн. В Івано-Франківській області відповідно – 0,39 і 0,600, у Львівській – 0,4 і 0,778 ткм/грн.

Розглядаючи залежність рівня фондовіддачі автотранспортних підприємств різних форм власності від величини коефіцієнта використання пробігу, слід зазначити, що лише серед державних підприємств немає таких, в яких би величина останнього переважала 0,6 (табл. 2.2.7). Цю групу склали автогосподарства колективної форми власності, питома вага яких – 10,3 % від загальної чисельності досліджуваних автогосподарств регіону. В основному це автогосподарства з найвищим рівнем фондовіддачі основних виробничих фондів. У зв'язку з цим в усіх інших виділених у таблиці групах спостерігається значно нижчий рівень фондовіддачі вантажних автогосподарств колективної форми власності порівняно з державними. Нижчою є у них і продуктивність рухомого складу.

**Таблиця 2.2.7**

Характеристика рівня фондовіддачі автогосподарств за формами власності Івано-Франківської, Львівської та Тернопільської областей у 1996 р.

| Показники | Групи автопідприємств за величиною коефіцієнта використання пробігу | | | | | | | | | | В середньому по автогос-подарствах регіону | |
|---|---|---|---|---|---|---|---|---|---|---|---|---|
| | до 0,15 | | 0,16 – 0,30 | | 0,31 – 0,45 | | 0,46 – 0,60 | | понад 0,61 | | | |
| | дер-жавна | колек-тивна | дер-жавна | колек-тивна | дер-жавна | колек-тивна | дер-жавна | колек-тивна | дер-жавна | колек-тивна | дер-жавна | колек-тивна |
| Кількість викона-них тонно-кіломет-рів у розрахунку на 1 грн. основних ви-робничих фондів | 0,395 | 0,144 | 0,722 | 0,594 | 0,910 | 0,611 | 1,071 | 0,836 | – | 1,108 | 0,624 | 0,709 |
| Кількість перевезе-них тонн вантажів у розрахунку на 1 грн. основних виробни-чих фондів | 0,011 | 0,013 | 0,030 | 0,030 | 0,058 | 0,043 | 0,158 | 0,101 | – | 0,117 | 0,027 | 0,043 |
| Річна продуктив-ність однієї спис-ко-вої автомобіле-тон-ни: | | | | | | | | | | | | |
| а) тис. тонн; | 0,109 | 0,056 | 0,184 | 0,184 | 0,310 | 0,289 | 0,594 | 0,487 | – | 0,615 | 0,138 | 0,251 |
| б) тис. тонно-кіло-метрів | 2,086 | 1,161 | 3,307 | 3,307 | 4,683 | 4,005 | 5,731 | 5,316 | – | 5,512 | 3,187 | 4,127 |

Так, інтенсивність його використання у вказаних чотирьох групах у 1,2 раза нижча, ніж у державних АТП. Отже, якщо в цілому основні виробничі фонди ефективніше використовуються в автогосподарствах з колективною формою власності порівняно з державними підприємствами автотранспорту, то це досягається за рахунок господарств з найвищим коефіцієнтом використання пробігу. В державних автогосподарствах, коефіцієнт використання пробігу яких не перевищує 0,6, ефективність використання основних засобів вища.

Зіставлення показників коефіцієнта використання пробігу і фондовіддачі основних виробничих фондів, а також дослідження характеру зв'язку між ними свідчить, що цей зв'язок можна виразити кількісно, застосувавши кореляційний метод – рівняння прямої.

$$y_x = a + bx \quad ,(2.2.4)$$

де $y_x$ – середня величина фондовіддачі основних виробничих фондів, ткм/грн.;

$x$ – коефіцієнт використання пробігу;

$b$ – коефіцієнт регресії, який відображає зміну фондовіддачі внаслідок зміни величини коефіцієнта використання пробігу.

Для визначення щільності зв'язку і правильності типу вибраного рівняння регресії розраховується коефіцієнт кореляції за такою формулою:

$$r_{xy} = \frac{\overline{xy} - \overline{x} \cdot \overline{y}}{\sigma_x \cdot \sigma_y} \quad , \quad (2.2.5)$$

де $r_{xy}$ – лінійний коефіцієнт кореляції;

$x$ – середнє значення факторної ознаки;

$y$ – середнє значення результативної ознаки;

$\sigma_x$, $\sigma_y$ – середньоквадратичні відхилення.

Оскільки різниця між його значенням і кореляційним відношенням не перевищує 0,1 (табл. 2.2.8), то гіпотезу про прямолінійну форму кореляційної залежності можна вважати підтвердженою. На графіку (рис. 2.2.2) цю залежність у розрізі областей «підтверджує характер форми зв'язку» [123, с. 90] отриманих емпіричних ліній регресії.

**Таблиця 2.2.8**

Показники кореляційної залежності фондовіддачі основних виробничих фондів вантажних АТП Івано-Франківської, Львівської та Тернопільської областей від величини коефіцієнта використання пробігу за 1996 р.

|  | Область | | | По регіону |
|---|---|---|---|---|
|  | Івано-Франківська | Львівська | Тернопільська |  |
| Рівняння зв'язку | $y_x = 0{,}1819 + 0{,}5092x$ | $y_x = 0{,}5752 + 0{,}7825x$ | $y_x = 0{,}1837 + 0{,}4221x$ | $y_x = 0{,}2901 + 0{,}7266x$ |
| Середня помилка коефіцієнта регресії | 0,0453 | 0,1286 | 0,0746 | 0,0751 |
| Гранична помилка коефіцієнта регресії | 0,0485 | 0,1826 | 0,0739 | 0,0879 |
| Довірчі межі коефіцієнта регресії | $0{,}4607 \leq b \leq 0{,}5577$ | $0{,}5999 \leq b \leq 0{,}9651$ | $0{,}3482 \leq b \leq 0{,}4960$ | $0{,}3482 \leq b \leq 0{,}4960$ |
| Лінійний коефіцієнт кореляції | 0,7135 | 0,8449 | 0,6781 | 0,7592 |
| Кореляційне відношення | 0,6829 | 0,7554 | 0,7375 | 0,7049 |
| Коефіцієнт детермінації | 0,5091 | 0,7139 | 0,4598 | 0,5764 |
| Фактичне значення F-критерію | 11,4 | 46,2 | 13,6 | 64,6 |
| Критичне значення F-критерію | 4,35 | 4,08 | 4,17 | 3,97 |
| Коефіцієнт еластичності | 0,521 | 0,351 | 0,454 | 0,483 |

Отримані коефіцієнти кореляції свідчать про високу щільність зв'язку між коефіцієнтом використання пробігу і фондовіддачею основних виробничих фондів. Чим ближче знаходиться величина коефіцієнта кореляції до одиниці, тим щільніший зв'язок між досліджуваними ознаками. Обчислені коефіцієнти детермінації показують, що рівень фондовіддачі в цілому по регіону на 57,6 % залежить від ступеня використання пробігу рухомого складу.

Отримані внаслідок дослідження показники залежності можуть бути використані для виявлення резервів зростання фондовіддачі основних виробничих фондів за рахунок підвищення рівня використання пробігу транспортних засобів, який в останні роки внаслідок різкого спаду виробництва дещо знизився.

З отриманих рівнянь можна прийти до висновку, що приріст коефіцієнта використання пробігу на 0,1 веде до збільшення обсягу транспортної продукції з однієї гривні основних фондів по підприємствах Івано-Франківської області на

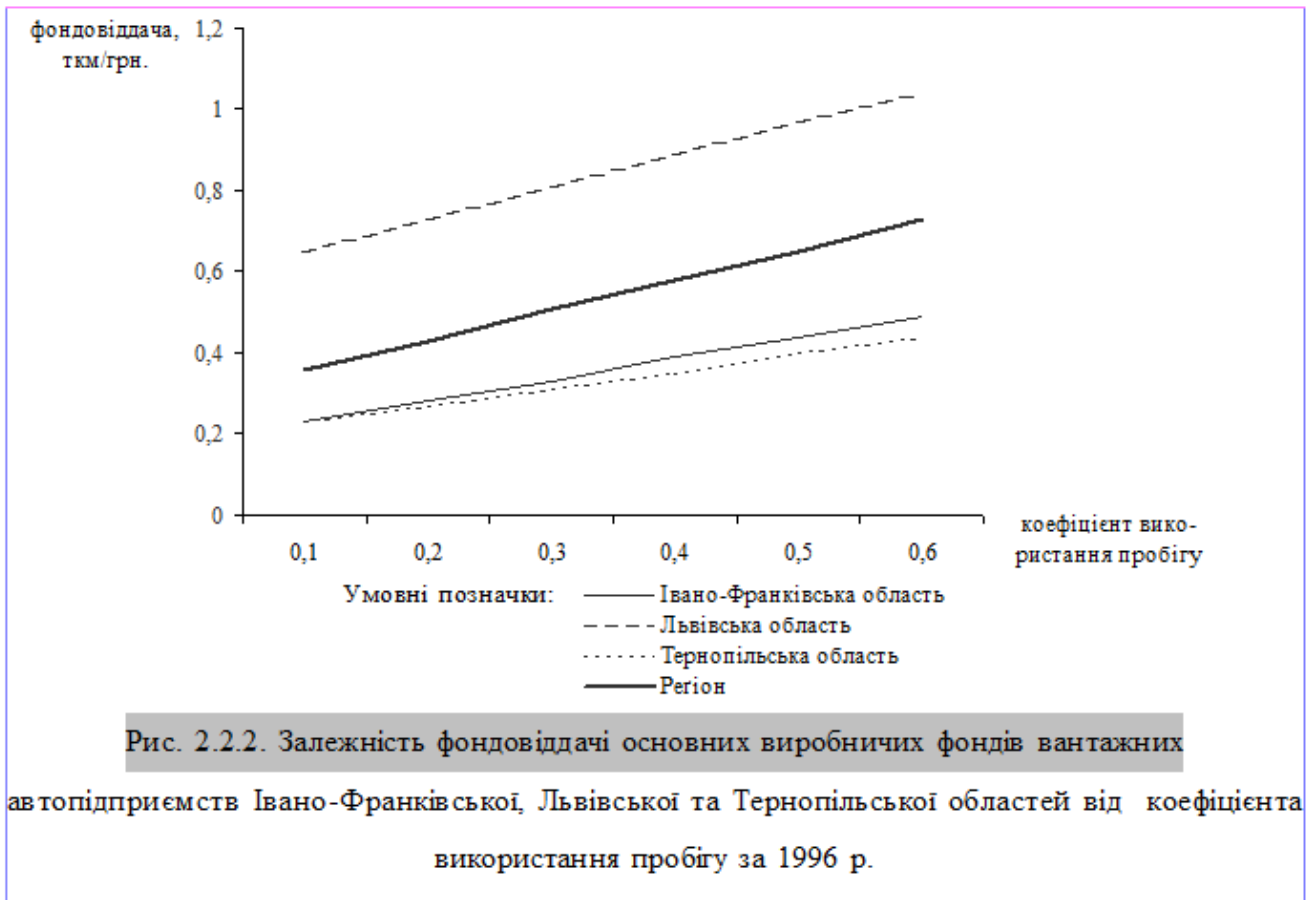

Рис. 2.2.2. Залежність фондовіддачі основних виробничих фондів вантажних автопідприємств Івано-Франківської, Львівської та Тернопільської областей від коефіцієнта використання пробігу за 1996 р.

0,509 ткм, Львівської – на 0,782 ткм, Тернопільської – на 0,422 ткм. Беручи до уваги те, що гранична помилка коефіцієнта регресії в цілому по регіону становить 0,0879, то фондовіддача досліджуваних автогосподарств при зростанні коефіцієнта використання пробігу на 0,1 підвищується не менше, як на 0,639 ткм, і не більше, як на 0,815 ткм.

### 2.2.3. Вплив виробітку рухомого складу на фондовіддачу основних виробничих фондів вантажних автотранспортних підприємств

Рівень фондовіддачі значною мірою залежить від ступеня використання активної частини основних виробничих фондів – транспортних засобів. Підвищення їх продуктивності дозволяє виконувати виробничу програму меншою кількістю одиниць транспортних засобів, що, в свою чергу, позначається на підвищенні рівня фондовіддачі.

За останні роки у зв'язку із різким спадом виробництва помітно знизились обсяги транспортної роботи і, відповідно, показники фондовіддачі та продуктивності транспортних засобів автогосподарств регіону. Так, за 1992 – 1996 рр. фондовіддача основних фондів автогосподарств регіону знизилась на 84,6 %, а продуктивність спискової автомобіле-тонни – на 69,4 %.

Економічна природа порівнюваних показників (виробітку однієї спискової автомобіле-тонни і фондовіддачі) вказує на існування між ними закономірної залежності. В міру підвищення виробітку спискової автомобіле-тонни зростає і рівень фондовіддачі основних виробничих фондів. Показовим є те, що ця закономірність простежується й для інших показників. Так, наприклад, в автопідприємствах з найвищим рівнем річного виробітку рентабельність основних фондів у 11,4 раза переважає значення цього ж показника в автогосподарствах, виробіток яких не перевищує 1500 ткм на спискову автомобіле-тонну (табл. 2.2.9).

**Таблиця 2.2.9**

Вплив рівня річного виробітку спискової автомобіле-тонни на фондовіддачу основних виробничих фондів вантажних АТП
Івано-Франківської, Львівської і Тернопільської областей за 1996 р.

| Показники | Групи автопідприємств за рівнем річного виробітку однієї спискової автомобіле-тонни, ткм | | | | | В середньому по автопідприємствах регіону |
|---|---|---|---|---|---|---|
| | до 1500 | 1501 – 3000 | 3001 – 4500 | 4501 – 6000 | понад 6000 | |
| Кількість виконаних тонно-кілометрів у розрахунку на 1 грн. вартості основних виробничих фондів | 0,152 | 0,483 | 0,670 | 0,814 | 1,310 | 0,632 |
| Кількість перевезених тонн вантажів у розрахунку на 1 грн. вартості основних виробничих фондів | 0,007 | 0,011 | 0,031 | 0,040 | 0,094 | 0,038 |
| Дохід у розрахунку на 1 грн. вартості основних виробничих фондів, тис. грн. | 0,150 | 0,170 | 0,273 | 0,284 | 0,294 | 0,213 |
| Рентабельність основних виробничих фондів, ткм/грн. | 0,008 | 0,013 | 0,030 | 0,034 | 0,091 | 0,029 |
| Річна продуктивність однієї спискової автомобіле-тонни: | | | | | | |
| а) тис. тонн; | 0,041 | 0,051 | 0,199 | 0,229 | 0,560 | 0,212 |
| б) тис. тонно-кілометрів | 0,874 | 2,281 | 3,815 | 5,138 | 7,769 | 4,127 |

Аналізуючи залежність фондовіддачі основних виробничих фондів від річного виробітку спискової автомобіле-тонни, неважко зауважити значну розбіжність у приростах фондовіддачі, вираженої в тонно-кілометрах та у гривнях доходу. Якщо рівень останньої в автогосподарствах з річним виробітком понад 6000 ткм у 1,96 раза вищий, ніж в автопідприємствах, що ввійшли до першої групи вказаної таблиці, то рівень фондовіддачі, вираженої в тонно-кілометрах, для вказаних груп підприємств – у 8,62 раза. Основною причиною, що породжує вказану розбіжність, є значна заборгованість перед автопідприємствами замовників транспортних послуг. Отже, вантажообіг (виконані тонно-кілометри), взятий нами в якості вимірника фондовіддачі, забезпечує більш точну оцінку ефективності використання основних фондів.

Спад промислового виробництва в Україні призвів до різкого зниження обсягів транспортних перевезень, виробітку рухомого складу. Протягом 1992 – 1996 рр. кількість автогосподарств із середньорічним виробітком спискової автомобіле-тонни до 1500 ткм зросла з 5,2 % до 38,1 % (табл. 2.2.10). Причому найбільшу питому вагу такі автопідприємства становлять в Івано-Франківській та Тернопільській областях – відповідно 47,5 і 41,2 %. Частка автопідприємств, річна продуктивність спискової автомобіле-тонни в яких перевищувала 6000 ткм, зменшилася протягом цього ж періоду з 62,8 % до 5,2 %.

Отже, найбільш високого рівня фондовіддачі досягли автогосподарства Львівської області, де він складає 0,778 ткм/грн., що на 30 % більше, ніж в автогосподарствах Івано-Франківської області і на 94 % більше, ніж у Тернопільській області.

Досліджуючи ефективність використання основних виробничих фондів автотранспортних підприємств державної та колективної форм власності, згрупованих за рівнем річного виробітку однієї спискової автомобіле-тонни (табл. 2.2.11), виявлено тенденцію, подібну до описаної у пункті 2.2.1. В цілому по регіону в автогосподарствах колективної форми власності і, зокрема, по кожній виділеній у таблиці групі підприємств спостерігається вищий рівень інтенсивності використання рухомого складу, ніж у державних АТП. Результат цього – вищий рівень фондовіддачі основних виробничих фондів в автопідприємствах з колективною формою власності. Так, в АТП з колективною формою власності

**Таблиця 2.2.10**

Залежність фондовіддачі вантажних автотранспортних підприємств від рівня річного виробітку спискової автомобіле-тонни за 1996 р.

| Групи автопідприємств за рівнем річного виробітку однієї спискової автомобіле-тонни, ткм | Область | | | | | | | | Регіон | | |
|---|---|---|---|---|---|---|---|---|---|---|---|
| | Івано-Франківська | | | Львівська | | | Тернопільська | | | | |
| | кількість підприємств у % до підсумку | фондовіддача | | кількість підприємств у % до підсумку | фондовіддача | | кількість підприємств у % до підсумку | фондовіддача | | кількість підприємств у % до підсумку | фондовіддача |
| | | ткм/грн. | у % до середнього значення | | ткм/грн. | у % до середнього значення | | ткм/грн. | у % до середнього значення | | ткм/грн. | у % до середнього значення |
| до 1500 | 47,5 | 0,133 | 22,2 | 20,5 | 0,233 | 29,9 | 41,2 | 0,104 | 25,9 | 38,1 | 0,152 | 24,1 |
| 1501 – 3000 | 19,1 | 0,136 | 22,7 | 41,0 | 0,538 | 69,2 | 32,4 | 0,355 | 88,3 | 28,9 | 0,483 | 76,4 |
| 3001 – 4500 | 16,7 | 0,772 | 128,7 | 17,9 | 0,790 | 101,5 | 14,7 | 0,457 | 113,7 | 16,5 | 0,670 | 106,0 |
| 4501 – 6000 | 16,7 | 0,792 | 132,8 | 10,3 | 1,079 | 138,7 | 8,8 | 0,631 | 157,0 | 11,3 | 0,814 | 128,8 |
| понад 6001 | – | – | – | 10,3 | 1,404 | 180,5 | 2,9 | 0,771 | 191,8 | 5,2 | 1,310 | 207,3 |
| Разом: | 100,0 | – | – | 100,0 | – | – | 100,0 | – | – | 100,0 | – | – |
| Середня фондовіддача | – | 0,600 | – | – | 0,778 | – | – | 0,402 | – | – | 0,632 | – |

**Таблиця 2.2.11**

Характеристика рівня фондовіддачі автогосподарств регіону за формами власності у 1996 р.

| Показники | Групи автогосподарств за рівнем річного виробітку однієї спискової автомобіле-тонни, ткм. | | | | | | | | | | В середньому по автогосподарствах регіону | |
|---|---|---|---|---|---|---|---|---|---|---|---|---|
| | до 1500 | | 1501 – 3000 | | 3001 – 4500 | | 4501 – 6000 | | понад 6001 | | | |
| | дер-жавна | колек-тивна | дер-жавна | колек-тивна | дер-жавна | колек-тивна | дер-жавна | колек-тивна | дер-жавна | колек-тивна | дер-жавна | колек-тивна |
| Кількість виконаних тонно-кілометрів у розрахунку на 1 грн. основних виробничих фондів | 0,094 | 0,160 | 0,398 | 0,557 | 0,603 | 0,785 | 0,704 | 1,001 | 1,301 | 1,771 | 0,624 | 0,709 |
| Кількість перевезених тонн вантажів у розрахунку на 1 грн. основних виробничих фондів | 0,006 | 0,012 | 0,009 | 0,014 | 0,020 | 0,036 | 0,038 | 0,054 | 0,090 | 0,158 | 0,027 | 0,043 |
| Річна продуктивність однієї спискової автомобіле-тонни: а) тис. тонн; | 0,038 | 0,058 | 0,042 | 0,065 | 0,188 | 0,247 | 0,197 | 0,343 | 0,250 | 0,558 | 0,138 | 0,251 |
| б) тис. тонно-кілометрів | 0,496 | 1,090 | 2,169 | 2,386 | 3,744 | 3,860 | 4,870 | 5,172 | 6,638 | 7,808 | 3,187 | 4,127 |

обсяг транспортної роботи (у тонно-кілометрах) в розрахунку на одну спискову автомобіле-тонну в 1,29 раза вищий, ніж у державних автогосподарствах; в тоннах перевезених вантажів – у 1,82 раза. А отже, й фондовіддача основних виробничих фондів АТП колективної форми власності, виражена у тонно-кілометрах, у 1,14 раза і в тоннах перевезених вантажів у 1,59 раза перевищує відповідні значення державних автогосподарств.

Вплив виробітку рухомого складу на фондовіддачу не викликає сумнівів. Для того, щоб виразити цю закономірність у кількісній формі, необхідно зіставити показники рівня фондовіддачі і виробітку спискової автомобіле-тонни. Важливо при цьому встановити не лише зв'язок між досліджуваними ознаками, але й виразити цей зв'язок у вигляді рівняння регресії. Аналіз даних табл. 2.2.9 переконує в існуванні між ними прямолінійної залежності. Тому для визначення впливу рівня виробітку спискової автомобіле-тонни на ступінь використання основних виробничих фондів застосовується рівняння прямої:

$$y_х = a + bx, \quad (2.2.6)$$

де $y_х$ – середня величина фондовіддачі, ткм/грн. (при заданих значеннях $x$);

$x$ – річний виробіток однієї спискової автомобіле-тонни, ткм;

*a, b* – параметри рівняння.

Для визначення щільності зв'язку і правильності типу вибраного рівняння регресії (про що йшлося вище) обчислюється лінійний коефіцієнт кореляції.

Отримані коефіцієнти кореляції (табл. 2.2.12) характеризують сильну щільність зв'язку між річним виробітком у розрахунку на одну спискову автомобіле-тонну і рівнем фондовіддачі основних виробничих фондів. Оскільки різниця між величинами лінійного коефіцієнта кореляції і кореляційного відношення не перевищує 0,1, то правильність обраного типу рівняння регресії доведена. Підтвердженням цього слугують й отримані емпіричні лінії регресії (рис. 2.2.3).

Аналізуючи одержані рівняння зв'язку, можна прийти до висновку, що в цілому по регіону приріст річного виробітку однієї автомобіле-тонни на 1000 ткм призводить

**Таблиця 2.2.12**

Показники кореляційної залежності фондовіддачі основних виробничих фондів вантажних АТП Івано-Франківської, Львівської і Тернопільської областей від рівня річного виробітку спискової автомобіле-тонни за 1996 р.

| | Область | | | По регіону |
|---|---|---|---|---|
| | Івано-Франківська | Львівська | Тернопільська | |
| Рівняння зв'язку | $y_х = 0,0422 + 0,1621x$ | $y_х = 0,2329 + 0,2151x$ | $y_х = 0,0943 + 0,1036x$ | $y_х = 0,0983 + 0,1838x$ |
| Середня помилка коефіцієнта регресії | 0,0204 | 0,0354 | 0,01831 | 0,0189 |
| Гранична помилка коефіцієнта регресії | 0,0293 | 0,0654 | 0,0165 | 0,0313 |
| Довірчі межі коефіцієнта регресії | $0,1328 \le b \le 0,1914$ | $0,1497 \le b \le 0,2805$ | $0,0871 \le b \le 0,1201$ | $0,1525 \le b \le 0,2151$ |
| Лінійний коефіцієнт кореляції | 0,7225 | 0,8765 | 0,4013 | 0,8153 |
| Кореляційне відношення | 0,6321 | 0,9664 | 0,4923 | 0,8885 |
| Коефіцієнт детермінації | 0,5220 | 0,7683 | 0,2424 | 0,6647 |
| Фактичне значення F-критерію | 27,3 | 127,7 | 11,1 | 207,5 |
| Критичне значення F-критерію | 3,49 | 3,23 | 3,32 | 3,07 |
| Коефіцієнт еластичності | 0,889 | 0,737 | 0,720 | 0,827 |

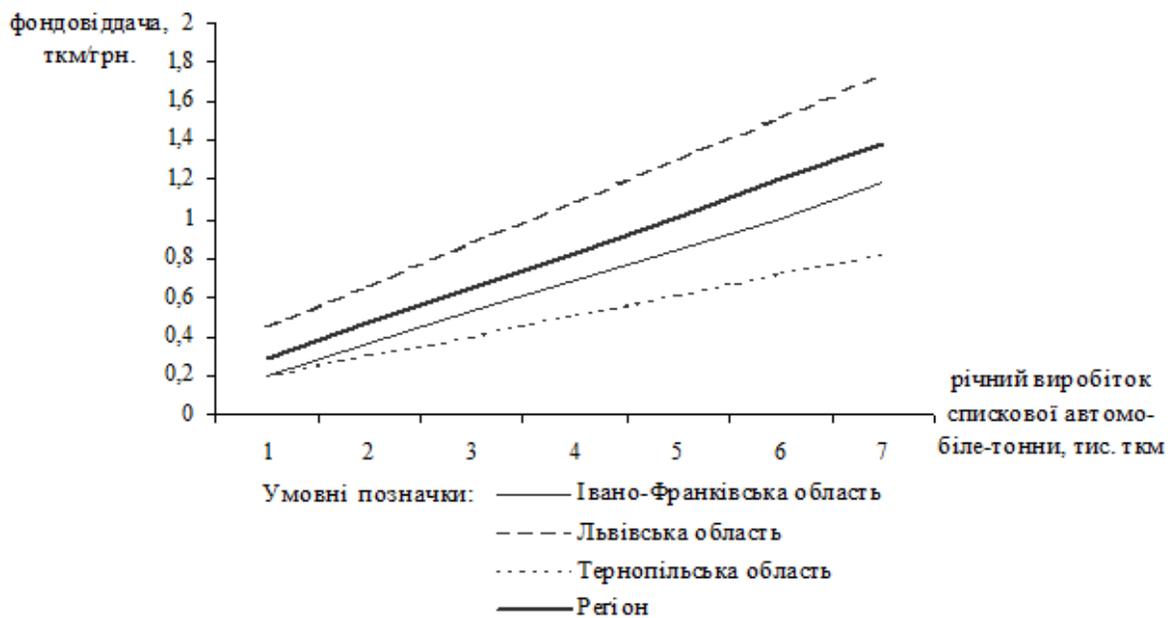

Рис. 2.2.3. Залежність фондовіддачі основних виробничих фондів вантажних автогосподарств від рівня річного виробітку спискової автомобіле-тонни до збільшення обсягу транспортної продукції з кожної гривні основних виробничих фондів на 0,184 ткм, в тому числі по автогосподарствах Львівської області – на 0,215 ткм, Івано-Франківської – на 0,162 ткм, Тернопільської – на 0,104 ткм. Зважаючи на граничну помилку коефіцієнта реґресії (табл. 2.2.12), можна стверджувати, що загалом по трьох областях приріст фондовіддачі в результаті зміни річного виробітку спискової автомобіле-тонни коливатиметься в межах від 0,153 до 0,215 ткм.

Отже, значення рівнянь зв'язку полягає не лише у тому, що вони дають можливість встановити кількісну характеристику залежності ступеня використання основних виробничих фондів від рівня виробітку спискової автомобіле-тонни. За їх допомогою можна прогнозувати з відповідною екстраполяцією підвищення ступеня використання основних фондів досліджуваних автогосподарств за рахунок зростання виробітку рухомого складу. Знаючи параметри одержаного рівняння і підставляючи в нього відповідні значення аргумента (фактора), можна передбачити значення функції (результативної ознаки). Визначивши таким чином необхідний приріст виробітку вантажних автомобілів, можна знайти раціональний варіант найефективнішого використання основних виробничих фондів.

Обчислені коефіцієнти детермінації (табл. 2.2.12) виражають ступінь загальної варіації фондовіддачі основних фондів, що зумовлена зміною виробітку рухомого складу. Так, по автотранспортних підприємствах Івано-Франківської області 52,2 % загальної варіації фондовіддачі зумовлено зміною виробітку спискової автомобіле-тонни, по автогосподарствах Львівської області – 76,8 %, Тернопільської – 24,2 %, в цілому по регіону – 66,5 %.

Обчислені по досліджуваних автотранспортних підприємствах регіону коефіцієнти еластичності показують, що при збільшенні річного виробітку однієї спискової автомобіле-тонни на 1 % від середнього рівня рівень фондовіддачі підвищується на 0,83 %, в т.ч. в Івано-Франківській області – на 0,89 %, Львівській – на 0,74 %, Тернопільській – на 0,72 %.

## 2.3. Аналіз резервів підвищення ефективності використання основних виробничих фондів вантажного автомобільного транспорту

Аналіз ефективності використання основних виробничих фондів вантажних автотранспортних підприємств Івано-Франківської, Львівської та Тернопільської областей засвідчив залежність рівня фондовіддачі від багатьох факторів. Характер взаємозв'язків, що існують між ними, проявляється у посиленні, чи послабленні дії інших чинників. Їх вплив на фондовіддачу є комплексним, і його не можна розглядати як суму ізольованих впливів.

Попередній аналіз дозволив встановити залежність рівня фондовіддачі від багатьох факторів, дія яких тісно переплітається, простежити загальний характер зв'язку з ними. Проте застосовані методи дослідження не дозволили повною мірою розкрити суть даного явища, оцінити силу впливу кожного у сукупності з іншими на рівень фондовіддачі. Застосування методу багатофакторного кореляційно-реґресійного аналізу, компенсуючи ці прогалини, дозволяє оцінити ступінь впливу на фондовіддачу основних виробничих фондів кожного із включених у модель факторів при фіксованому значенні інших, а також при будь-яких можливих їх поєднаннях, з певним ступенем точності знайти теоретичне значення цього показника [див., напр., 123, 150, 220, 236,]. Ефективність кореляційно-реґресійного аналізу виявляється саме при побудові багатофакторних моделей. Дослідження економічних явищ на основі вивчення лише парних зв'язків між ознаками хибує певною неточністю. Реально ж коефіцієнти парної кореляції можуть виявитися завищеними, оскільки вони не відображають впливу інших чинників і цим перебільшують щільність зв'язку між явищами та ступінь їх впливу на результативну ознаку. І навпаки, низькі значення коефіцієнта парної кореляції при досить складному зв'язку не завжди означають слабкий вплив фактора на досліджуване явище. У цьому легко переконатися, зіставляючи між собою коефіцієнти парної та часткової кореляції. Тому при вивченні економічних явищ слід застосовувати багатофакторний кореляційно-реґресійний аналіз як інструмент проникнення в сутність механізму формування закономірностей явищ, що складаються під впливом багатьох, часто суперечливих тенденцій. В сучасних умовах саме це зумовлює підвищення ролі цього аналізу. Нині діюча система господарювання ставить перед автомобільним транспортом завдання доведення аналізу діяльності автогосподарств до такого рівня, який дозволяв би не лише характеризувати загальну закономірність розвитку досліджуваного явища, а й давав би точну кількісну оцінку наявних внутрішньо-виробничих резервів. Можливості кореляційно-реґресійного аналізу у вивченні закономірностей розвитку явищ великі. Побудовані багатофакторні кореляційні моделі поруч з кількісною характеристикою сили впливу окремих факторів на фондовіддачу можуть успішно використовуватися для оцінки роботи окремих автотранспортних підприємств щодо їх об'єктивних можливостей. Ці ж моделі можна використовувати і при вивченні ефективності використання основних виробничих фондів окремого автопідприємства у зіставленні з фондовіддачею інших автогосподарств та виявленні причин їх відхилень. Тобто кореляційно-реґресійний аналіз дозволяє виявити резерви підвищення рівня фондовіддачі і отримати відповідь на запитання, якою мірою і за яких умов вони можуть бути досягнуті в інших автотранспортних підприємствах.

Застосування кореляційно-реґресійного аналізу висуває до вхідної інформації ряд вимог. Передусім, це вимога однорідності і достатності досліджуваної сукупності. Для дотримання цих вимог були відібрані 97 вантажних автотранспортних підприємств Івано-Франківської, Львівської та Тернопільської областей, які перебувають приблизно в однакових економічних, природно-кліматичних та інших умовах.

З метою визначення щільності зв'язку між досліджуваними факторами складено матрицю коефіцієнтів парної кореляції (табл. 2.3.1). За величиною парних коефіцієнтів можна судити про силу впливу факторів на рівень фондовіддачі. Серед досліджуваних факторів найбільшим рівнем впливу на фондовіддачу відзначається коефіцієнт використання автомобіле-тонно-днів перебування у господарстві. Щільність його зв'язку з результативною ознакою (фондовіддачею) становить 0,966. Проте слід зазначити, що за величиною парних коефіцієнтів кореляції можна судити про щільність зв'язку між результативною і факторною ознаками лише наближено, оскільки між самими чинниками також існують складні взаємозв'язки, вплив яких взаємопереплітається. В результаті цього невисокий парний коефіцієнт кореляції у багатофакторному рівнянні може виявитися істотним. Тому не слід відсіювати фактори лише на основі парних коефіцієнтів кореляції. З даних табл. 2.3.1 видно, що зв'язок між самими факторами значно нижчий від зв'язку факторів з результативною ознакою. Це свідчить про правильний відбір чинників, про їх відносно невелику колінеарність.

**Таблиця 2.3.1**

Матриця коефіцієнтів парної кореляції

|  | Фондовіддача | Коефіцієнт використання автомобіле-тонно-днів перебування у господарстві | Коефіцієнт використання пробігу | Річний виробіток спискової автомобіле-тонни |
|---|---|---|---|---|
| Фондовіддача | – | 0,966 | 0,705 | 0,889 |
| Коефіцієнт використання автомобіле-тонно-днів перебування у господарстві | 0,966 | – | 0,086 | 0,234 |
| Коефіцієнт використання пробігу | 0,705 | 0,086 | – | 0,276 |
| Річний виробіток спискової автомобіле-тонни | 0,889 | 0,234 | 0,276 | – |

Одним з важливих етапів багатофакторного кореляційного аналізу є вибір форми зв'язку між досліджуваними ознаками. У багатофакторних моделях вибрати рівняння регресії значно складніше, ніж при парній кореляції, оскільки дії різних факторів взаємно переплітаються і їх неможливо перевірити графічно.

Важливого значення набуває якісний аналіз кожного із факторів з результативною ознакою. Визначення форми кореляційної залежності вимагає чіткого уявлення про сутність досліджуваних процесів. На думку багатьох економістів, для визначення багатофакторних зв'язків можна обмежитися рівнянням лінійної регресії, оскільки на невеликому інтервалі будь-яку функцію можна розглядати як пряму. Дослідження економічних процесів за допомогою лінійної залежності не допускає великої похибки, зате значно спрощує інтерпретацію отриманих результатів.

Кореляційну залежність фондовіддачі основних виробничих фондів автотранспортних підприємств в розрізі областей можна виразити рівняннями, наведеними в табл. 2.3.2. Обчислення параметрів рівняння множинної регресії і окремих статистичних оцінок здійснено з допомогою комп'ютерної техніки.

**Таблиця 2.3.2**

Рівняння багатофакторної кореляційної залежності рівня фондовіддачі основних виробничих фондів вантажних АТП Івано-Франківської,

Львівської та Тернопільської областей у 1996 р.

| Область | Рівняння зв'язку | Коефіцієнт детермінації | Критичне значення коефіцієнта детермінації при рівні істотності 0,05 | Фактичне значення F-критерію | Критичне значення F-критерію при рівні істотності 0,05 |
|---|---|---|---|---|---|
| Івано-Франківська | $y = -0{,}0708 + 0{,}7033x_1 + 0{,}0671x_2 + 0{,}1585x_3$ | 0,8418 | 0,318 | 13,162 | 3,10 |
| Львівська | $y = 0{,}0964 + 0{,}2338x_1 + 0{,}2726x_2 + 0{,}2062x_3$ | 0,5871 | 0,152 | 83,640 | 2,88 |
| Тернопільська | $y = 0{,}1271 + 0{,}1668x_1 + 0{,}0849x_2 + 0{,}1067x_3$ | 0,6037 | 0,220 | 6,78 | 2,91 |
| За регіоном | $y = -0{,}0296 + 0{,}5238x_1 + 0{,}1239x_2 + 0{,}1735x_3$ | 0,6705 | 0,06 | 133,188 | 2,71 |

Отримані коефіцієнти регресії відображають, на скільки одиниць в середньому зміниться фондовіддача основних виробничих фондів вантажних АТП регіону при зміні кожного фактора на одиницю його виміру при фіксованому значенні інших факторів.

Щільність зв'язку між результативною ознакою та сукупністю факторних ознак характеризується за допомогою сукупного коефіцієнта детермінації. Він «характеризує частку варіації результативної ознаки, яка лінійно пов'язана з варіацією включених в рівняння регресії факторних ознак» [220, с. 94].

Таким чином, на основі обчислень величини сукупного коефіцієнта детермінації можна зробити висновок про те, що в обстеженій сукупності вантажних автогосподарств 67,05 % варіації фондовіддачі лінійно пов'язані з різними рівнями коефіцієнта використання автомобіле-тонно-днів перебування у господарстві, коефіцієнта використання пробігу та річного виробітку спискової автомобіле-тонни, а на 32,95 % – із впливом інших факторів, що не увійшли у модель. В істотності цього зв'язку легко переконатися, зіставивши фактичні значення сукупного коефіцієнта детермініції з критичними, що відповідають рівню істотності 0,05. Оскільки величини останніх значно менші від фактичних, то істотність цього зв'язку не викликає сумніву.

Здійснена за допомогою F-критерію перевірка адекватності одержаних рівнянь регресії підтвердила його достатню апроксимаційну здатність. Фактичне значення F-критерію по автотранспортних підприємствах регіону становить 133,1 і значно перевищує його критичне значення 2,71 з рівнем істотності 0,05 (табл. 2.3.2).

Багатофакторний кореляційно-регресійний аналіз фондовіддачі має велике практичне значення не лише для визначення щільності зв'язку та сили впливу окремих чинників на результативну ознаку. Характеризуючи закономірності змін певного явища, рівняння множинної регресії може використовуватися з метою виявлення резервів підвищення ефективності використання основних виробничих фондів вантажних автотранспортних підприємств. Підвищення рівня фондовіддачі за рахунок зміни факторів, що впливають на її величину, здійснюється з допомогою коефіцієнтів регресії.

Визначення резервів підвищення рівня фондовіддачі ґрунтується на обчисленні середнього значення кожного фактора. Після цього всю сукупність досліджуваних АТП кожної області поділяють на дві групи: із значеннями вищими та із значеннями нижчими від середньої величини відповідного фактора. Для кожної групи обчислюється своє середнє значення фактора. Визначити зміну середнього значення кожного фактора можна двома способами: шляхом підтягування автогосподарств з нижчими значеннями до середнього рівня, або шляхом підтягування всієї сукупності автопідприємств до середньопрогресивного значення фактора. Перший спосіб забезпечує мінімальний резерв збільшення фактора, другий – його оптимальний резерв збільшення. Підставляючи отримані значення резервів підвищення фондовіддачі в рівняння множинної регресії, одержимо величину резервів підвищення фондовіддачі за рахунок впливу різних факторів. У табл. 2.3.3 наведені дані про величину резервів підвищення фондовіддачі вантажних автогосподарств Івано-Франківської, Львівської та Тернопільської областей за 1996 р.

**Таблиця 2.3.3**

Оптимальні резерви зростання фондовіддачі підприємств
вантажного автомобільного транспорту Івано-Франківської,
Львівської та Тернопільської областей

| Фактор | Область | | | | | | Регіон | |
|---|---|---|---|---|---|---|---|---|
| | Івано-Франківська | | Львівська | | Тернопільська | | | |
| | ткм/грн. | % | ткм/грн. | % | ткм/грн. | % | ткм/грн. | % |
| Коефіцієнт використання автомобіле-тонно-днів перебування у господарстві | +0,037 | 9,7 | +0,060 | 6,8 | +0,005 | 1,5 | +0,023 | 4,0 |
| Коефіцієнт використання пробігу | +0,005 | 1,3 | +0,016 | 1,8 | +0,006 | 1,8 | +0,005 | 1,0 |
| Річна продуктивність спискової автомобіле-тонни, ткм | +0,145 | 38,2 | +0,150 | 16,9 | +0,147 | 43,7 | +0,135 | 23,8 |
| Разом: | +0,187 | 49,2 | +0,226 | 25,5 | +0,158 | 47,0 | +0,163 | 28,7 |

Отже, реалізація резервів у межах оптимального рівня дозволила б підвищити фондовіддачу основних виробничих фондів в цілому по автопідприємствах досліджуваного регіону на 0,163 ткм/грн., тобто на 28,7 %. Значні резерви підвищення фондовіддачі спостерігаються в автогосподарствах Івано-Франківської та Тернопільської областей. Їх реалізація дозволила б підвищити рівень фондовіддачі відповідно на 49,2 % та 47,0 %. Реалізація резервів за умови досягнення всіма автопідприємствами регіону максимального значення відібраних факторів дозволила б підвищити рівень фондовіддачі на 114,1 %. Проте, аналізуючи резерви зростання фондовіддачі, слід зважати на те, що кожен з факторів відібраний за умови найефективнішої його дії за певних обставин. Отримані показники резервів можуть використовуватись в перспективному плануванні підвищення ефективності використання основних виробничих фондів вантажних автотранспортних підприємств.

Параметри рівняння множинної кореляції можуть також використовуватися з метою більш об'єктивної оцінки виробничої діяльності окремих автотранспортних підприємств, які працюють приблизно в однакових умовах. З їх допомогою можна зіставляти рівні фондовіддачі як між окремими автотранспортними підприємствами, так і з середньогруповими областей чи регіону. Це має велике практичне значення при перспективному плануванні зростання фондовіддачі з урахуванням однакової напруженості їх роботи. До того ж за допомогою коефіцієнтів регресії можна визначити ступінь розходжень у рівнях фондовіддачі, зумовлених неоднаковим рівнем використання наявних виробничих можливостей окремими автотранспортними підприємствами. Для цього в рівняння регресії слід підставити фактичні значення відібраних факторів кожного автотранспортного підприємства. Отримана теоретична величина фондовіддачі характеризує такий її рівень, який може бути досягнутий при заданих значеннях факторів за умови, що ефективність використання останніх відповідає середньому рівню досліджуваної сукупності підприємств.

Порівнюючи фактичне значення фондовіддачі автогосподарства з розрахунковим теоретичним, визначають коефіцієнт ефективності, який характеризує ступінь використання автопідприємством своїх потенційних можливостей.

Результати, отримані внаслідок кореляційно-регресійного аналізу, засвідчують наявність значних резервів для підвищення ефективності використання основних виробничих фондів в автотранспортних підприємствах Івано-Франківської, Львівської та Тернопільської областей. Їх реалізація сприятиме зростанню загальної ефективності виробничої діяльності вантажних автогосподарств.

# РОЗДІЛ 3
# АДАПТИВНА СПРЯМОВАНІСТЬ ЯКІСНИХ ЗМІН В УПРАВЛІННІ І РЕАЛІЗАЦІЇ ВИРОБНИЧОГО ПОТЕНЦІАЛУ ВАНТАЖНИХ АВТОГОСПОДАРСТВ ТА АКТИВІЗАЦІЇ ІННОВАЦІЙ В УМОВАХ ТРАНСФОРМАЦІЇ ЕКОНОМІЧНИХ ВІДНОСИН

## 3.1. Маркетинг як інструмент забезпечення збалансованості поточних і стратегічних інтересів вантажних автотранспортних підприємств на ринку транспортних послуг

Трансформація економічних відносин в сучасній Україні охоплює всі сфери діяльності суб'єктів господарювання. Розвиток народного господарства країни «безпосередньо залежить від швидкості трансформації первинного економічного аґента-виробника, його опанування сучасним арсеналом методів та форм конкурентної та неконкурентної співпраці» [30, с. 48]. У галузі вантажних автомобільних перевезень це знаходить свій вияв у формуванні нових для нашої економіки підходів в організації діяльності автотранспортних підприємств, забезпеченні відповідних інфраструктурних змін. Необхідність цього зумовлюється й постійним загостренням конкурентної боротьби за замовника, що ведеться не тільки між автогосподарствами, але і приватними власниками вантажівок, які з'являються на ринку транспортних послуг.

У цих умовах підвищення ефективності використання основних виробничих фондів є назрілим у переліку пріоритетних завдань, від успішного вирішення яких залежить результативність діяльності автотранспортних підприємств, а в окремих випадках й доцільність функціонування. У зв'язку з цим на перший план висувається необхідність формування в автогосподарствах такого організаційно-економічного середовища, в якому раціональне використання основних виробничих фондів, зокрема рухомого складу, виступало б одним з ґарантів економічної стійкості та конкурентоспроможності транспортної продукції вантажних автопідприємств. В сучасних умовах найбільш доцільним і результативним засобом досягнення цього бачиться перехід автогосподарств на таку організацію діяльності, що в своїй основі базується на принципах маркетингу і транспортної логістики.

Якщо маркетинг займається виявленням, формуванням і стимулюванням попиту з метою задоволення усього спектру потреб ринку в автотранспортних послугах, то логістика покликана притаманними їй способами і методами задовольнити сформований маркетингом попит шляхом найбільш ефективної організації транспортно-експедиційного обслуговування з мінімальними витратами. Під останніми «розуміють не просто меншу суму витрат на виробництво продукції чи послуг, ніж у конкурентів, а спроможність підприємства розробляти, продукувати і збувати продукцію чи послуги більш ефективно, ніж конкуренти» [1, с. 30]. Це досягається як за рахунок поліпшення техніко-економічних показників, серед яких коефіцієнти використання пробігу парку вантажівок, вантажопідіймальності, часу перебування в наряді, так і шляхом розширення набору послуг, зниження витрат на технічне обслуговування, поточний ремонт і ін. (про що детальніше мова піде далі). Проте, як засвідчують результати та аналіз особливостей і характеру діяльності вантажних автотранспортних підприємств Івано-Франківської, Львівської і Тернопільської областей, цим засобам ефективної організації збутової діяльності з боку керівництва автогосподарств не надається належна увага. І якщо принципи логістичного підходу в тій чи іншій мірі враховуються і реалізуються у процесі організації та виконання транспортної роботи, то маркетингово орієнтована діяльність лише починає зароджуватись. А поки що «в умовах зниження попиту на перевезення підприємства всупереч ринковій логіці зберігають всі робочі місця, порівняно високий рівень оплати праці (особливо управлінського персоналу), не вживають ніяких серйозних заходів для зниження матеріальних витрат і диверсифікації виробництва. Логіка все та ж: всі витрати «загнати» в тариф, «накрутити» рентабельність, а якщо тариф клієнтам не під силу – «вибивати» бюджетні дотації» [9, с. 37].

Необхідність маркетингу в забезпеченні ефективного використання основних виробничих фондів автопідприємств, в першу чергу їх активної частини, зумовлене:

1) зростанням ролі вантажного автотранспорту в умовах переходу до ринкових відносин, характерною ознакою яких є поява значної кількості порівняно невеликих підприємств різних форм власності, які, по-перше, мають невеликі потреби в одноразовому споживанні сировини, товарів, напівфабрикатів;

по-друге, висувають більш жорсткі вимоги до перевізників щодо строків доставки вантажів через невелику власну потужність споживання і неможливість створення значних запасів; по-третє, не мають альтернативних шляхів доставки товарів, наприклад, залізничних. До того ж доставка вантажу на віддаль до 200 км здійснюється в середньому у 12 разів швидше, ніж залізницею, при завезенні і вивезенні вантажів автомобілями зі станцій, або ж у 5 разів швидше порівняно з прямим перевезенням залізницею, а на віддаль 500 км – у 7 разів швидше, ніж залізницею з участю автотранспорту, і в 3 рази швидше, якщо використовуються рейкові під'їзні шляхи при прямих перевезеннях залізницею [див.: 81, с. 10];

2) загостренням проблеми реалізації транспортної продукції як наслідку зростання конкуренції на ринку вантажних автоперевезень та значним розривом між обсягами попиту і пропозиції.

Вплив ринкових чинників відображається і на підвищенні вимог до зовнішнього вигляду, марки і типу використовуваного рухомого складу, необхідності малотоннажного, рефрижераторного автомобільного парку або спеціально обладнаного для створення необхідних режимів і умов перевезення тих чи інших видів вантажів. «В міру зміцнення економічних зв'язків між покупцем і продавцем транспортних послуг зростає роль економічних розрахунків» [82, с. 12], в основі яких – маркетингова стратегія.

Ефективне використання рухомого складу в комплексі маркетингових завдань, зумовлених вдосконаленням транспортного обслуговування споживачів, базується на врахуванні системи взаємопов'язаних принципів (див. рис. 3.1.1). Їх впорядкування та формалізація є необхідною умовою побудови стрункої системи маркетингових цілей.

Досягнення цілей маркетингу здобувається через реалізацію властивих йому функцій:

– дослідження та ідентифікація ринку вантажних перевезень;
– розмежування ринку транспортних послуг за співвідношенням попиту і пропозиції;

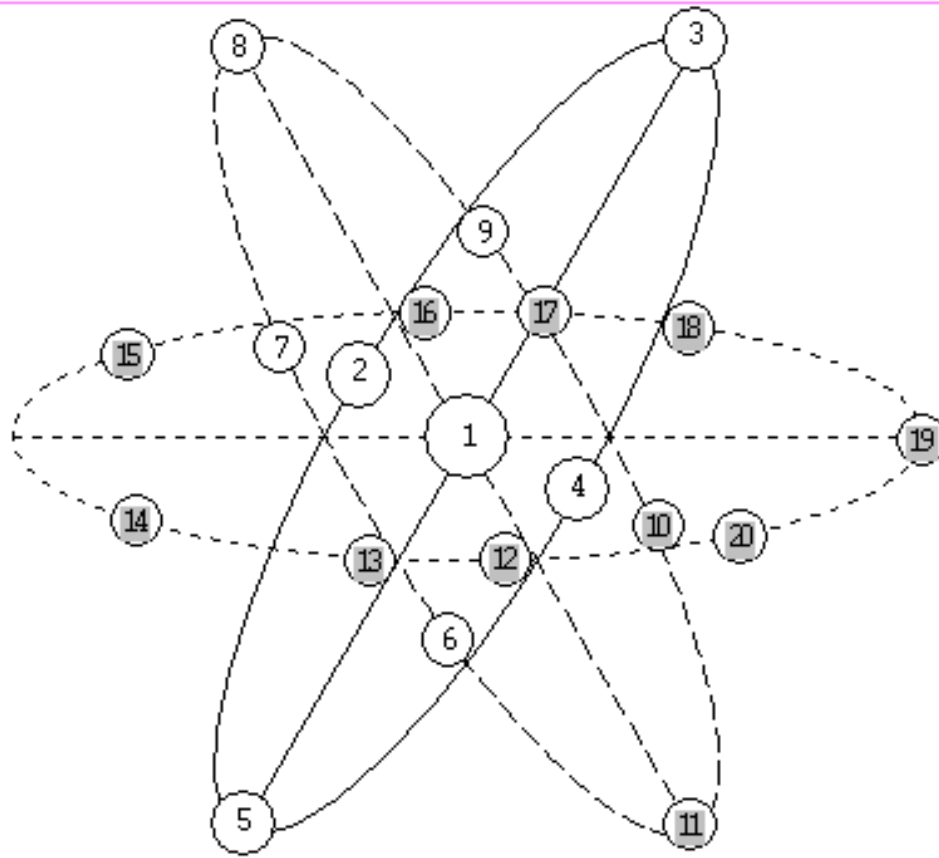

**Основоположний:** 1– задоволення попиту на перевезення вантажів.

**Визначальні:** 2 – врахування якості транспортного обслуговування; 3 – відповідність провізної спроможності рухомого складу величині та коливанням попиту; 4 – своєчасність перевезень; 5 – віднесення до сфери послуг.

**Забезпечуючі:** 6 – фінансова достатність розвитку системи вантажних перевезень; 7 – оптимізація структури парку автомобілів; 8 – диференціація тарифів; 9 – досконалість правового урегулювання; 10 – системна безпечність перевезень; 11 – комерціалізація і конкурентність.

**Узгоджувальні:** 12 – реалізація логістичного підходу; 13 – загальна координація функціонування різних видів транспорту; 14 – комплексність технологічного забезпечення; 15 – відповідність спеціалістів; 16 – уніфікація звітності; 17 – взаємна відповідальність учасників перевезень; 18 – єдність управління; 19 – автоматизація управління; 20 – циклічна замкнутість управління.

Рис. 3.1.1. Система принципів удосконалення транспортного обслуговування споживачів

– формалізація і забезпечення переваг транспортних послуг стосовно конкурентів;

– розроблення і реалізація маркетингової пропозиції.

Головною функцією є, звичайно, розробка маркетингової пропозиції, яка передбачає різні варіанти реалізації транспортних послуг потенційним споживачам. Їй передує виконання робіт щодо вивчення ринку вантажних автоперевезень, яке ґрунтується на таких засадах: системності, комплексності, регулярності, об'єктивності, точності, економічності, оперативності, ретельності [див.: 53, с. 45]. Дотримання вказаних принципів забезпечує належний рівень достовірності ринкових досліджень і зумовлюється орієнтацією автопідприємств на конкретний ринок транспортних послуг з притаманним йому конкурентним середовищем. У зв'язку з цим автогосподарства більш гостро відчувають потребу в детальній і різнобічній інформації з ринку транспортних послуг і всіх змінах у попиті. Швидке реагування на найменші зміни кон'юнктури ринку стає їх життєвою необхідністю і є можливим за умови ефективного функціонування інформаційного потоку.

Виходячи на ринок транспортних послуг, вантажне автотранспортне підприємство повинне розуміти, що воно не може обслуговувати всіх його споживачів навіть за умови достатньої виробничої потужності. Адже споживачі, виявляючи попит на транспортне обслуговування, по-різному використовують цей вид послуг і керуються при цьому різними мотивами. А тому в процесі дослідження ринку одним із складових елементів є його сеґментація (розподіл потенційних споживачів) за цими мотивами поведінки та іншими ознаками, максимальне врахування яких необхідне для вироблення маркетингової пропозиції. Під поняттям сеґмента ринку розуміють «деяку кількість споживачів, які однаково реаґують на один і той же склад спонукальних стимулів маркетингу» [160, с. 10]. Виходячи з цього, потенційних споживачів транспортних послуг можна охарактеризувати за трьома ознаками: 1) географічною (територіальною); 2) галузевою; 3) видами вантажу.

Модель багатофакторного сеґментування ринку автотранспортних послуг щодо перевезень вантажів подано на рисунках 3.1.2 – 3.1.4.

Початковий етап моделювання попиту на ринку автотранспортних послуг

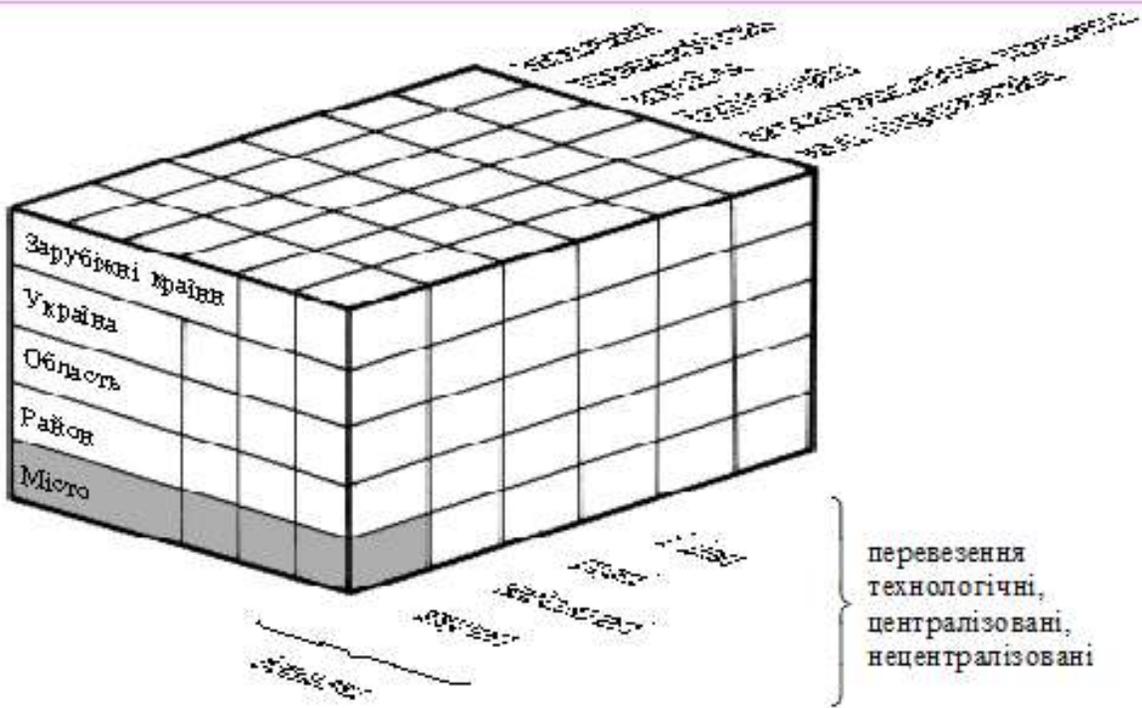

Рис. 3.1.2. Сегментація ринку відповідно до попиту на вантажні автоперевезення

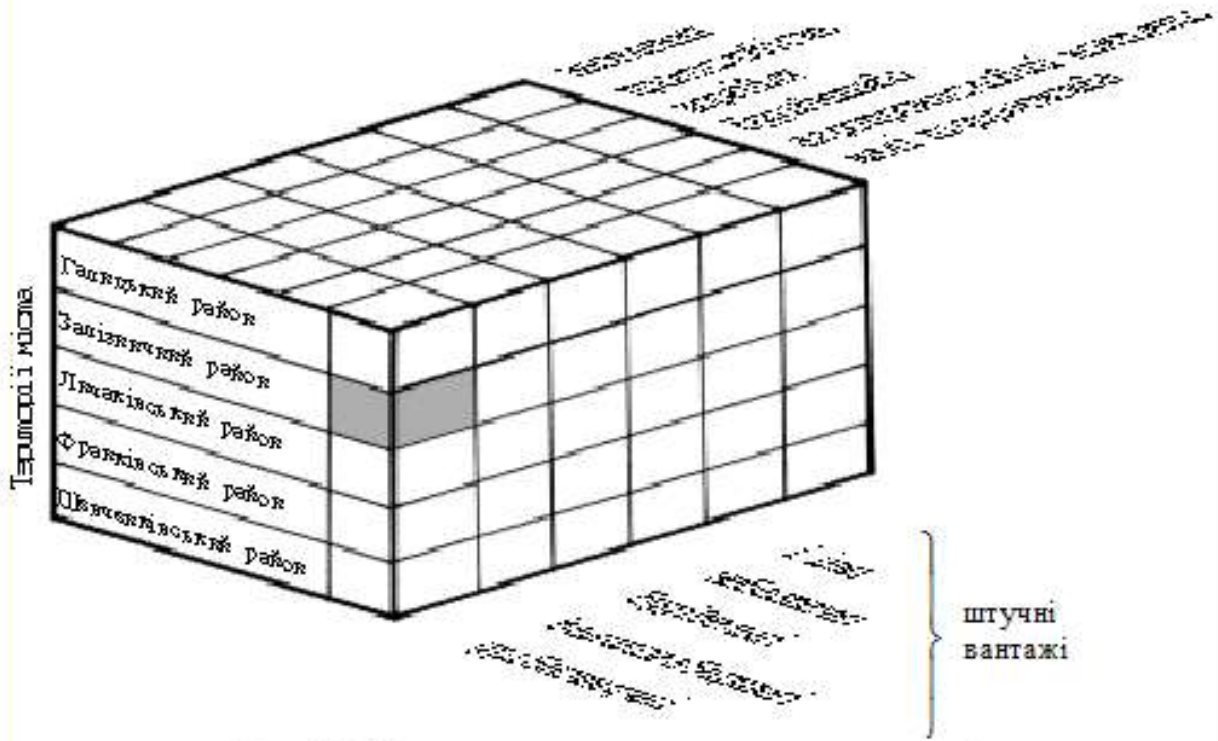

Рис. 3.1.3. Ринок попиту на автоперевезення на прикладі означеної території і конкретних вантажів зображено на рис. 3.1.2. Заретушований сеґмент ринку відображає попит на перевезення штучних вантажів у межах міста.

На рис. 3.1.3 наведено наступний етап сеґментації ринкового попиту по території міста та конкретних штучних вантажах. На цьому рисунку заретушовано сеґмент попиту малих підприємств на перевезення дрібноштучних вантажів у Залізничному районі м. Львова.

Наступним елементом сеґментації ринку є моделювання реальної пропозиції транспортних послуг. У нашій моделі – це пропозиція автоперевезень дрібноштучних вантажів у Залізничному районі м. Львова для малих підприємств (див. рис. 3.1.4).

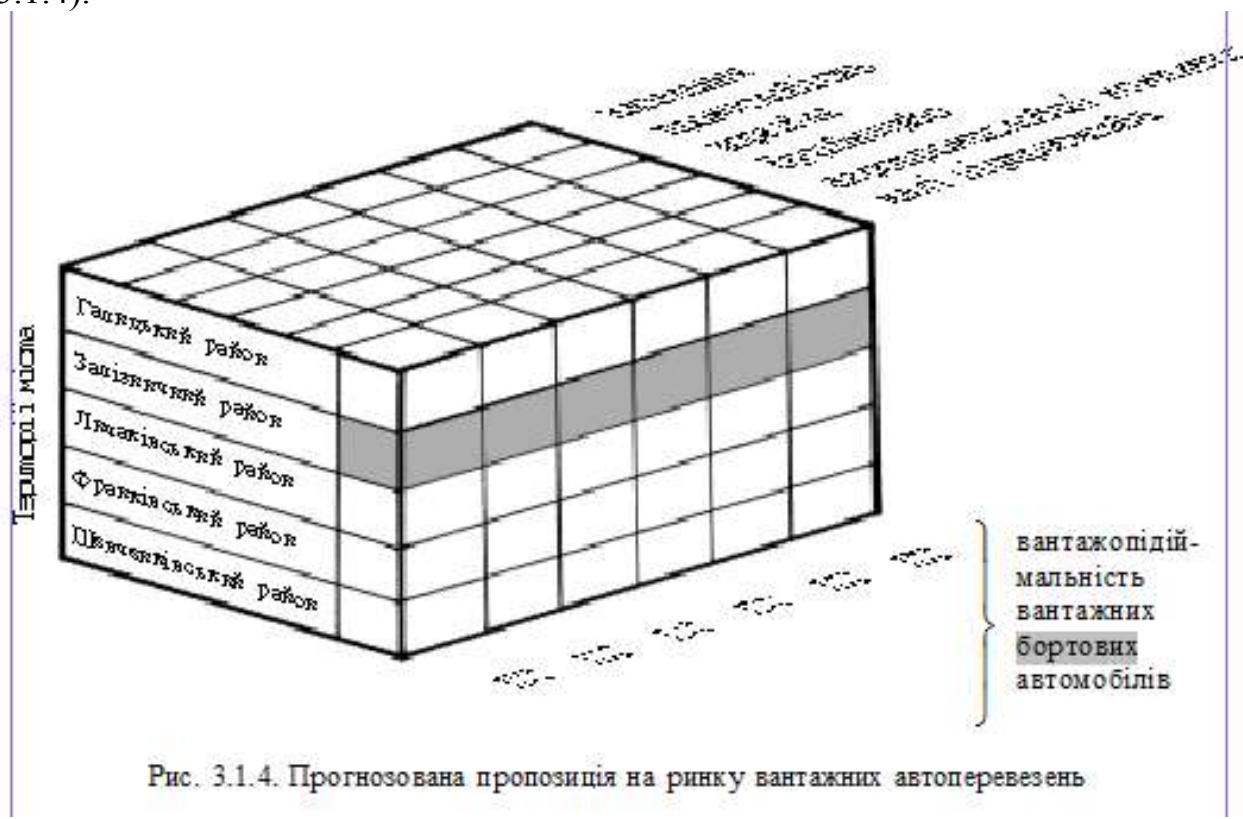

Рис. 3.1.4. Прогнозована пропозиція на ринку вантажних автоперевезень

На основі інформації, отриманої в результаті зіставлення попиту і пропозиції на ринку автотранспортних послуг, розробляється відповідне рішення щодо поведінки автогосподарства на тому чи іншому ринку або його окремому сеґменті.

Користуючись запропонованою схемою сеґментації ринку вантажних автоперевезень, можна встановити зв'язок між окремими територіями, галузями в розрізі попиту і пропозиції як на території відправника, так і на території отримувача. А тому наступна «оцінка попиту має важливе значення для аналізу можливостей ринку, планування маркетингових зусиль і аналізу їхньої ефективності» [189, с. 96].

У практиці діяльності підприємств виділяють методи оцінки поточного та методи оцінки прогнозованого попиту на транспортні послуги. Аналіз поточного попиту слід здійснювати методом оцінки загальної місткості ринку або методом оцінки територіальної місткості ринку.

Загальна місткість ринку вантажних автоперевезень – це максимально можливий обсяг надання транспортних послуг автогосподарствами всіх форм власності протягом визначеного періоду часу при певному рівневі маркетингової діяльності з урахуванням тенденцій розвитку ринку.

Застосування методу оцінки територіальної місткості ринку обумовлюється, як правило, діяльністю автогосподарства, пов'язаною не з одним, а з кількома територіальними ринками. Цей метод передбачає виявлення потенційних споживачів транспортних послуг на кожному ринку, підсумовування обсягів надання транспортних послуг та вирішення на цій основі проблеми оптимального розподілу маркетингових витрат.

Рівень досконалості маркетингової стратегії вантажного АТП значною мірою залежить від обґрунтованості вибору та реалізації методів щодо прогнозування попиту на транспортні послуги. На сьогодні відомо чимало методів прогнозування попиту, які можна об'єднати в кілька груп.

Перша група охоплює методи, за допомогою яких знаходиться зміна величини попиту відповідно до емпіричних даних без їх особливої обробки і попереднього вивчення загальної тенденції попиту (метод простої середньої, метод ланцюгових індексів, метод Персонса, метод відносних чисел тощо). До цієї ж групи методів слід зарахувати алгоритм розрахунку попиту і пропозиції на ринку транспортних послуг за галузями виробництва стосовно територій, де працює конкретний виробник транспортної продукції [див.: 160, с. 11 – 12]. Визначення місткості сегмента ринку і частки виробника послуг та його конкурентів на ньому здійснюється за допомогою алгоритму, попередньо окресливши території цього сеґменту і знаючи, до яких галузей належать споживачі автотранспортних послуг. Алгоритм розрахунку базується на тому, що потреба в транспортних послугах окремої галузі є похідною від виробництва і залежить від кількості працівників галузі, їхньої продуктивності праці і частки вантажів, що перевозяться в галузі, стосовно вантажів, перевезених автотранспортом України за попередній період. Знаючи попит на автоперевезення по окремих галузях, можна визначити потребу в перевезеннях вантажів усіх споживачів конкретного сеґменту. Вибір автопідприємством окремого чи кількох сеґментів ринку залежить від таких чинників:

– географічного (територіального) розташування виробника транспортних послуг;
– чисельності галузевих споживачів послуг, що розташовані на цій же території;
– структури рухомого складу, яким оперує виробник послуг;
– власних виробничих можливостей в порівнянні з конкурентами.

Другу групу методів прогнозування попиту в автомобільних перевезеннях вантажів утворюють ті, в яких спочатку з емпіричних даних виключається загальна тенденція розвитку часового ряду (тренду), а потім усуваються сезонні коливання (метод розрахунку еволюції сезонної хвилі з використанням ряду Фур'є, метод ітерації, методи аналітичного та комбінованого виявлення загальної тенденції внутрішніх коливань і ін.).

Особливої уваги заслуговує метод прогнозування величини попиту різних видів вантажних автоперевезень, в основі якого – арифметичне складання двох функцій [див.: 191, с. 87 –89].

Суть цього методу полягає в попередньому розкладанні динамічного ряду на дві складові, одна з яких $f(x)_1$ відображає загальний розвиток виробництва і є многочленом $n$-го порядку, що розраховується за даними попередніх років з допомогою арифметичних операцій, а друга складова $f(x)_2$ – тригонометрична функція сезонних коливань. У загальному виразі вона матиме такий вигляд:

$$f(x)_2 = \pm R \cos nx, \text{ або } f(x)_2 = \pm R \sin nx, \qquad (3.1.1)$$

що залежить від того, в якому кварталі спостерігався екстремум функції ($R$ – коефіцієнт амплітуди коливань; $n$ – коефіцієнт періоду коливання). Тому загальна функція матиме вигляд:

$$F(x) = f(x)_1 + f(x)_2 \qquad (3.1.2)$$

Для задавання функції загального розвитку виробництва $f(x)_1$ найбільш прийнятним є многочлен першого та другого порядку. Причому многочлен другого порядку посилює дію факторів у період кризи (наприклад, різкого спаду виробництва).

У цілому прогнозування попиту за цим методом передбачає здійснення кількох етапів:

1) виявлення загальної тенденції розвитку виробництва, яке проводиться методом змінної середньої і має вигляд (на основі квартального поділу):

$$y_t = \frac{y_{t-2} + y_{t-1} + y_{t+1} + y_{t+2}}{5} \qquad (3.1.3)$$

2) вивчення функції тренду на основі отриманих даних попередньо виконаного етапу:   $f(x)_1 = ax + b,$   або   $f(x)_1 = ax^2 + bx + c,$   (3.1.4) де $a, b, c,$ – коефіцієнти спаду чи зростання виробництва;

3) вибір функції сезонних коливань, який залежить від того, на який квартал (місяць) припадає максимум чи мінімум. Обчислення здійснюються за формулами:

$$f(x)_2 = - R \sin x, \qquad (3.1.5)$$

де

$$R = \frac{\sum |y_i - \bar{y}_i|}{n}, \qquad (3.1.6)$$

де $y_i$ та $\bar{y}_i$ – значення функції динамічного ряду в точках екстремумів та значення тренду; $n$ – кількість точок екстремумів;

4) визначення многочлена динамічного ряду на основі формули (3.1.2);

5) прогнозування величини попиту, що здійснюється простою математичною підстановкою відповідних значень, що нас цікавлять, у многочлен динамічного ряду.

Використання наведених методів у прогнозуванні попиту на вантажні автомобільні перевезення в сучасних умовах є необхідністю у розробці автогосподарством конкурентної маркетингової пропозиції. Реалії сьогодення спонукають виробника транспортних послуг постійно стежити за динамікою попиту і запитів у сегментах ринку та, відповідно, вчасно вносити корективи у виробничу діяльність.

Успішна реалізація маркетингової пропозиції, а особливо у довгостроковій перспективі, перебуває у прямій залежності від реалізації комплексу заходів щодо управління виробничими ресурсами автотранспортних підприємств [див., напр., 105; 112]. Найбільшою мірою це стосується парку вантажних автомобілів як головної складової основних виробничих фондів. У цьому плані важливим є забезпечення адекватності маркетингових рішень вибору підприємством технічної номенклатури рухомого складу з врахуванням перспективного розмаїття типів конструкцій і параметрів, які, з одного боку, орієнтовані на максимальне задоволення потреб споживачів, а з другого – в реалізації політики енерґо- ресурсозбереження в процесі оновлення рухомого складу.

Сучасні маркетингові методи та методології дозволяють здійснювати системний аналіз параметрів конструкції вантажного автомобіля, його типів, а також відповідності принципам енерґо- ресурсозбереження [див., напр., 239; 240]. Такий підхід, як засвідчує досвід зарубіжних країн, «є передумовою успіху не тільки технічної, але й фінансової політики автопідприємств» [239, с. 4].

Діяльність маркетингово орієнтованих автогосподарств не обмежується лише заходами, спрямованими на вивчення ринку, прогнозування та формування пропозиції транспортних послуг. Не менш важливими є вибір каналів розподілу, формування системи маркетингових комунікацій та стимулювання збуту транспортної продукції [див., напр., 53; 117; 189].

Будь-яка система каналів розподілу вантажного автогосподарства формується під впливом місцевих можливостей і конкретних умов, в яких здійснюються виробничі функції. Їх формування розпочинається з визначення цілей, що можуть бути досягнуті на конкретному цільовому ринку шляхом вирішення завдань щодо досягнення наміченого рівня транспортного обслуговування споживачів, виконання посередниками необхідних функцій і ін. Виробник послуг, розробляючи цілі, узгоджує їх з тими обмеженнями, які зумовлені вимогами споживачів, особливостями та характером послуг, посередниками (напр., транспортними агентствами), конкурентами, політикою підприємства, зовнішніми чинниками.

Формування системи маркетингових комунікацій має на меті комплекс заходів для постійного інформування споживачів стосовно пропонованих пакетів послуг. Важливе значення в цьому процесі відводиться рекламі як засобу цілеспрямованого інформативного впливу опосередкованого характеру на споживача для просування пакету транспортних послуг на ринок вантажних автоперевезень. Без ефективної реклами, розробленої «з урахуванням економічної, психологічної і соціальної складових» [106, с. 6], неможливе формування широких ринків транспортних послуг, перетворення потенційних потреб у попит. Мета рекламної діяльності випливає із стратегії розвитку автопідприємства, його маркетингової політики.

Незадовільний стан маркетингової діяльності у вантажних автотранспортних підприємствах Івано-Франківської, Львівської та Тернопільської областей підтверджується аналогічною ситуацією у сфері рекламування власної транспортної продукції. Зрозуміло, кризовий стан економіки країни та породжені ним проблеми, негативний вплив яких особливо відчутний на мікрорівні, спричинюється до того, що питанням маркетингу не надається належна увага через нагальність у вирішенні інших фінансових, організаційних, виробничих проблем. Проте, як стверджує світовий досвід, перспективи економічної стабілізації та наступного розвитку досягають, власне, маркетингово орієнтовані підприємства. «Без маркетингу не може стати продуктивною ніяка інша праця» [8, с. 3].

Сучасний етап розвитку економічних відносин вимагає розглядати підприємство у постійному взаємозв'язку з постачальниками виробничих ресурсів і споживачами готового продукту. В умовах загострення конкуренції успіх будь-якого підприємства залежить від швидкості реагування на постійні зміни зовнішньої інфраструктури, пристосовуючись до них «за рахунок відповідної зміни своєї внутрішньої структури та парадигми економічної поведінки» [210, с. 62]. Ось чому підприємство повинне володіти такими механізмами управління, які забезпечували б адаптацію до ринкових умов з притаманною їм конкуренцією. Забезпечення конкурентоспроможності автогосподарства повинне бути «наслідком цілеспрямованої політики довгострокового характеру» [30, с. 48].

Реалізація цього означає, що саме споживач повинен визначати напрям розвитку підприємства. У зв'язку з цим з'являється необхідність формування на підприємствах таких організаційно-економічних умов, що забезпечують його ефективне функціонування. Організаційно-економічні умови, що передбачають використання сукупності методів і алгоритмів управління підприємством у ринковому середовищі, повинні забезпечувати досягнення цілей функціонування, узагальнення яких для всіх підприємств і виробничих систем може бути сформульоване у вигляді тези: «Створити споживача».

Формування кола постійних споживачів є основою завоювання підприємством стійкого становища на ринку виробників транспортних послуг, наслідком і причиною ефективного використання основних виробничих фондів.

## 3.2. Реалізація логістичного підходу в системі засобів ефективного використання основних фондів вантажних автопідприємств

Сучасний етап переходу до ринкових відносин характеризується корінними змінами у взаємовідносинах між власниками вантажного автотранспорту і обслуговуваною клієнтурою. Стихійний ринок товарів спричинює непередбачуваність у характері вантажопотоків, зумовлює непродуктивне використання транспортних засобів, перевантаженість мережі автомобільних доріг, погіршення екологічного стану навколишнього середовища.

Сьогодні в Україні перевезення вантажів здійснюють майже мільйоном вантажних автомобілів різних форм власності, з яких 23,5 % державних, 20,0 % приватних, 56,5 % належать до колективної форми власності [див.: 96, с. 10]. Такий значний за чисельністю парк вантажних автомобілів повною мірою задовольняє потреби суспільства і виробництва України у вантажних перевезеннях. Більше того, через невеликий попит на перевезення автогосподарства мають надмірні провізні можливості, які використовуються недостатньо. Так, у 1996 р. «незатребуваність перевізних можливостей цього виду транспорту стала причиною низького рівня його використання та складного фінансового становища: при кількісному зменшенні вантажного парку автотранспортних підприємств на чверть, експлуатувався лише кожний п'ятий автомобіль, потенційні можливості якого реалізовані наполовину» [217, с. 8]. Якщо у 1991 р. кількість вантажних автомобілів перевищувала потреби народного господарства на 20 %, то сьогодні майже 80 % рухомого складу не має роботи. У Львівській області – 76, Тернопільській – 82 %. А поки що втрати власників вантажів і транспортних підприємств через невикористані можливості обчислюються значними сумами коштів. Отже, «сучасний етап функціонування транспортної системи характеризується кризовим станом і розглядається як перехідний до ринкових відносин. За багатьма ознаками він нагадує енерґетичну кризу, яка мала місце в першій половині 70-х років у капіталістичних країнах. Ця криза дала потужний поштовх удосконаленню техніки та технологій перевезень в Західній Європі і США. Виходові з такої ситуації сприяло використання логістики при плануванні, розподілі та управлінні перевезеннями і їх забезпеченням» [97, с. 4].

Як відомо, основним законом в ринкових умовах є відповідність кількості і якості автомобільного парку попитові у перевезеннях. Якщо попит на перевезення зростає, то парк рухомого складу автопідприємств повинен відповідно збільшуватись за чисельністю та змінюватись за структурою і навпаки. Якщо ж така зміна не відбувається, то це обтяжує і транспортну систему, і народне господарство держави.

«Транспорт, як ніяка інша галузь, відображає стан економіки країни в цілому. В нових умовах виробництва власник товару повинен сам пропонувати його, а не чекати попиту. Головне при цьому – якість товару, в нашому випадку – якісне автотранспортне обслуговування.

В умовах ринку оптимізація співвідношення між протилежними вимогами – максимальним виконанням вимог ринку при дотриманні низьких загальних витрат – вирішується питаннями логістики. Головне транспортне завдання логістики – створення необхідних умов для раціонального використання автотранспорту, організації перевезень, що дозволили б з мінімальними затратами доставляти вантажі в необхідній кількості у заданий пункт призначення в обумовлений строк» [96, с. 10 – 11].

Перша невдала спроба з використанням елементів логістики на українських теренах мала місце у середині 80-х років, коли на автотранспорті впроваджувались комплексні системи управління якістю перевезень та ефективним використанням ресурсів [див.: 226]. Вони передбачали виконання взаємопов'язаних організаційних, технічних, економічних, соціальних та ідеологічних засобів, спрямованих на забезпечення належного рівня обслуговування споживачів і ефективного використання рухомого складу. При цьому теоретично була реалізована частина основних принципів побудови логістичних систем управління, але не враховувалась оцінка кінцевих результатів роботи транспорту, що спричинило розрив між техніко-експлуатаційними та економічними показниками. До того ж не були вирішені питання організаційних структур управління перевезеннями. Так, в Івано-Франківську, Львові, Тернополі і інших великих містах управління перевезеннями здійснювалось як транспортними підприємствами, так і місцевими органами влади, що порушувало головний логістичний принцип його єдності. Недосконалість та статичність інформаційного забезпечення пояснювались не тільки відсутністю мобільних обчислювальних комплексів, а й усередненням показників оцінки якості. Вплив цих недоліків неґативно позначався на результативності прийнятих рішень та ефективності функціонування транспорту, а тому комплексні системи у свій час не зазнали подальшого розвитку і припинили своє існування.

 У сучасних умовах формування логістичного підходу з метою забезпечення ефективного функціонування транспортного комплексу України зумовлюється сумою взаємопов'язаних причин макро- та мікроекономічного рівнів. Розширення міжнародних зв'язків України, інтеґрація її в європейські економічні структури «вимагає розвитку загальної інфрастуктури, що охоплює такі її галузі як: комуні-каційні мережі і системи; енерґетику; охорону навколишнього середовища; науку і технологію. Названі галузі безпосередньо стосуються створення і руху матеріальних потоків між просторово розмежованими виробниками і споживачами» [134, с. 61].

 Однією із концепцій, що пов'язана з управлінням матеріальними потоками, є саме логістична концепція, що отримує в сучасних умовах все більше визнання і здобуває нові виміри. На думку експертів, до 2000-го року понад 70 % підприємств та фірм організовуватимуть свою діяльність на основі логістичної концепції. Все частіше порушується питання, що стосуються єврологістики, міжнародної логістики і логістичних суперструктур, макрологістики. У зв'язку з цим нові функції і структури отримують підприємства і організації транспорту як найбільш причетні до формування і руху матеріальних потоків.

Застосування логістики полегшує прийняття ефективних рішень на основі комплексного підходу до всіх ланок виробничо-транспортного процесу. Перш за все це стосується перевезень, вантажно-розвантажувального процесу та пристосування інформації до процесу управління і прийняття рішень [див.: 257, s. 129 – 130]. Так, на початку 90-х років у результаті «впровадження розробленої Держ­автотрансдіпроект і ГІОЦ ВО «Укртрансінформсистема» Єдиної системи плану­вання і управління транспортом» [112, с. 9] (що розглядається як новий етап формування макрологістичної системи України) вдалося досягнути підвищення ефективності транспортного процесу. Зокрема, позитивний вплив логістичного підходу при автоперевезеннях сипких вантажів виявився у значному зростанні прибутковості перевезень, коефіцієнта використання пробігу, продуктивності автомобіля і ін. (див. рис. 3.2.1).

Реалізація вантажними автопідприємствами логістичного підходу може здійснюватись двома способами: 1) як окрема складова ланка ланцюга зовнішньої логістичної системи (на рівні промислового підприємства, регіону, галузі або на національному чи міждержавному рівнях); 2) як завершена мікрологістична система на рівні управління автогосподарством.

Необхідність розробки логістичної системи на мікрорівні, як правило, зумов­люється задіяністю автогосподарства у зовнішніх логістичних системах, але необов'язково. В обох випадках реалізація логістичного підходу автопід­приємством має на меті підвищення результативності виробничої діяльності через ефективне використання в першу чергу основних виробничих фондів, зокрема рухомого складу як їх найважливішої складової.

Мікрологістична система управління охоплює транспортне обслуговування як один з елементів виробничої діяльності, до функцій якої входить планування транспортного обслуговування клієнтів та витрат на його виконання, створення необхідних умов і контроль вантажних перевезень. У загальному вигляді структура мікрологістичної системи подається на рис. 3.2.2.

Серед завдань, що вирішуються мікрологістичною системою і складають

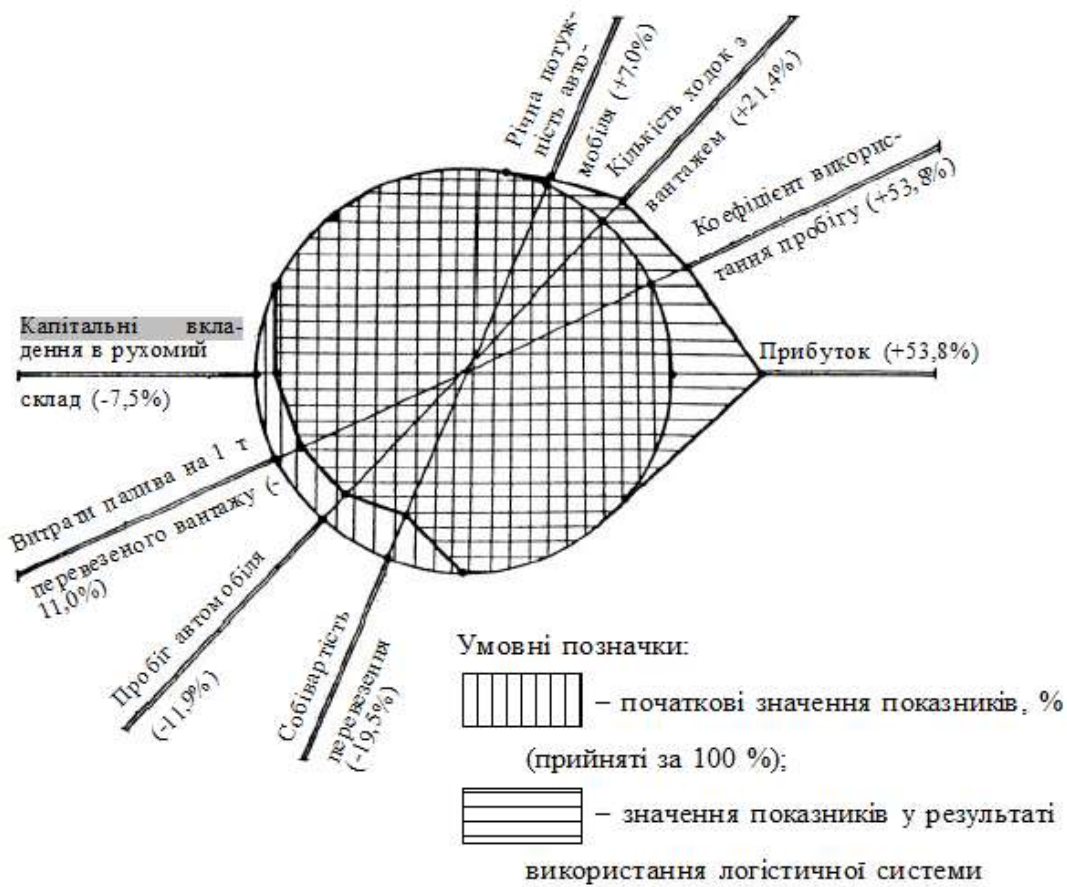

Рис. 3.2.1. Вплив логістичного підходу на техніко-економічні показники перевезень сипких вантажів [див.: 112, с.10]

основу розробки стратегії вантажного автопідприємства, можна виділити три групи:

І. Завдання, що пов'язані з формуванням ринкових зон транспортного обслуговування, прогнозом матеріалопотоку, обробкою матеріалопотоку обслуговуючою системою (склад постачальника чи споживача і т.п.) та іншими роботами, що стосуються оперативного регулювання і управління матеріалопотоком.

ІІ. Завдання, що вимагають розробки системи організації транспортного процесу (план перевезень, план розподілу видів діяльності, план формування вантажопотоків, графік руху транспортних засобів і ін.).

ІІІ. Завдання, що виникають з необхідності управління запасами на підприємствах, складських комплексах; розміщення запасів і їх обслуговування транспортними засобами та інформаційними системами.

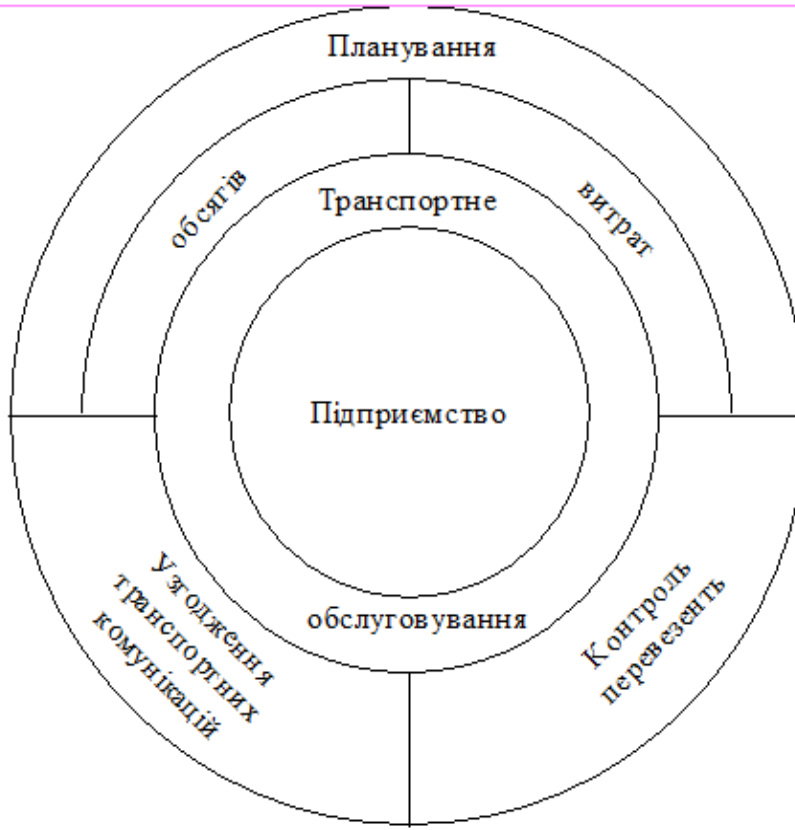

Рис. 3.2.2. Структура мікрологістичної системи

Оптимізація та вирішення цих завдань залежать від конкретної ситуації, вихідних даних, умов і вимог до ефективної роботи логістичної системи, обґрунтуванням яких служить «розроблення стратегії і логістичної концепції побудови моделі транспортного обслуговування споживачів і фірм, що базується на раціональних маршрутах перевезень і складанні графіків доставки продукції споживачам» [145, с. 108]. Специфіка кожного завдання вимагає використання відповідних методів їх вирішення. Наприклад, завдання щодо оптимізації як тривалості руху матеріалопотоку, так і загальних витрат, зумовлених організацією і рухом матеріальних потоків, можуть успішно вирішуватись постановкою задач математичного програмування. Серед них: транспортна задача в класичному варіанті, транспортна задача з додатковими обмеженнями, транспортна задача з обмеженнями пропускної здатності комунікацій, задача про призначення (задача вибору), багатоіндексні транспортні задачі, транспортні задачі у вигляді сітьових графіків, задача про максимальний потік, задача про найкоротший шлях, нелінійна транспортна задача, розподільча задача [див.: 194, с. 100]. Чіткість і грамотність постановки транспорної задачі чи її систем є надзвичайно важливим питанням, оскільки від достовірності результатів розв'язку залежить ефективність використання парку рухомого складу автопідприємств.

Основні положення логістики, що характерні для підприємств виробників і споживачів продукції (пріоритет споживача, високий рівень сервісу, скорочення часу на виконання замовлення і т.п.), повною мірою стосуються і автотранспортних підприємств. У сучасних умовах конкуренції на ринку автотранспортних послуг все частіше їх діяльність характеризується політикою комплексного підходу до вирішення транспортних і суміжних з ними проблем на іншому, дедалі вищому якісному рівні. Практика досліджуваних автогосподарств засвідчує, що така політика приносить позитивні результати, якщо вона достатньо диференційована і базується на таких її складових, як надання нетрадиційних нових додаткових послуг, політика в сфері комунікацій, політика укладання контрактів.

Політика вантажних автопідприємств щодо надання різних додаткових видів послуг об'єднує всі рішення і відповідні їм дії, що спрямовані на комплексне здійснення транспортного процесу [див., напр., 132]. Це означає, що організація перевезень вантажів з урахуванням відстані їх транспортування, кількості і строків доставки планується у поєднанні з додатковими послугами і потребами попиту. Зокрема, «за кордоном у розвинених країнах в доходах від транспортних послуг частка роботи від перевезень складає 40 %, інші припадають на експедиційні операції, зберігання» [63, с. 24] тощо. В Україні ж питома вага доходів від експедиційної діяльності не перевищує 5 %, у підприємствах досліджуваного регіону вона складає 5,7 %.

Політика вантажних автогосподарств, спрямована на розширену диверсифікацію своєї діяльності, дозволить підвищити потенціал залучення клієнтури, збільшити прибутковість, прискорити впровадження найновіших транспортних технологій, закріпити своє становище на ринку транспортних послуг. Це пояснюється тим, що підприємства-виробники не менш зацікавлені у тому, щоб звільнитися від багатьох додаткових логістичних функцій і зосередити свою увагу на основній профілюючій діяльності з метою зниження витрат і підвищення гнучкості в роботі. Останнє є особливо важливою умовою для тих підприємств, ринки яких охоплюють значні території, що в разі несвоєчасної переорієнтації товарних потоків у відповідь на коливання попиту може завдати значних втрат. Так, наприклад, згідно з даними американського журналу «Traffic Management» «із 350 обстежених підприємств різних галузей економіки США 70 % передали транспортним фірмам функції щодо виконання і оформлення розрахунків за перевезення вантажів. Приблизно 20 – 22 % підприємств відмовились на користь транспортників від роботи, пов'язаної з визначенням ціни за перевезення, складськими операціями і вибором оптимального маршруту доставки товарів. Фірми-перевізники вважали вигідним перекласти на себе виконання контрольних функцій за вантажами, що знаходяться в дорозі. Вони почали також займатись організацією електронного обміну даними між всіма учасниками логістичного процесу і зберіганням інформації. Як результат приблизно 12 – 15 % фірм-продуцентів звільнились від виконання вказаних функцій. І, врешті, від 7 до 11 % фірм, що обстежувались, передали перевізникам функції щодо здійснення контролю за товарно-матеріальними запасами, виконання замовлень і експлуатації парку транспортних засобів» [127, с. 284].

Політика вантажних автогосподарств у сфері комунікацій спрямована на інформування клієнтів щодо пропонованих пакетів послуг, а також здійснення необхідного впливу на клієнтуру з метою залучення її до одержання якомога більших обсягів послуг. Другою метою цієї політики є сприяння розширенню і удосконаленню взаємодії транспортних підприємств і вантажовідправників на основі використання обчислювальної техніки і, головним чином, за допомогою електронного обміну даними.

Інтенсифікація господарських зв'язків між вантажними автотранспортними підприємствами і іншими учасниками логістичного процесу об'єктивно зумовила зростання попиту на інформацію і одночасно ускладнила обмін нею. Ось чому в Україні «на всіх видах транспорту повинна бути впроваджена система інформації, зорієнтована на ринковий механізм взаємодії підприємств транспорту, споживачів їх послуг і наскрізну систему обробки даних на всіх рівнях управління» [83, с. 23], що базується на використанні сучасної комп'ютерної техніки.

Досвід розвинених країн засвідчує наявність прямого зв'язку між економічним рівнем розвитку та темпами інформатизації суспільства як вияв і врахування «первинного значення інформаційно-цільового чинника в процесах організації соціально-економічної діяльності» [17, с. 70]. На сьогодні в департаментах, об'єднаннях і підприємствах транспорту України «створені і функціонують автоматизовані системи різного рівня і призначення, що використовуються для обробки і надання інформації, необхідної в управлінні технологічними процесами і виробничо-господарською діяльністю підприємств і організацій» [120, с. 2]. Проте у зв'язку з їх низькою ефективністю в проект Національної програми інформатизації були включені завдання по створенню галузевих систем: інформаційно-аналітичної та інформаційно-довідкової. «Згідно із завданнями Програми було введено в дію першу чергу галузевої інформаційно-аналітичної системи (ІАС – ТРАНСПОРТ) і фрагмент галузевої інформаційно-довідкової системи (ІДС – ТРАНСПОРТ)» [там само]. Однак сподівання на їх результативність у сучасних умовах є незначними. У сфері вантажних автоперевезень використання цих систем з метою ефективної реалізації логістичних підходів в управлінських процесах надзвичайно обмежені у зв'язку з незадовільно низьким рівнем забезпечення первинних ланок транспортного процесу (автогосподарств) сучасними засобами комп'ютерної техніки. Так, питома вага обчислювальної техніки (далеко не завжди в нових її зразках) у структурі досліджуваних вантажних автотранспортних підприємств Івано-Франківської, Львівської та Тернопільської областей на кінець 1996 р. становила 0,54 %. Найвище її значення у Тернопільській області – 0,7 %, найнижче – у Львівській – 0,5 % (див. табл. 1.2.2). Незначна частка цієї групи основних виробничих фондів є однією з комплексу причин низького рівня комп'ютеризації транспортного процесу.

Низький рівень комп'ютеризації, а відповідно, і інформатизації управлінських рішень автогосподарств регіону зумовлюється і тим, що лише 31 % з них володіє засобами обчислювальної техніки. У цьому зв'язку вкрай критична ситуація спостерігається в автопідприємствах споживчої кооперації, серед яких лише 2 % господарств, до того ж лише в Тернопільській області, використовують обчислювальну техніку. Її питома вага в структурі основних виробничих фондів досліджуваних автопідприємств Західного регіону становить 0,0081 %.

Рівень комп'ютеризації транспортних фірм у США на кінець 80-х років складав 26,8 % і до 1996 р. передбачалось його підвищення до 35,8 % [див.: 127, с. 287].

Ефективність процесу транспортування і забезпечення ґарантованого транспортного обслуговування підприємств народного господарства України може бути досягнуте через налагодження належного рівня інформатизації логістичної діяльності на всіх рівнях управління ним, «у їх якісній зміні, зокрема у впровадженні інформаційних систем нового покоління» [214, с. 170].

Узагальнюючи, слід зазначити, що побудова логістичних систем вантажних перевезень повинна базуватись на таких принципах:

– комплексність розгляду елементів логістичної системи від зародження попиту на перевезення до його задоволення;

– обґрунтування оптимального рівня обслуговування і визначення способів його досягнення з урахуванням ефективного використання ресурсів;

– забезпечення відповідності провізної здатності рухомого складу до попиту на перевезення;

– оцінка кінцевих результатів роботи за величиною прибутку;

– організація перевезень і суміжного обслуговування клієнтів єдиною структурою, що здатна приймати компетентні рішення, узгоджені з інтересами виробництва;

– достатність інформаційного забезпечення з використанням обчислювальної техніки, що виступає порадником в процесі прийняття рішень;

– забезпечення кадрами, що володіють знаннями та досвідом логістичної діяльності.

У сучасних умовах дотримання принципів логістичного підходу є важливою умовою забезпечення організаційно-економічної стійкості вантажних автопідприємств. Найближчим часом саме розвиток логістики виявлятиме суттєвий вплив на транспортну політику автогосподарств, зумовлюватиме іманентні структурні зміни в характері їх діяльності на основі раціонального використання всіх видів ресурсів: трудових, інформаційних, фінансових, матеріальних, а отже, на формування «ефективного суб'єкта виробничої діяльності в сучасних і доцільних формах» [164, с. 69].

Організаційно-економічна стійкість є важливою характеристикою загальної результативності прийняття та реалізації управлінських рішень, налагодженості і взаємодії всіх ланок транспортного процесу, а також виступає важливим чинником забезпечення конкурентоспроможності підприємства, закріплення і розширення позицій на ринку транспортних послуг.

Отже, логістика як система чіткої організації якісного транспортного обслуговування споживачів і логістика як засіб підвищення ефективності використання виробничих ресурсів автопідприємств, в першу чергу, досягається через реструктуризацію їх основних фондів, зокрема рухомого складу, та реорганізацію підходів до їх використання.

## 3.3. Лізинг у системі вантажних автоперевезень як засіб формування ефективної структури рухомого складу автогосподарств України

Успіхи ринкових перетворень у сфері вантажних автоперевезень неможливі без пожвавлення господарської діяльності, що, в свою чергу, вимагає оновлення матеріальної бази і модернізації основних фондів, підвищення ефективності їх використання автопідприємствами різних форм власності.

На сьогодні більшість автогосподарств використовують вантажні автомобілі, які застаріли як морально, так і фізично, що є «наслідком зменшення інвестиційних можливостей переважної більшості суб'єктів господарювання» [193, с. 77]. До того ж структура наявного парку рухомого складу вантажного автотранспорту України та досліджуваних автогосподарств Івано-Франківської, Львівської і Тернопільської областей далека від оптимальної як за продуктивністю, вантажопідіймальністю, типом кузова, так і за екологічністю, про що йшлося вище у попередніх розділах дослідження.

Через брак необхідної суми власних коштів на рахунках автопідприємств, а також через неможливість отримання середньо- та довгострокових банківських кредитів інвестиційна діяльність не досягла належного рівня. Причиною цього є слабка інвестиційна зорієнтованість економічної політики держави, за якої «вигідним має стати капітал, що бере участь у довготермінових реальних проектах» [67, с. 74].

У зв'язку з тим, що вітчизняні виробничі потужності можуть забезпечувати потреби транспортного комплексу у вантажівках лише на 15 – 17 % [див.: 81, с. 11], сьогодні «оновлення парку рухомого складу шляхом імпорту автомобілів є важливим напрямком довгострокового управління розвитку автотранспорту України» [239, с. 4]. До того ж паливна економічність вантажних транспортних засобів, якими укомплектовані вітчизняні автогосподарства, в 1,5 рази нижча порівняно з автомобілями провідних зарубіжних фірм. Ситуація ускладнюється ще й тим, що при зростанні попиту на міжнародні автоперевезення «наші зарубіжні партнери в Європі докладають максимум зусиль для того, щоб в односторонньому порядку освоїти спільний транспортний ринок, забуваючи про умови паритету» [227, с. 5]. Протекціоністська політика західноєвропейських держав сприяє цьому. Загальний стан рухомого складу, що переважає в автогосподарствах України, «не відповідає європейським вимогам, зокрема за екологічними параметрами» [91, с. 18].

Таке становище так чи інакше орієнтує керівників вантажних автопідприємств України на придбання іноземних автомобілів, які, проте, для більшості вітчизняних підприємств є поки що недоступними. Насамперед – через високу ціну.

Це ставить вітчизняні автогосподарства перед розв'язанням дилеми, що за своєю природою є замкненим колом: 1) необхідність перекомплектації наявного витратного парку автомобілів, що за своєю структурою відповідав би існуючому попиту на автоперевезення; 2) відсутність власних джерел фінансування, однією з причин якої є нераціональна структура застарілого рухомого складу, що зводить нанівець інвестиційну та інноваційну діяльність у сфері вантажних автоперевезень.

Ця ситуація змушує керівників автогосподарств шукати інші шляхи залучення інвестицій. Одним з них є лізинг вантажних автомобілів.

Немає потреби говорити про те, яку роль в економіці відіграють інвестиції. Джерелами коштів при цьому можуть бути або власні грошові засоби підприємств, або позичені. А лізинг якраз і виступає в ролі джерела позичених коштів – комерційного кредиту в натуральній формі.

Лізинг як «важливий інструмент стимулювання оновлення виробництва» [64, с. 27] поширений в країнах Центральної та Західної Європи, Північної Америки. Постійний розвиток форм і методів організації лізингу в розвинених країнах засвідчує результативність притаманних йому іманентних переваг. Важливість лізингу та його ролі в процесах європейської інтеграції засвідчує завершення роботи над «створенням інформаційної системи для забезпечення вільного руху орендованих машин і устаткування в рамках єдиного спільноєвропейського ринку» [40, с. 51].

Серед об'єктів, що використовуються в лізингових операціях, значна частка припадає на вантажні автомобілі (див.: табл. 3.3.1). Так, у Болгарії в 1994 р. 71,4 % лізингових операцій припадало на вантажівки, в Іспанії – 32,1 %, Люксембурзі – 31,5 %, Франції – 26,2 % (що забезпечує їй перше місце в Європі за абсолютними обсягами – 2324 млн. екю), Португалії – 22,6 %, Угорщині, Словенії, Словаччині – по 20,0 %, Нідерландах – 19,0 %, Великобританії – 12,1 %. В середньому на лізинг вантажних автомобілів серед 23 країн Європи припадало 13,8 % від загального обсягу лізингових операцій [див.: 190, с. 38 –39]. У США ця частка становить 86,7 % [140, с. 37].

Обсяги лізингу транспортних засобів в Україні значно скромніші, що є прямим наслідком несформованості вітчизняного ринку лізингових послуг, на якому існує всього біля 20 різних лізингових компаній «та й ті функціонують нерідко лише формально» [94, с. 5; див. також: 245, с.4]. І причина цього криється навіть не у відсутності до грудня 1997 р. Закону України «Про лізинг». Бо, наприклад, у Німеччині та США немає спеціальних законів, що регламентують лізингові угоди [див.: 216, с. 38-39; 231, с.14], хоча частка лізингу в загальній сумі інвестицій цих країн складає відповідно 15,8 і біля 30 % [див.: 139, с.9; 190, с.36].

Нерозвиненість лізингового ринку в Україні пояснюється радше внутріеконо-

**Таблиця 3.3.1**

Частка лізингових операцій, що припадають
на вантажні автомобілі у країнах Європи*

| Країна | Частка лізингових операцій, що припадає на вантажні автомобілі |
|---|---|
| Австрія | 11,2 |
| Бельгія | 15,7 |
| Болгарія | 71,4 |
| Великобританія | 12,1 |
| Греція | 10,5 |
| Данія | 18,7 |
| Ірландія | 11,0 |
| Іспанія | 32,1 |
| Італія | 16,4 |
| Люксембург | 31,5 |
| Нідерланди | 19,0 |
| Німеччина | 6,8 |
| Норвегія | 14,9 |
| Португалія | 22,6 |
| Словаччина | 20,0 |
| Словенія | 20,0 |
| Угорщина | 20,0 |
| Фінляндія | 5,2 |
| Франція | 26,2 |
| Чехія | 9,2 |
| Швейцарія | 4,6 |
| Швеція | 6,0 |
| В середньому по 22 країнах Європи | 13,8 |

* Побудовано на основі даних: 190, с. 38 – 39.

мічними причинами, законодавчою неузгодженістю та «браком кваліфікованих спеціалістів, що володіють тонкощами організації лізингових відносин» [254, с. 91], відсутністю заохочувальних чинників, в т.ч. і податкових пільг для суб'єктів лізингу. Згідно з висновками, яких дійшли експерти Bank of Ireland International Services Ltd., у результаті досліджень розвитку лізингу в Україні (що здійснювались від імені Світового банку), реальні умови в нашій державі на даний час є несприятливими, лізинг не функціонує належним чином і найближчі перспективи такого функціонування не видаються втішними. Адже обсяг лізингових операцій в Україні складає біля 100 млн. американських доларів, з яких тільки 10 % фінансуються за рахунок українського капіталу [див.: 7, с. 15].

Недивно, що на ринку лізингу вантажних автомобілів України панівне становище посідають лізингові компанії провідних іноземних автомобілебудівних фірм, в основному європейських. З 1991 по 1997 рр. їх кількість зросла від однієї до восьми, а лізингова діяльність здійснюється як прямими представництвами, так і фірмами-торговими агентами виробників. Сьогодні інтерлізинг вантажівок в Україні представлений такими марками автомобілів: французькою «Renault», голландською «DAF», італійською «Iveco», німецькою «Mercedes-Benz», шведськими «Scania» і «Volvo», чеською «Tatra», американською «International». За словами представника компанії DAF Trucks в Україні А. Кузнєцова, «рівень технології і якості у всіх виробників практично однаковий. Представництва конкурують у якості обслуговування і умовах укладання контрактів. Успіх тієї чи іншої марки визначається умовами лізингу. Перевізник, що обирає фірму для укладання лізингового контракту, сьогодні орієнтується перш за все на умови фінансування» [249, с. 84].

У наведеній нижче таблиці подаються основні відомості, що стосуються особливостей надання у лізинг вантажних автомобілів іноземного виробництва.

У міру насичення українського ринку лізинговими послугами та підвищення вимог до сервісу відчутно зростає різноманітність послуг, що надаються лізинговими компаніями. В першу чергу це виявляється у комплексності лізингових угод. Лізингові компанії, здаючи в оренду вантажівки, часто перебирають на себе функції не тільки щодо ремонту та страхування, але й доставки орендованих машин, надання маркетингових і постачальницько-збутових послуг. Це дозволяє лізинговим компаніям зберігати високі темпи росту при відносному зниженні рівня лізингових ставок.

Так, окрім розглянутих в таблиці умов надання в лізинг вантажних автомобілів, кожна з фірм намагається зацікавити українських споживачів специфікою сервісного обслуговування, наданням додаткових послуг. Представництва концернів Daimler-Benz AG та Renault, наприклад, надають послуги з експедування вантажів, забезпечують візовою підтримкою. До того ж за рахунок концерну Renault здійснюється навчання водіїв на французькому полігоні. Конкуренція між

**Таблиця 3.3.2**

Особливості імпортного лізингу в Україні*

| Фірма-виробник автомобілів та її лізингова компанія в Україні | Рік заснування діяльності в Україні | Схема лізингу | Строки лізингу нових автомобілів, років | Авансова плата від суми контракту, % | Строк першого внеску після отримання автомобіля в лізинг, міс. | Річна процентна лізингова ставка, % | Гарантія на автомобіль, років | Особливі умови |
|---|---|---|---|---|---|---|---|---|
| Концерн Renault, фірма "С.Т.Р. Рено В.И. Україна" | 1991 | без банківської гарантії | 4 | 15 | 3 – 6 | 14 від лізингової плати | 1 | обов'язкове КАСКО-страхування |
| Компанія DAF Trucks N.V, представництво агентської фірми Traxim B.V. | 1993 | без банківської гарантії, якщо сума контракту менша 1,8 млн. німецьких марок | 4 | 20 | 3 | 7,75 від суми контракту | 2 | обов'язкове КАСКО-страхування |
| Фірма Scania, представництво "Сканія кредит АБ" | 1993 | гнучка | 4 | 20 | 3 | 6,61 від суми контракту | 1 | індивідуальні скидки залежно від обсягів лізингу та представлених гарантій і ін. |
| Концерн Iveco, пряме представництво концерну Iveco | 1995 | гнучка | 4 | 15 – 20 | 3 | 15,5 від залишкової вартості | 1 | сплата 5% при розміщенні замовлення |
| Концерн Volvo, пряме представництво заводу Volvo | 1997 | гнучка | 5 | 15 – 25 | -- | 4,5 – 6 від суми контракту | 1 | індивідуальні скидки залежно від суми контракту, наявності банківської гарантії та ін. |
| Концерн Daimler-Benz AG, пряме представництво Daimler-Benz | 1997 | гнучка | 3,5 – 4,5 | 20 | 6 | 7 – 12 від залишкової суми | 2 | індивідуальні скидки, що залежать від строків лізингу, обсягів контракту, гарантій |
| Фірма Navistar-фірма MBL Motors Inc., компанія MBL Tecnology | – | без банківської гарантії | 3 – 4 | 15 – 30 | - | 13 – 16 від лізингової плати | 1 | індивідуальні скидки, що залежать від обсягів лізингу |

* Побудовано на основі даних: 68, с. 13; 249, с. 84 – 85.

фірмами точиться і в сфері технічного обслуговування автомобілів. Нині в Україні найкраще розвинена мережа станцій технічного обслуговування фірм Tatra,

Scania, концернів Iveco, Daimler-Benz. Американська фірма MBL Tecnology допомагає перевізникам в пошуку кредитних ресурсів.

Лізингові компанії, що представлені сьогодні на українському ринку вантажних автомобілів, поряд з інвестуванням найбільш перспективних автогосподарств виконують роль своєрідного індикатора попиту і пропозиції, до певної міри реґулюючи ці процеси.

Разом з тим, розрізнені дії кожного з представників інтерлізингу в Україні швидше нагадують «розвідку боєм», аніж узгоджені дії єдиної команди. Скидається на те, що континентальні національні лізингові об'єднання (такі як міжнародна асоціація Leaseurope – Європейське федеративне об'єднання компаній з лізингу устаткування, в складі якого 25 країн і понад 1200 лізингових компаній; спеціалізована асоціація ECATRA – Європейська асоціація орендодавачів легкових та вантажних автомобілів та ін.) не ризикують виходити на український ринок, залишаючись в ар'єрґарді і очікуючи завершення «чорнової» роботи щодо формування в Україні ринку лізингових послуг, здійснюваних поодинокими лізинговими фірмами. Як не дивно, але саме міжнародні лізингові організації покликані надавати допомогу у вирішенні правових, податкових, фінансових та інших проблем, уреґульовувати питання, пов'язані з практикою лізингових операцій в рамках процесу інтеґрації європейських країн. Очевидно, на заваді цьому стає, по-перше, нерозвиненість вітчизняного лізингового ринку. Головна ознака – відсутність до недавнього часу української національної асоціації лізингових компаній, яка виступала б посередником між іноземними лізинговими організаціями та Урядом України. Лише в кінці квітня 1997 р. відбулась установча конференція асоціації «Укрлізинг». По-друге, «кризовий стан української економіки, високий рівень невизначеності щодо перспектив та напрямів реформування господарської системи, нестабільність законодавства, обмеженість валютних коштів у більшості українських підприємств, а також відсутність будь-яких ґарантій щодо недоторканості приватної власності стримують європейський капітал від активного виходу на потенційно дуже місткий український ринок» [29, с. 33].

Недосконалість реґіональної політики в сфері лізингу, імпортного зокрема, утруднюють і затримують розвиток та повноцінне фукціонування в Україні цієї своєрідної інвестиційної галузі підприємницької діяльності. Прийняття під кінець 1997 р. Закону «Про лізинг», покликаного врегулювати правові відносини між суб'єктами лізингових операцій, спричинило й нові недоречності. «Сам закон суперечить цілому букету інших документів. Влада ж настійно не поспішає ані виправляти помилки, ані, навіть, видавати пояснення до прийнятих законодавчих актів» [39, с. 34]. А ті нормативні акти, що з'являються, не дозволяють запрацювати лізинговому механізмові. Незрозумілою залишається позиція Кабінету Міністрів, програмою діяльності якого у транспортному комплексі на 1996 р. передбачалось «запровадити систему лізингової форми придбання рухомого складу» [193, с. 18] і який «не підтримав низки положень, які зменшували рівень фінансових навантажень на учасників лізингових угод. Чого варті хоча б нотаріальні послуги, які сягають 5 % від вартості договору! Закон орієнтовано на стару систему бухобліку з її основоположним принципом амортизації» [94, с. 5]. Або ж інструкція з бухгалтерського обліку орендних операцій зі змінами і доповненнями, затверджена наказом Міністерства фінансів України від 6.03.1998 р. № 50 [див.: 101; 215], «не дає відповіді майже на всі складні та незрозумілі питання, а в окремих випадках навіть не узгоджується з чинним законодавством» [93, с. 126]. У результаті можливості застосування документа в Україні є мінімальними, і навіть ті компанії (як вітчизняні, так і зарубіжні), що до 1997 р. могли розвивати свою лізингову діяльність, сьогодні стоять перед перспективою закриття.

Закон «Про лізинг» суперечить чинному закону «Про оподаткування прибутку підприємств», згідно з яким об'єкти оперативного і фінансового лізингу можуть використовуватись протягом строку, що не перевищує строку їх амортизації [див. 85, ст. 18.1], тобто за суттю для фінансового лізингу нижня межа строку встановлювалась нульовою.

Незамортизовану за угодою фінансового лізингу вартість об'єкта оренди згідно з Законом «Про лізинг» необхідно викупити. А це в середньому становить 40 % при виконанні мінімального обсягу лізингу в 60 % вартості [див. 84, ст. 4], що може виявитись надзвичайно складною фінансовою проблемою для лізингоодержувача.

При продовженні строків відповідного контракту можна вести мову про оперативний лізинг. Але при цьому виникає проблема захисту лізингодавця.

Пріоритетність амортизації, що виявляється в поділі лізингу на оперативний і фінансовий згідно із Законом України «Про лізинг», породжує недоречності іншого характеру. За міжнародними стандартами обліку для об'єктів лізингу приймаються економічні строки служби основних засобів, що обумовлюється в угоді відповідно до виробничих умов їх експлуатації. Визначальним при цьому повинне бути поняття ризику, оскільки прийняття лізингоодержувачем на себе всіх ризиків (в т.ч. пов'язаних із власністю), є найважливішою специфічною ознакою фінансового лізингу. На жаль, правова база України не має напрацьованого юридичного трактування поняття економічних ризиків у контексті фінансових відносин [див. 39, с.34].

Останні зміни, внесені до Закону «Про оподаткування прибутку підприємств» від 30.12.1997 р., виявляють негативну дію на розвиток імпортного лізингу в Україні. Згідно з внесеними змінами до відома лізингодавця (нерезидента України) доводилося, що податок на прибуток обчислюється за період, який уже минув; і те, що лізингодавець отримує не 100 % належної йому суми відповідно до лізингової угоди, а лише 85 %. Інші 15 % (податок на дохід нерезидента в Україні) будуть утримані в бюджет. Або інший варіант – заплатити в бюджет за нерезидента 15 % прибутку, що залишився у розпорядженні лізингової компанії після сплати всіх обов'язкових податків та інших витрат.

Складна ситуація виникає при сплаті податку на додану вартість (ПДВ), що ним учасники лізингу обкладаються на загальних підставах [див. 86, ст. 2], але який повинен був би «сплачуватись в момент зміни власника, тобто коли його амортизація сягне (або буде близькою) 100 % » [43, с. 34].

Не позбавлене законодавчої неузгодженості надання послуг з імпортного лізингу автотранспортних засобів. По-перше, згідно з Законом «Про податок на додану вартість» лізингові операції відносяться до об'єктів оподаткування, а Указ про критичний імпорт виділяє транспортні засоби серед категорії товарів, що не обкладаються ПДВ. Щоправда, в цьому документі не уточнюється, що автотранспортні засоби можуть бути об'єктом лізингу. По-друге, залишається неврегульованим питання про те, що виступає об'єктом оподаткування імпортного лізингу. Згідно із Законом України «Про податок на додану вартість» – лізингова плата, що, в свою чергу, порушує норми інших законодавчих актів України, які звільняють імпорт послуг від сплати ПДВ.

Неврегульованість правового поля України в сфері імпортного лізингу виявляється і в особливостях митного законодавства. Так, при оформленні об'єктів лізингу лізингоодержувач, який юридично не є власником отримуваного майна, змушений виплачувати значні суми ще не зароблених грошей у вигляді митного збору. А це – 10 % від митної вартості. За найскормнішими підрахунками сума митного збору складає понад 20 тис. грн. за одну вантажну машину. У Молдавії, наприклад, якщо транспортні засоби чи якесь інше устаткування отримують у лізинг, лізингоодержувача звільняють від усіх митних зборів [див. 68, с. 13].

Перепоною в реалізації міжнародних лізингових проектів в Україні є проблема сублізингу, що, на жаль, не знайшло свого відображення в Законі «Про лізинг». На офіційному рівні пояснення цього зводиться до мотивації ціни на лізингові послуги, що за умови сублізингу можуть значно зрости. А в результаті міжнародний лізинговий бізнес не може ефективно впроваджуватись безпосередньо через угоди з українськими лізинговими компаніями, які опісля самі могли б виступати суб'єктами лізингу в ролі лізингодавця. Поширення такої схеми лізингу на початковому етапі його становлення цілком виправдане з огляду на те, що «лізингові компанії, які виникли на пострадянській території, малочисельні, слабкі і не володіють в повній мірі технікою лізингування» [39, с. 34].

Позитивного вирішення проблеми сублізингу очікують і вітчизняні лізингові компанії, що працюють на українському ринку. Особливо гостро це стосується вантажного автотранспорту, оскільки лізингоодержувач інколи (за відсутності замовлень) вимушений іти на сублізинг, щоб уникнути збитків.

Нерозуміння специфіки лізингу при прийнятті відповідних нормативних документів проявляється в елементарних речах. Так, зокрема часто звертають увагу на подорожчання об'єкта лізингу в порівнянні з ціною купівлі. При цьому в якості аргумента виступає загальна сума лізингової плати. Але сумування здійснюється без врахування фактора часу. Тільки при ставці 22 % річних по кредитуванню даної операції відбудеться збільшення початкової вартості об'єкта лізингу. В Постанові Кабінету Міністрів України говориться про те, що вартість кредитних ресурсів лізингового фонду складає 50 % облікової ставки Національного банку України. Рішення приймалося, коли її величина становила 16 % і при наступному її підвищенні вартість кредитування не може бути пропорційно зміненою, оскільки її верхня межа сьогодні становить 22 % (50 % від 44 %).

Поступ України до цивілізованого ринку лізингових послуг ускладнюється мінливістю і нестабільністю вітчизняного законодавства. Багато міжнародних лізингових проектів «вступають у конфлікт з українським законодавством, яке змінюється значно швидше, ніж завершується строк дії контракту» [198, с. 11]. Однак, незважаючи на всю суму означених причин, що, безумовно, мають характер неґації, очевидною є потреба якнайшвидшого їх подолання, бо лізинг є одним із безперечних і ефективних засобів оздоровлення економіки країни. Стосовно України орендні відносини в системі євролізингу мають характер нових інтеграційних зв'язків, спрямованих на пожвавлення і розвиток економічних взаємовідносин України та Європейського Союзу. «За оцінками європейських дослідників лізингової діяльності ріст промислового виробництва, наприклад, на 5,7 % за рік корелював з 7 % ростом обсягу лізингових операцій» [48, с. 35].

Домінування іноземних фірм у сфері лізингу вантажівок на ринку транспортних засобів України в окремих моментах негативно впливає на формування аналогічних вітчизняних лізингових компаній, а отже, існує нагальна потреба у якомога швидшому їх становленні. Перешкодою на цьому шляху є те, що українські банки без особливого ентузіазму дивляться як на створення власних, так і фінансову підтримку позабанківських лізингових компаній. У зв'язку з цим цілком прийнятним бачиться створення таких організаційних структур при автомобілебудівних заводах України. В першу чергу це стосується Кременчуцького автозаводу, що випускає автомобілі марки «КрАЗ». Тим більше, що «вже є конкретна домовленість про виробництво спільного автомобіля компаній IVECO та КрАЗ. Такі ж наміри щодо КрАЗу має й італійська фірма FIAT» [138, с. 12]. Лізингові компанії також могли б бути створені на базі тих українських підприємств (наприклад, «Крим-автоГАЗсервіс», «Чернігівавто-деталь» і ін.), що складають вантажні автомобілі з комплектуючих Горьківського автомобільного заводу [див.: 47, с. 37].

Організаційного завершення потребує робота над створенням при корпорації «Укравтотранс» республіканської лізингової компанії з філіалами в областях, засновником якої виступила Українська біржа.

Розвиток республіканської мережі лізингових компаній є важливою складовою транспортного процесу, без якої в сучасних умовах неможливо сформувати завершений цикл на ринку автотранспортних послуг. Ось чому роль держави повинна проявлятися у «селективному сприянні інвестиціям (національним і зарубіжним) передусім у реальний сектор економіки з метою його ефективної реструктуризації у процесі ринкових перетворень» [67, с. 79]; на мікрорівні пріоритетним має стати «формування стійкої інвестиційної мотивації підприємств – і приватизованих, і державних» [там само].

Переваги ж, які отримують вітчизняні автопідприємства-рентери від лізингу вантажівок, виявляють позитивний вплив в усіх сферах господарської діяльності.

Використовуючи лізинг, споживач відкриває для себе не тільки нове джерело інвестування, але й нові можливості. Для вантажних автопідприємств – це, по-перше, диверсифікація умов і джерел фінансування: лізинг дозволяє уникнути залучення кредитів з інших джерел фінансування, послаблюючи залежність суб'єктів інвестування від банків, отримуючи додаткову свободу маневру; дозволяє отримати значно більший обсяг інвестицій, що для підприємств часто може мати принципове значення.

По-друге, повне фінансування угоди з поступовою і гнучкою формою оплати вартості об'єкту лізингу. Лізинг передбачає 100 % кредитування угоди з боку лізингодавця, і, як правило, не потребує негайного початку платежів; тоді як при банківському кредитуванні відбувається в основному лише часткове покриття вартості необхідного для споживача майна, інша частина вартості оплачується за рахунок власних коштів підприємств. Порядок здійснення лізингових платежів гнучкіший, ніж за кредитними угодами, «оскільки суми лізингових платежів, методи їх нарахування, способи, форми та терміни сплати визначають на договірній основі» [80, с. 29], що дозволяє оптимально поєднати інтереси всіх сторін. До того ж порівняно з кредитуванням лізинг передбачає триваліші строки виплати, що відповідає періоду експлуатації інвестованого майна. Загалом лізинг, на відміну від кредиту, дає змогу створити автопідприємствам надійніші умови господарювання.

По-третє, лізинг доступний малим і середнім підприємствам (які, власне, і складають основну масу господарств вантажного автотранспортного комплексу України), в той час як отримання банківських кредитів на сприятливих умовах для них є досить проблематичним; часто лізингові компанії не вимагають від лізинго-одержувача ніяких додаткових ґарантій, оскільки забезпеченням угоди виступає орендоване майно, яке може бути забране лізингодавцем в разі невиконання рентером своїх зобов'язань.

По-четверте, для автогосподарств-рентерів лізинг як особливий вид інвестування забезпечує можливість використання найновіших сучасних зразків вантажних автомобілів та іншої техніки високої вартості без значних одноразових витрат, а також можливість їх придбання за залишковою вартістю після завершення лізингової угоди. Так, наприклад, залишкова вартість вантажівки, отриманої за лізинговою схемою шведського представництва фірми Scania в Україні, становить лише 4 % [див. 249, с. 85].

Лізинг вантажних автомобілів здатний забезпечити безперервні і швидкі темпи оновлення рухомого складу автогосподарств. До того ж орендні зв'язки між лізингодавцем і лізингоодержувачем значно міцніші і стабільніші, ніж зв'язки між продавцем і покупцем. Угоди можуть укладатися на основі постійних лізингових домовленостей. Тому лізингоодержувач отримує більш оперативну реакцію на свої запити. Тимчасово володіючи автотранспортними засобами, рентер вимушений постійно думати про заміну, підшукуючи більш досконалу за якістю і продуктивністю техніку. У зв'язку з цим окремою умовою лізингової угоди може виступати надання лізингоодержувачу можливості обміну попередньо лізингованої техніки на більш сучасні її зразки. Угоди, що передбачають таку можливість, називають «револьверним лізингом», тобто лізингом з автоматичною заміною автотранспортної техніки, що експлуатувалась, новою.

По-п'яте, суттєвою особливістю цієї форми підприємницької діяльності є розподіл функцій власності, а саме – відокремлення використання майна від володіння ним, що згідно з чинним законодавством дозволяє орендарю за відповідних умов уникнути додаткових витрат, пов'язаних із страхуванням майна, технічним обслуговуванням і ремонтом, виплатою відповідних податків та зборів, які лягають на лізингодавця. До того ж «майно за лізинговою угодою не зараховується на баланс лізингоодержувача, що не збільшує його активів і звільняє від сплати податку на майно; його вартість не включається в залишок кредитної заборгованості. Це поліпшує фінансові показники підприємства-орендаря і відповідно дозволяє йому залучити додаткові кредитні ресурси (у зв'язку з цим сучасний лізинг часто класифікують як «позабалансове фінансування»)» [139, с. 14].

Ще одним наслідком розподілу функцій власності, за даними французьких соціологів, є бажання рентера використовувати лізинговане майно інтенсивніше і якісніше, порівняно з власником такого ж майна. Причина тут криється у тому, що лізингоодержувач, враховуючи сплачувані ним лізингові внески, намагається зрівнятися в прибутках з аналогічним власником. Крім цього, лізинг створює у рентера психологічне відчуття тимчасового володіння майном, а отже, стимулює його бажання експлуатувати майно, поки є така можливість. Власник же, розглядаючи майно як таке, що від нього невіддільне, виявляє дещо нижчу господарську активність у його використанні.

По-шосте, лізингові операції знижують вплив фактора економічного ризику, що загострюється в міру становлення ринкових механізмів господарювання; сприяють більш обґрунтованому плануванню розвитку підприємства. Споживач, що уклав лізинговий контракт, має більшу, порівняно з покупцем, свободу дій. Він може повернути лізингове майно, викупити його, обміняти на більш сучасне, доповнити наявне заново лізингованим, здати в суборенду. Всі ці дії рентер має можливість узгодити з реальними виробничими, фінансовими умовами, кон'юнктурою ринку.

По-сьоме, лізинг іноземних вантажівок дозволяє освоювати нові ринки транспортних послуг, в т.ч. і міжнародні, підвищувати свій вплив у цій сфері діяльності.

Аналіз можливостей вдосконалення лізингових угод засвідчує, що для лізингоодержувача їх ефективність може виражатися не тільки в прямій матеріальній вигоді, але й у підвищенні іміджу автогосподарства, в налагодженні цінних контактів і ін.

Належна законодавча підтримка і заохочення лізингу вантажних автомобілів в Україні сприятиме швидкому реформуванню та розвитку системи вантажних автоперевезень як на внутрішньому, так і на зовнішньому ринках, а в загальному підсумку – побудові ринкової інфраструктури транспортного комплексу держави, розширенню її інтеграційних зв'язків.

З народногосподарської точки зору лізинг виступає засобом оновлення та розвитку виробництва, впровадження досягнень науково-технічного прогресу, створення нових робочих місць. Зокрема, у щорічному зверненні Президента України Л.Д. Кучми до народу, Верховної Ради України, виголошеному 12 травня 1998 р., вказувалось на важливість формування інститутів фінансового лізингу для активізації інвестиційного процесу в Україні [див.: 89, с. 3]. Якщо в Україні «запрацює механізм лізингу – можна очікувати бурхливого розвитку малого і середнього бізнесу» [113, с. 12].

Лізинг дозволяє ліквідувати невідповідність між платоспроможним попитом і пропозицією основних засобів, одночасно стабілізуючи розміри процентних ставок за кредити, оскільки при лізингу вони фіксуються на момент укладання угоди.

Поширення лізингу допомогло б «здійснити необхідну структурну перебудову економіки України у бік енерґозбереження, як це відбулося з економікою США після енерґетичної кризи 1974 – 1975 рр.» [140, с. 39].

Значення лізингу як одного з інструментів санації підприємств виявляється у «можливості залучення до процесу виробництва найсучасніших технологій при відсутності не тільки необхідних для здійснення капіталовкладень фінансових ресурсів, а й достатнього для виходу на фінансовий ринок кредитного забезпечення» [225, с. 34], що є особливо важливим в умовах, коли 30 % усіх господарських структур працюють збитково [див., напр., 217].

Позитивний вплив лізингу, що проявляється як на мікро-, так і на макроекономічному рівнях, переконує у необхідності створення сприятливих умов з боку держави у становленні та розвитку в Україні цього виду підприємницької діяльності.

# ЛІТЕРАТУРА

156. Орлов П., Орлов С. Державна амортизаційна політика // Економіка України. – 1998. – № 8. – С. 31 – 36.

157. Осетинський А. Про Закон України «Про лізинг» / Вищий арбітражний суд України, лист № 01 –8/104 від 23.03.1998 // Діло. – 1998. – № 20. – С. 30 – 31.

158. Остапенко В. Инфляция и воспроизводство основных фондов // Вопросы экономики. – 1994. – № 7. – С. 45 – 55.

159. Павловський М. Шлях України: Шлях вліво, шлях вправо – хибний шлях... – К.: Техніка, 1996. – 152 с.

160. Панченко Є.Г., Дубняк О.І. Деякі аспекти маркетингу на автомобільному транспорті // Автошляховик України. – 1993. – № 1. – С. 6 – 12.

161. Пащенко Ю., Давиденко А. Розвиток і модернізація транспортної системи України // Економіка України. – 1993. – № 9. – С. 28 – 34.

162. Петрова Е.В., Алексеева И.М. Статистика автомобильного транспорта: Учебник. – 4-е изд. – М.: Финансы и статистика, 1988. – 215 с.

163. Піндайк Р.С., Рубінфелд Д.Л. Мікроекономіка: Пер. з англ. А. Олійник, Р. Скільський. – К.: Основи, 1996. – 646 с.

164. Плотников О. Економіка сучасної України: виміри деградації // Політична думка. – 1997. – № 3. – С. 59 – 70.

165. Повышение эффективности грузовых автомобильных перевозок: Сборник статей / Отв. ред. А. Баублис. – Вильнюс, 1986. – 179 с.

166. Пода А.К., Вашків О.П., Куц Л.Л. Мікроекономіка. Збірник задач для економічних спеціальностей: Навч. посібник. – К.: ІСДО, 1995. – 123 с.

167. Положення про порядок нарахування амортизаційних відрахувань по основних засобах у народному господарстві // Бухгалтерский учет. – 1991. – № 4. – С. 74 – 78.

168. Положення про Міністерство транспорту України // Зібрання постанов Уряду України. – 1993. – № 6. – Ст. 101 – 102.

169. Положення про порядок застосування норм прискореної амортизації активної частини (машин, устаткування, транспортних засобів) основних виробничих фондів. Затверджене наказом Мінфіну і Мінекономіки України від 6 червня

215. Смєян Ю., Холощак О. Бухгалтерські проводки по операціях з оперативної і фінансової оренди // Галицькі контракти. – 1998. – № 17. – С. 128 – 129.
216. Смирнов А.Л. Лизинговые операции. – М.: Издательство АО «Консалт-банкир», 1995. – 136 с.
217. Соціально-економічне становище України у 1996 році / Повідомлення Міністерства статистики //Урядовий кур'єр. – 1997. – № 26 – 27. – С. 7 – 10.
218. Справочник по организации и планированию грузовых автомобильных перевозок / И.Г. Крамаренко и др.: Под ред. И.Г. Крамаренко. – К.: Техніка, 1991. – 206 с.
219. Справочник по рентабельности эксплуатации грузовых автомобилей. – К.: Техніка, 1980. – 160 с.
220. Статистика: Підручник / А.В. Головач, А.М. Єріна, О.В. Козирєв та ін.; За ред. А.В. Головача, А.М. Єріної, О.В. Козирєва. – К.: Вища школа, 1993. – 623 с.
221. Статистика рынка товаров и услуг: Учебник / Под ред. И.К. Беляевского. – М.: Финансы и статистика, 1995. – 432 с.
222. Супруненко В., Жадан А. Індексація балансової вартості основних фондів підприємств і організацій у 1996 році // Економіка, фінанси, право. – 1996. – № 12. – С. 38 – 42.
223. Суторміна В.М., Федосов В.М., Рязанова Н.С. Фінанси зарубіжних корпорацій. – К.: Либідь, 1993. – 248 с.
224. Сухова Л.Ф. Модели и методы оптимизации размещения грузового автомобильного транспорта. – М.: Транспорт, 1991. – 127 с.
225. Терещенко О. Державне сприяння реструктуризації та санації підприємств // Економіка України. – 1998. – № 2. – С. 27 – 34.
226. Типовой проект по разработке КС УКПП и ЭИР: РД 238 УССР 84001-42-81 Минтранс УССР. – К., 1981. – 218 с.
227. Ткаченко А.Н., Игнатенко А.С., Марунич В.С. Международным перевозкам – законодательную основу // Автошляховик України. – 1997. – № 1. – С. 2 – 5.
228. Транспорт і зв'язок України: Стат. збірник / Державний комітет статистики України; Відповідальний за випуск О.В. Голуб. – К.: Інформаційно-видавничий центр Держкомстату України, 1997. – 150 с.
167

# ЗМІСТ